\documentclass{aa}
  \usepackage{graphicx}
  \usepackage{times}

\graphicspath{{plots/}}

\def\AA{\mbox{\r A}}
\def\keV{{\rm ke\kern-.3ex V}}

\def\ol{\overline}
\def\dnud{\Delta\nu_{\rm D}}
\def\d{{\rm d}}

\def\qn{q_{\rm N}}

\def\q0{q_{0}}
\def\qinf{q_{\infty}}

\def\chil{\overline{\chi}_{\rm l}}
\def\chiross{\chi_{\rm Ross}}
\def\etal{\overline{\eta}_{\rm l}}
\def\taur{\tilde\tau_{\rm Ross}}
\def\tauross{\tau_{\rm Ross}}

\def\Teff{T_{\rm ef\/f}}
\def\logg{\log g}
\def\Msun{M_\odot}
\def\Mdot{\dot M}
\def\vinf{v_\infty}
\def\vturb{v_{\rm t}/v_\infty}
\def\logLx{\log(L_{\rm X}/L_{\rm bol})}
\def\Rrs{R_{\ast}/R_{\odot}}

\def\HI{H\,{\sc i}}
\def\HII{H\,{\sc ii}}
\def\HeI{He\,{\sc i}}
\def\HeII{He\,{\sc ii}}
\def\CIII{C\,{\sc iii}}
\def\CIV{C\,{\sc iv}}
\def\NIII{N\,{\sc iii}}
\def\NIV{N\,{\sc iv}}
\def\NV{N\,{\sc v}}
\def\OIV{O\,{\sc iv}}
\def\OV{O\,{\sc v}}
\def\OVI{O\,{\sc vi}}
\def\NeIII{Ne\,{\sc iii}}
\def\SiIII{Si\,{\sc iii}}
\def\SiIV{Si\,{\sc iv}}
\def\SiV{Si\,{\sc v}}
\def\PV{P\,{\sc v}}
\def\SIV{S\,{\sc iv}}
\def\SV{S\,{\sc v}}
\def\SVI{S\,{\sc vi}}
\def\FeIV{Fe\,{\sc iv}}
\def\FeV{Fe\,{\sc v}}
\def\FeVI{Fe\,{\sc vi}}

\newdimen\mywidth
\newcounter{mycount}

\def\romanlist{
\settowidth\mywidth{--(iii)--}
\begin{list}
  {(\roman{mycount})}
  {\usecounter{mycount}
   \leftmargin\mywidth
   \labelwidth\mywidth
  }
}

\def\arabiclist{
\settowidth\mywidth{--(0)--}
\begin{list}
  {(\arabic{mycount})}
  {\usecounter{mycount}
   \leftmargin\mywidth
   \labelwidth\mywidth
   \topsep0ex
  }
}

\def\mycaption#1{\caption{\footnotesize{#1}}}

\begin{document}
  \thesaurus{07(02.12.1;  
              08.01.3;  
              08.05.1;  
              08.13.2;  
              08.09.2 $\alpha$~Cam;  
              13.25.5)} 

\title{Radiation-driven winds of hot luminous stars}
\subtitle{XIII.\settowidth\mywidth{XIII. }
A description of NLTE line blocking and blanketing\\
\hspace*\mywidth
towards realistic models for expanding atmospheres}

\author{A.W.\,A.\,Pauldrach, T.\,L.\,Hoffmann, M.\,Lennon}

\offprints{A.W.\,A.\,Pauldrach\\
  (http://www.usm.uni-muenchen.de/people/adi/adi.html)}

\institute{
Institut f\"ur Astronomie und Astrophysik der Universit\"at M\"unchen,
Scheinerstra{\ss}e~1,
D-81679~M\"unchen,
Germany}

\date{Received (date) / Accepted (date)}
\titlerunning{NLTE line blocking and blanketing}
\maketitle

  \begin{abstract}

Spectral analysis of hot luminous stars requires adequate model
atmospheres which take into account the effects of NLTE and radiation
driven winds properly.  Here we present significant improvements of
our approach in constructing detailed atmospheric models and synthetic
spectra for hot luminous stars.  Moreover, as we regard our solution
method in its present stage already as a standard procedure, we make
our program package {\em WM-basic} available to the community
(download is possible from the URL given below).

The most important model improvements towards a realistic description
of stationary wind models concern:

\romanlist
\item
A sophisticated and consistent description of line blocking and
blanketing.  Our solution concept to this problem renders the {\em
line blocking influence on the ionizing fluxes} emerging from the
atmospheres of hot stars --- mainly the spectral ranges of the EUV and
the UV are affected --- in identical quality as the {\em synthetic
high resolution spectra} representing the observable region.  In
addition, the line blanketing effect is properly accounted for
in the energy balance.
\item
The atomic data archive which has been improved and enhanced
considerably, providing the basis for a detailed multilevel NLTE
treatment of the metal ions (from~C to~Zn) and an adequate
representation of line blocking and the radiative line acceleration.
\item
A revised inclusion of EUV and X-ray radiation produced by cooling
zones which originate from the simulation of shock heated matter.
\end{list}

This new tool not only provides an easy to use method for O-star
diagnostics, whereby physical constraints on the properties of stellar
winds, stellar parameters, and abundances can be obtained via a
comparison of observed and synthetic spectra, but also allows the
astrophysically important information about the ionizing fluxes of hot
stars to be determined automatically.
Results illustrating this are discussed by means of a basic model grid
calculated for O-stars with solar metallicity.
To further demonstrate the astrophysical potential of our new method
we provide a first detailed spectral diagnostic determination of the
stellar parameters, the wind parameters, and the abundances by an
exemplary application to one of our grid-stars, the O9.5Ia
O-supergiant $\alpha$~Cam.


\end{abstract}

\section{Introduction}

Spectral analyses of hot luminous stars are of growing astrophysical
interest as they provide a unique tool for the determinination of the
properties of young populations in galaxies. This objective, however,
requires spectral observation of individual objects in distant
galaxies.  That this is feasible has already been shown by \cite{st96}
who detected galaxies at high redshifts ($z \sim 3.5$) and found that
the corresponding optical spectra show the typical features usually
found in the UV spectra of hot stars. In order to determine stellar
abundances and physical properties of the most UV-luminous stars in at
least the Local Group galaxies via quantitative UV spectroscopy
another principal difficulty needs to be overcome: the diagnostic
tools and techniques must be provided. This requires the construction
of detailed atmospheric models and synthetic spectra for hot luminous
stars. It is a continuing effort of several groups to develop a
standard code for solving this problem. (Basic papers of the different
groups are \cite{}Hillier and Miller~(1998), Schaerer and
de~Koter~(1997) and Pauldrach et al.~(1994, 1994a, 1998).)

The most important output of this kind of model calculation are the
ionizing fluxes and synthetic spectra emitted by the atmospheres of
hot stars. As these spectra consist of hundreds of not only strong,
but also weak wind-contaminated spectral lines which form the basis of
a quantitative analysis, and as the energy distribution from hot stars
is also used as input for the analysis of emission line spectra
(e.\,g., of gaseous nebulae) which depend sensitively on the structure
of the emergent stellar flux, a sophisticated and well tested method
is required to produce these data sets accurately. However, this is
not an easy task, since modelling hot star atmospheres involves
replicating a tightly interwoven mesh of physical processes: the
equations of radiation hydrodynamics including the energy equation,
the rate equations for all important ions (from H to Zn) including the
atomic physics, and the radiative transfer equation at all transition
frequencies have to be solved simultaneously.

The most complicating effect in this system is the overlap of
thousands of spectral lines of different ions. Especially concerning
this latter point we have made significant progress in developing a
fast numerical method which accounts for the blocking and blanketing
influence of all metal lines in the entire sub- and supersonically
{\it expanding atmosphere}. As we have found from previous model
calculations that the behaviour of most of the spectral lines depends
critically on a detailed and consistent description of {\em line
blocking and line blanketing}, (cf.~\cite{}Pauldrach et al.,~1994;
Sellmaier et al.,~1996; Taresch et al.,~1997; Haser et al.,~1997),
special emphasis has been given to the correct treatment of the
Doppler-shifted line radiation transport, the corresponding coupling
with the radiative rates in the rate equations, and the energy
conservation.

In Section~\ref{sec:nltemodel} we will demonstrate that the realistic
and consistent description of line blocking and blanketing and the
involved modifications to the models lead to changes in the energy
distributions, ionizing continua, and line spectra with much better
agreement with the observed spectra when compared to previous, not
completely consistent models. This will obviously have important
repercussions for quantitative analysis of hot star spectra.

In the next two sections we will first summarize the general concept
of our procedure and then discuss the current status of our treatment
of hydrodynamical expanding atmospheres.

  \section{The general method}
\label{sec:general}

The basis of our approach in constructing detailed atmospheric models
for hot luminous stars is the concept of {\em homogeneous, stationary,
and spherically symmetric radiation driven winds}, where the expansion
of the atmosphere is due to scattering and absorption of
Doppler-shifted metal lines (\cite{lus70}). In contrast to previous
papers of this series (cf.~\cite{}Pauldrach et al.,~(1994, 1994a,
1998) and papers referenced therein) the above approximations are now
the most significant ones for the present approach --- others of
similar importance have meanwhile been dropped (see below). These
approximations are, however, quite restrictive, since only the
time-averaged mean of the observed spectral features can be described
correctly by our method. Nevertheless we are confident that it is
reasonable to continue with the stationary, spherically symmetric
approach and to improve its inherent physics since the detailed
comparison with the observations, which is the only way to demonstrate
the reliability of this concept, leads to promising results
(cf.~Section~4).

Before we describe the latest improvements in detail we first
summarize the principal features of our procedure of simulating the
atmospheres of hot stars. (For particular points a comprehensive
discussion is also found in the papers cited above.)

\begin{figure*}
\vspace{0.5cm}
%
%

\unitlength1mm
\normalsize

\begin{picture}(150,220)(-10,-5)

\thinlines
\put(40,168){\begin{picture}(90,45)
  \put(2,45){\makebox(0,0)[l]{\bf Hydrodynamics}}
  \put( 2,18){\framebox(50,20){} }
    \put( 6,34){\makebox(0,0)[l]{line force: $g_{\rm L}(k,\alpha,\delta)$}}
    \put( 6,28){\makebox(0,0)[l]{continuum force: $g_{\rm C}(r)$}}
    \put( 6,22){\makebox(0,0)[l]{temperature: $T(r)$}}
    \put(60,34){\makebox(0,0)[l]{[$k_0$, $\alpha_0$, $\delta_0$]} }
    \put(60,28){\makebox(0,0)[l]{[Thomson force]} }
    \put(60,22){\makebox(0,0)[l]{[$T(r)=\Teff$]} }
  \put( 2, 0){\framebox(30,10){$\rho(r)$, $v(r)$} }
  \thicklines\put(50, 0){\framebox(30,10){$\Mdot$, $\vinf$} }
  \thicklines\put(17,18){\line(0,-1){7}}\put(17,10){\vector(0,-1){0}}
  \put(40,5){\makebox(0,0){$\Longrightarrow$}}
  \end{picture} }

\put(40,113){\begin{picture}(90,45)
  \put(2,42){\makebox(0,0)[l]{\bf Spherical grey model}}
  \put( 2,21){\framebox(50,14){} }
    \put( 10,31){\makebox(0,0)[l]{$\rho(r)$, $v(r)$}}
    \put( 6,25){\makebox(0,0)[l]{LTE continuum opacities}}
    \put(60,25){\makebox(0,0)[l]{[LTE]} }
  \put( 2, 0){\framebox(30,10){$T_{\rm g}(r)$, $g_{\rm C}(r)$} }
  \thicklines\put(50, 0){\framebox(38,10){} }
  \put(52, 7.5){\makebox(0,0)[l]{LTE continuum force} }
  \put(52, 2.5){\makebox(0,0)[l]{LTE temperature} }
  \thicklines\put(17,21){\line(0,-1){10}}\put(17,10){\vector(0,-1){0}}
  \put(40,5){\makebox(0,0){$\Longrightarrow$}}
  \end{picture} }

\thicklines \put(40, 55.5){\framebox(90,41){}} \thinlines
\put(40,58){\begin{picture}(90,45)
  \put(2,42){\makebox(0,0)[l]{\bf Spherical NLTE model}}
  \put( 2,21){\framebox(30,14){} }
    \put( 17,31){\makebox(0,0){$Z$, $\rho(r)$, $v(r)$, $T_{\rm g}(r)$}}
    \put( 17,25){\makebox(0,0){$R_{ij}$, $C_{ij}$, $\chi_\nu$, $\eta_\nu$}}
  \put( 2, 0.5){\framebox(30,15){} }
    \put( 17,11){\makebox(0,0){$n_i(r)$, $H_\nu(r)$, $T(r)$}}
    \put( 17,5){\makebox(0,0){$g_{\rm L}(r)$, $g_{\rm C}(r)$}}
  \thicklines\put(50, 0.5){\framebox(38,15){} }
  \put(52,13){\makebox(0,0)[l]{line force} }
  \put(52, 8){\makebox(0,0)[l]{continuum force} }
  \put(52, 3){\makebox(0,0)[l]{temperature} }
  \thicklines\put(17,21){\line(0,-1){4.5}}\put(17,15.5){\vector(0,-1){0}}
  \put(40,8){\makebox(0,0){$\Longrightarrow$}}
  \end{picture} }

\put(40, 8){\begin{picture}(90,41)
  \put(2,39){\makebox(0,0)[l]{\bf Spectrum (formal solution)}}
  \put( 2,22){\framebox(30,10){} }
    \put( 17,27){\makebox(0,0){$n_i(r)$, $\chi_\nu$, $\eta_\nu$, $v(r)$}}
  \put( 2, 3){\framebox(30,10){$H_\nu^{\rm em}$} }
  \thicklines\put(17,22){\line(0,-1){8}}\put(17,13){\vector(0,-1){0}}
  \thicklines\put(50, 3){\framebox(38,10){} }
  \put(52,10){\makebox(0,0)[l]{line and continuum} }
  \put(52, 5.2){\makebox(0,0)[l]{\quad\quad spectrum} }
  \put(40,8){\makebox(0,0){$\Longrightarrow$}}
  \end{picture} }

\thicklines
\put(36,134.0){\oval(40,137)[l]}
\put(36,65.5){\line(1,0){3}}
\put(36,202.5){\line(1,0){2}}
\put(39,202.5){\vector(1,0){0}}
\put(36,157.0){\oval(30,78)[l]}
\put(36,118.0){\line(1,0){3}}
\put(36,196.0){\line(1,0){2}}
\put(39,196.0){\vector(1,0){0}}

\end{picture}
\mycaption{
Schematic sketch of a complete model atmosphere calculation.
Starting procedures are presented in brackets.
For a discussion see the text.}
\label{fig:d2}
\end{figure*}

Fig.~\ref{fig:d2} gives an overview of the physics to be treated in
various iteration cycles. A complete model atmosphere calculation
consists of three main blocks,
\romanlist
\item {\em the solution of the hydrodynamics}
\item {\em the solution of the NLTE-model (calculation of the
  radiation field and the occupation numbers)}
\item {\em the computation of the synthetic spectrum}
\end{list}
which interact with each other.

In the first step the {\bf hydrodynamics} is solved in dependence of
the stellar parameters (effective temperature $\Teff$, surface gravity
$\logg$, stellar radius $R_*$ (defined at a Rosseland optical depth of
$2/3$), and abundances $Z$ (in units of the corresponding solar
values)) and of pre-specified force multiplier parameters ($k$,
$\alpha$, $\delta$), which are used for describing the radiative line
acceleration. In addition, the continuum force is approximated by the
Thomson force, and a constant temperature structure ($T(r)=\Teff$) is
assumed in this step. In a second step the hydrodynamics is solved by
iterating the complete continuum force $g_{\rm C}(r)$ (which includes
the opacities of all important ions) and the temperature structure
(both are calculated using a spherical grey model), and the density
$\rho(r)$ and the velocity structure $v(r)$ in a pre-iteration cycle.
In a final outer iteration cycle these structures are iterated again
together with the line force $g_{\rm L}(r)$ obtained from the
spherical NLTE model.  (New force multiplier parameters, which are
depth dependent if required, are deduced from this calculation.)
\footnote{ This latter step is currently not available for the
download version of the code; it will be made available for version
2.0.}

The main part of the code consists of the solution of the {\bf
NLTE-model}.  In this step the radiation field (represented by the
Eddington-flux $H_\nu(r)$ and the mean intesity $J_\nu(r)$), the final
temperature structure, occupation numbers $n_i(r)$, and opacities
$\chi_\nu$ and emissivities $\eta_\nu$ are computed using detailed
atomic models for all important ions. For the solution of the
radiative transfer equation the influence of the spectral lines
(i.\,e., the UV and EUV {\it line blocking}) is properly taken into
account in addition to the usual consideration of continuum opacities
and source functions which consist of Thomson-scattering and free-free
and bound-free contributions of all important ions.  Moreover, the
{\it shock source functions} produced by radiative cooling zones which
originate from a revised simulation of shock heated matter are
included also.  For the calculation of {\it the final NLTE
temperature structure} the {\it line blanketing} effect, which is a
direct consequence of line blocking, is considered by demanding
luminosity conservation and the balance of microscopic heating and
cooling rates. The {\it rate equations} which yield the occupation
numbers contain collisional ($C_{ij}$) and radiative ($R_{ij}$)
transition rates, as well as low-temperature dielectronic
recombination and Auger-ionization due to K-shell absorption
(considered for C, N, O, Ne, Mg, Si, and S) of soft X-ray radiation
arising from shock-heated matter. (Further details concerning the
solution method of the NLTE-model are described in
Section~\ref{sec:nltemodel}.)

The last step consists of the computation of the {\bf synthetic
spectrum}. In dependence of the occupation numbers, the opacities and
the emissivities, a complete synthetic spectrum is computed using a
formal integral solution of the transfer equation in the observer's
frame. To accurately account for the variation of the line opacities
and emissivities due to the Doppler shift, the calculation is performed
on a properly adapted spatial microgrid which effectively resolves
individual line profiles. All lines which, through their Doppler
shift, can contribute to a given frequency point are considered.
(cf.~\cite{}Puls and Pauldrach~(1990), Pauldrach et al.~(1996)).

As results of the iterative solution of this system of equations we
obtain not only the {\em synthetic spectra and ionizing fluxes} which
can be used in order to determine stellar parameters and abundances,
but also the hydrodynamical structure of the wind (thus, constraints
for the {\it mass-loss rate} $\dot M$ and the {\it velocity structure}
$v(r)$ can be derived).

  \section{The consistent NLTE model}
\label{sec:nltemodel}

The construction of realistic models for expanding atmospheres
requires a correct and {\em extremely consistent} description of the
main part of the simulation, the {\em NLTE model}, which, in addition,
should not suffer from drastic approximations. From our continuing
effort to come to a reasonable approach of a solution to this problem
it turned out that the most crucial point in our present treatment is
an exact description of line blocking and blanketing.

The effect of line blocking --- mainly acting in between the \HeII\
and the \HI\ edge --- is that it influences the ionization and
excitation and the momentum transfer of the radiation field
significantly. This of course has important consequences for both the
spectral line formation and the dynamics of the expanding atmosphere.
Nevertheless, it is still not a common procedure to treat the line
opacities and emissivities in the radiative transfer equation and
their back-reaction on the occupation numbers via the radiative rates
{\em correctly}. We will therefore first discuss the effects of line
blocking and blanketing for {\em expanding atmospheres of hot stars}
in more detail.

The huge number of metal lines present in hot stars in the EUV and UV
attenuate the radiation in these frequency ranges drastically by
radiative absorption and scattering processes (an effect known as {\em
line blocking}). Only a small fraction of the radiation is re-emitted
and scattered in the outward direction; most of the energy is radiated
back to the surface of the star producing there a {\em backwarming}.
Due to the increase of the Rosseland optical depth ($\tau_{\rm Ross}$)
resulting from the opacities enhanced by the line blocking, and, in
consequence, of the temperature, the radiation is redistributed to
lower energies (this refers to {\em line blanketing}). In principle
these effects influence the NLTE model with respect to:
\romanlist
\item the radiative photoionization rates $R_{ik}$,
\item the radiative bound-bound rates $R_{ij}$,
\item the radiation pressure $g_{\rm rad}$,
\item the energy balance.
\end{list}
The terms of the first two items are directly connected to the
radiation field, and line blocking in general reduces them
considerably. Concerning the third item, the blocked incident
radiation reduces the radiative acceleration term in the inner part,
whereas it can be enhanced in the outer part due to multiple
scattering processes (cf.\cite{}~Puls (1987) and references therein).
In contrast to this, the energy equation --- last item --- is mostly
influenced by the impact of the line opacities, and this {\em
blanketing effect} results in an increased temperature (steeper
gradient) in the deeper layers of the photosphere.

Although the method for treating blanketing effects is well
established for cold stars, where the atmospheres are hydrostatic and
where the assumption of LTE is justified (cf.\cite{}~Kurucz, 1979 and
1992), the work to develop an adequate method for hot stars, where not
only NLTE effects are prominent, but where the atmospheres are also
rapidly expanding, is still under way. In this --- our --- case, in
addition to the four items given above, the solution of the radiative
transfer also has to account for the {\it lineshift} caused by the
Doppler effect due to the velocity field. The important effect of this
point is that the velocity field increases the frequential range which
can be blocked by a single line (see below). In the presence of a
velocity field the blocking effect is therefore more pronounced.

Concerning the basic requirements for calculating adequate line
opacities and source functions for expanding atmospheres of hot stars
we have to concentrate on the following points:

\arabiclist
\item
{\em consistent} NLTE occupation numbers,
\item a complete and accurate line list in connection with
detailed atomic models,
\item
a proper concept for treating the line blocking with due
regard to the lineshifts in the wind, in the course of which the method
for solving the complete radiative transfer including the spectral lines
has to be efficient with regard to computational time,
\item
a correct treatment of the influence of the blanketing effect
on the temperature structure,
\item
an adequate approximation of the EUV and X-ray radiation produced
by cooling zones of shock-heated matter.
\end{list}

\subsection{The concept of the solution of ionization and excitation}

It is obvious that ionization and excitation plays the major role in
calculating the emergent flux and spectrum of a hot star. Therefore, a
consistent and accurate description of the occupation numbers is extremely
important for a realistic solution of the NLTE model.

Figure~\ref{fig-d3} presents a sketch of our iteration scheme for the
calculation of the occupation numbers.

\begin{figure*}
\vspace{0.5cm}
\begin{center}
%
%

\unitlength1mm
\normalsize
\begin{picture}(160,205)
\thicklines
\put(60,195){\framebox(40,10){$Z$, $\rho(r)$, $v(r)$, $T_{\rm g}(r)$}}
\put(80,195){\line(0,-1){8}}
\put(80,185){\vector(0,-1){0}}
\put(0,20){\framebox(160,165){}}
\put(80,20){\line(0,-1){8}}
\put(80,10){\vector(0,-1){0}}
\put(60,0){\framebox(40,10){$n_i(r)$, $H_\nu(r)$, $T(r)$}}
\put(0,187){\bf Spherical NLTE model}
\put(70,160){\framebox(80,10){Solution of the rate equations for all
ions and levels}}
\put(30,160){\makebox(40,10){$n_i(r)$}}
\put(50,160){\line(0,-1){8}}
\put(50,150){\vector(0,-1){0}}
\put(70,130){\framebox(80,10){Radiative transfer}}
\put(71,146){Method I:~  opacity sampling for lines}
\put(71,142){Method II:~ continuum values and blocking factors}
\put(71,124){Method II:~ $J=J_{\rm cont}\cdot B_J$,~ $H=H_{\rm cont}\cdot B_H$}
\put(30,140){\makebox(40,10){$\eta(r,\nu)$, $\chi(r,\nu)$}}
\put(50,140){\line(0,-1){8}}
\put(50,130){\vector(0,-1){0}}
\put(30,120){\makebox(40,10){$J(r,\nu)$, $H(r,\nu)$}}
\put(50,120){\line(0,-1){8}}
\put(50,110){\vector(0,-1){0}}
\put(70,100){\framebox(80,10){Calculation of the temperature}}
\put(30,100){\makebox(40,10){$T(r)$}}
\put(50,100){\line(0,-1){18}}
\put(50,80){\vector(0,-1){0}}
\put(70,80){\framebox(80,10){Detailed radiative transfer (method II)}}
\put(30,70){\makebox(40,10){$I(p,z(\Delta\tau),\nu)$}}
\put(50,70){\line(0,-1){8}}
\put(50,60){\vector(0,-1){0}}
\put(30,50){\makebox(40,10){$J(r,\nu)$, $H(r,\nu)$}}
\put(50,50){\line(0,-1){8}}
\put(50,40){\vector(0,-1){0}}
\put(70,40){\framebox(80,10){Calculation of blocking factors (method II)}}
\put(70,30){\makebox(80,10){$B_J=J/J_{\rm cont}$,~ $B_H=H/H_{\rm cont}$}}
\put(30,30){\makebox(40,10){$B_J(r,\nu)$, $B_H(r,\nu)$}}
\put(30,148){\oval(20,86)[lb]}
\put(30,105){\oval(20,140)[lb]}
\put(20,150){\vector(0,1){0}}
\put(20,107){\vector(0,1){0}}
\put(30,105){\line(1,0){10}}
\put(0,150){\makebox(40,10){$R_{ij}$, $R_{ik}$, $C_{ij}$, $C_{ik}$}}
\put(40,160){\oval(40,10)[lt]}
\put(42,165){\vector(1,0){0}}
\end{picture}
\end{center}
\vspace{0.5cm}
\mycaption{
  Iteration scheme for the calculation of the NLTE occupation numbers.
  An {\em Accelerated Lambda Iteration} procedure is involved in the
  blocking and blanketing cycles. Note that two successive iteration
  cycles (I,II) of different quality are applied for the blocking and
  blanketing part of the model calculation (see text).}
\label{fig-d3}
\end{figure*}

In dependence of the abundances ($Z$), the density ($\rho(r)$) and
velocity ($v(r)$), and a pre-specified temperature structure
($T_g(r)$) (see section~\ref{sec:general}), the occupation numbers are
determined by the rate equations containing collisional ($C_{ij}$) and
radiative ($R_{ij}$) transition rates. The most crucial dependency of
the rates is not the density, which is nevertheless important for the
collisional rates and the equation of particle conservation, but the
velocity field which enters not only directly into the radiative rates
via the Doppler shift, but also indirectly through the radiation field
determined by the equation of transfer, which in turn is again
dependent on the Doppler shifted line opacities and emissivities.

For the calculation of the radiative bound-bound transition
probabilities $R_{ij}$ we make use of the Sobolev-plus-continuum
method (\cite{}Hummer and Rybicki,~1985; Puls and Hummer,~1988).  Only
for some weak second-order lines in the subsonic region of the
atmospheric layers where the continuum is formed might this be just a
poor approximation (cf.~\cite{}Sellmaier et al.,~1993). A more
important point of our procedure concerns the problem of {\it
self-shadowing} (cf.~\cite{}Pauldrach et al.,~1998). This problem
occurs because the rate equations are not really solved simultaneously
with the radiative transfer, but instead in the framework of the
accelerated lambda iteration ({\it ALI}), in which the radiation field
and the occupation numbers are alternately computed
(cf.~\cite{}Pauldrach and Herrero,~1988). Hence, the radiation field
which enters into a bound-bound transition probability is already
affected by the line itself, since the line has also been considered
for the blocking opacities. This procedure will lead to a systematic
error if a line transition dominates within a frequency interval (see
section~\ref{sec:opasamp}).  The solution for correctly calculating
the bound-bound rates even in these circumstances is quite simple and
has been described by Pauldrach et al.~(1998, section 3.2).

The spherical transfer equation yields the radiation field at 2,500
frequency points (see below) and at every depth point, including the
layers where the radiation is thermalized and hence the diffusion
approximation is a proper boundary condition. The solution includes
all relevant opacities. In particular, the effects of wind and
photospheric EUV line blocking on the ionization and excitation of
levels are treated on the basis of 4~million lines, with proper
consideration of the influence of the velocity field on the line
opacities and emissivities and on the radiative rates.

Regarding the latter point, the inclusion of line opacities and
emissivities in the transfer equation, two different concepts are
employed for iterating the occupation numbers and the temperature
structure until a converged radiation field $J_{\nu}(r)$ and
$H_{\nu}(r)$ is obtained. In a first step, a pre-iteration cycle with
an opacity sampling method is used (method~I). This procedure has the
advantage of only moderate computing time requirements, allowing us to
perform the major part of the necessary iterations with this method.
Its disadvantage, however, is that it involves a few substantial
approximations (cf.~section~\ref{sec:lineblock}). In a second step,
the final iteration cycle is therefore solved with the detailed
radiative line transfer (method~II). Although this procedure is
extremely time-consuming, it has the advantage that it is not affected
by any significant approximations. With this second method, blocking
factors $B_J(r,\nu)$ and $B_H(r,\nu)$ are calculated, defined as the
ratio of the radiative quantities obtained by considering the total
opacities and emissivities to those which include only the
corresponding continuum values (cf.~Pauldrach et al.,~1996).
$B_J(r,\nu)$ and $B_H(r,\nu)$ are then used as multiplying factors to
the continuum quantities calculated in the next NLTE-ALI-cycle with
the current continuum opacities, in order to iterate the radiative
rates $R_{ij}$ (both continuum and lines) and the resulting occupation
numbers until convergence (details are described in
section~\ref{sec:lineblock}).

In total, almost 1000 ALI iterations are required by the complete NLTE
procedure, divided into blocks of 30 iterations each. (One iteration
comprises calculation of the occupation numbers and the radiation
field.) Up to 31 of these iteration blocks are performed using the
opacity sampling method (method~I), updating the temperature structure
and the Rosseland optical depth after each third ALI-iteration, and
the total opacities and emissivities after each iteration block. The
following iterations are then all performed using method~II, updating
temperature, optical depth, and opacities and emissivities as before,
and additionally calculating the blocking factors after each iteration
block.  (Several iteration blocks using method~II can be executed,
but~1 is usually sufficient --- see below.) In this phase the
radiative transfer solved in the ALI-iterations is just based on
continuum opacities and emissivities, and the blocking factors are
applied to get the correct radiative quantities used for calculating
the radiative rates.

As a final result of the complete iteration cycle, the converged
occupation numbers, the emergent flux, and the final NLTE temperature
structure are obtained.

\subsection {The atomic models}

\begin{table*}
\mycaption{
  Summary of Atomic Data.
  Columns 2 and 3 give the number of levels in packed and
  unpacked form; in columns 4 and 5 the number of lines used in the
  rate equations and for the line-force and blocking calculations are
  given, respectively.}
\label{tab1}
\begin{minipage}{\columnwidth}
\begin{center}
\begin{tabular}{lrrrr}
&\multicolumn{2}{c}{\hrulefill\lower.5ex\hbox{levels}\hrulefill}
&\multicolumn{2}{c}{\hrulefill\lower.5ex\hbox{lines}\hrulefill} \\
Ion & packed & unpacked & rate eq. & blocking \\ \hline \\
    C  {\sc ii}   &   36 &   73 &          284 &     11005 \\
    C  {\sc iii}  &   50 &   90 &          520 &      4406 \\
    C  {\sc iv}   &   27 &   48 &          103 &       229 \\
    C  {\sc v}    &    5 &    7 &            6 &        57 \\[1ex]
    N  {\sc iii}  &   40 &   82 &          356 &     16458 \\
    N  {\sc iv}   &   50 &   90 &          520 &      4401 \\
    N  {\sc v}    &   27 &   48 &          104 &       229 \\
    N  {\sc vi}   &    5 &    7 &            6 &        57 \\[1ex]
    O  {\sc ii}   &   50 &  117 &          595 &     39207 \\
    O  {\sc iii}  &   50 &  102 &          554 &     24506 \\
    O  {\sc iv}   &   44 &   90 &          435 &     17933 \\
    O  {\sc v}    &   50 &   88 &          524 &      4336 \\
    O  {\sc vi}   &   27 &   48 &          102 &       231 \\[1ex]
    Ne {\sc iv}   &   50 &  113 &          577 &      4470 \\
    Ne {\sc v}    &   50 &  110 &          534 &      2664 \\
    Ne {\sc vi}   &   50 &  112 &          343 &      1912 \\[1ex]
    Mg {\sc iii}  &   50 &   96 &          529 &      2457 \\
    Mg {\sc iv}   &   50 &  117 &          589 &      3669 \\
    Mg {\sc v}    &   50 &  100 &          547 &      3439 \\
    Mg {\sc vi}   &   21 &   44 &           54 &       305 \\[1ex]
    Al {\sc iv}   &   50 &   96 &          529 &      2523 \\
    Al {\sc v}    &   50 &  117 &          588 &     18317 \\
    Al {\sc vi}   &   19 &   37 &           41 &       153 \\[1ex]
    Si {\sc iii}  &   50 &   90 &          480 &      4044 \\
    Si {\sc iv}   &   25 &   45 &           90 &       245 \\
    Si {\sc v}    &   50 &   98 &          531 &      3096 \\
    Si {\sc vi}   &   50 &  116 &          596 &      3889 \\[1ex]
    P  {\sc v}    &   25 &   45 &           90 &       245 \\
    P  {\sc vi}   &   14 &   26 &           41 &      1096 \\[1ex]
    S  {\sc v}    &   44 &   78 &          404 &       903 \\ \\
\hline
\end{tabular}
\end{center}
\end{minipage}
\begin{minipage}{\columnwidth}
\begin{center}
\begin{tabular}{lrrrr}
&\multicolumn{2}{c}{\hrulefill\lower.5ex\hbox{levels}\hrulefill}
&\multicolumn{2}{c}{\hrulefill\lower.5ex\hbox{lines}\hrulefill} \\
Ion & packed & unpacked & rate eq. & blocking \\ \hline \\
    S  {\sc vi}   &   18 &   32 &           59 &       142 \\
    S  {\sc vii}  &   14 &   26 &           39 &      1031 \\[1ex]
    Ar {\sc v}    &   40 &   86 &          328 &      3007 \\
    Ar {\sc vi}   &   42 &   93 &          400 &      1335 \\
    Ar {\sc vii}  &   47 &   87 &          483 &      2198 \\
    Ar {\sc viii} &   15 &   27 &           41 &       111 \\[1ex]
    Mn {\sc iii}  &   50 &  141 &          364 &    175593 \\
    Mn {\sc iv}   &   50 &  124 &          467 &    131821 \\
    Mn {\sc v}    &   50 &  124 &          508 &     61790 \\
    Mn {\sc vi}   &   13 &   25 &           35 &        87 \\[1ex]
    Fe {\sc ii}   &   50 &  148 &          405 &    227548 \\
    Fe {\sc iii}  &   50 &  126 &          246 &    199484 \\
    Fe {\sc iv}   &   45 &  126 &          253 &    172902 \\
    Fe {\sc v}    &   50 &  124 &          451 &    124157 \\
    Fe {\sc vi}   &   50 &  138 &          452 &     60458 \\
    Fe {\sc vii}  &   22 &   62 &           91 &     10123 \\
    Fe {\sc viii} &   42 &   96 &          300 &      4777 \\[1ex]
    Co {\sc iii}  &   50 &  141 &          469 &    200637 \\
    Co {\sc iv}   &   41 &   97 &           70 &    146252 \\
    Co {\sc v}    &   45 &  126 &          253 &    182780 \\
    Co {\sc vi}   &   43 &  113 &          317 &    124053 \\
    Co {\sc vii}  &   34 &   80 &          246 &     50270 \\[1ex]
    Ni {\sc iii}  &   40 &  102 &          281 &    131508 \\
    Ni {\sc iv}   &   50 &  146 &          528 &    183267 \\
    Ni {\sc v}    &   41 &   97 &           70 &    179921 \\
    Ni {\sc vi}   &   45 &  126 &          253 &    186055 \\
    Ni {\sc vii}  &   43 &  113 &          317 &    123386 \\
    Ni {\sc viii} &   34 &   80 &          246 &     43778 \\[1ex]
    Cu {\sc iv}   &   50 &  124 &          477 &     17466 \\
    Cu {\sc v}    &   50 &  146 &          527 &     30457 \\
    Cu {\sc vi}   &   50 &  126 &          246 &     10849 \\[2ex]
\hline
\end{tabular}
\end{center}
\end{minipage}

\end{table*}

It is obvious that the quality of the calculated occupation numbers
and of the synthetic spectrum is directly dependent on the quality of
the input data. We have therefore extensively revised and improved
the basis of our model calculations, the atomic models.

Up to now the atomic models of all of the important ions of the 149
ionization stages of the 26 elements considered (H~to~Zn, apart
from Li, Be, B, and Sc) have been replaced in order to improve the
quality. This has been done using the {\em Superstructure} program
(Eissner et al.,~1974; Nussbaumer and Storey,~1978), which employs the
configuration-interaction approximation to determine wave functions
and radiative data. The improvements include more energy levels
(comprising a total of about 5,000 observed levels, where the fine
structure levels have been ``packed'' together\footnote{ Note that
artificial emission lines may occur in the blocking calculations if
the lower levels of a fine structure multiplet are left unpacked but
the upper levels of the considered lines are packed --- the fine
structure levels of an ionization stage should either be all packed or
all unpacked.}) and transitions (comprising more than 30,000
bound-bound transitions for the NLTE calculations and more than
4,000,000 lines for the line-force and blocking calculations\footnote{
Note further that the consistency of the model calculation requires
that the wavelength of the bound-bound transition connecting packed
levels in the NLTE calculations to be identical to the wavelength of
the strongest component of the multiplet considered in the blocking
calculations in order to solve the line radiative transfer and
especially the problem of {\it self-shadowing}
properly.}$^,$\footnote{ Our line list also includes transitions to
highly excited levels above our limit of considering the level
structure; occupation numbers of these upper levels are estimated
using the two-level approximation on the basis of the (known)
occupation number of the lower level.}, and 20,000 individual
transition probabilities of low-temperature dielectronic recombination
and autoionization).

Additional line data were taken from the Kurucz~(1992) line list and
from the Opacity Project (cf.~Seaton et al.,~1994; Cunto and Mendoza,
1992); the latter was also a source for photoionization cross-sections
(almost 2,000 data sets have been incorporated). Collisional data have
become available through the IRON project (see Hummer et al.,~1993)
--- almost 1,300 data sets have been included.

Table~\ref{tab1} gives an overview of the ions affected by the
improvements. (Users of the program package {\em WM-basic} should
note that the model calculations will become inconsistent if the
atomic data sets are changed haphazardly by those who are not
familiar with the source code.)

\subsection{The treatment of line blocking}
\label{sec:lineblock}

As the thermal width of a UV metal line covers just a few m\AA, a
simple straightforward method would require considering approximately
$10^7$ frequency points in order to resolve the lines in the spectral
range affected by line blocking. Such a procedure would lead to a
severe problem concerning the computational time. The alternatives are
either to calculate the complete radiative transfer in the comoving
frame --- again a time consuming procedure ---, or to use a tricky
method which saves a lot of computation time through the application
of some minor approximations (method~I), dropping these approximations
in the final iteration steps (method~II) in order to come to a
realistic solution. Our treatment described here uses the second
approach.

Although frequently applied, a method using {\em opacity distribution
functions} ({\em ODF\/}s --- cf.~Labs, 1951; Kurucz,~1979), where the
opacities are rearranged within a rough set of frequency intervals in
such a way that a smoothly varying function is obtained which
conserves the statistical distribution of the opacities, is not
applicable in our case, since there is no appropriate way to treat the
lineshift in the wind, and due to the rearrangement of the opacities
the frequential position of the lines is changed. This, however,
prevents a correct computation of the bound-bound transitions used for
the solution of the statistical equilibrium equations.

The approach best suited for our purpose in the first step (method~I)
is the {\em opacity sampling} technique (cf.~Peytremann,~1974; Sneden
et al.,~1976) which compared to the ODF-method is computationally a
bit more costly, but does not suffer from the limitations mentioned
above. This method allows us to account for the {\it lineshift} in the
wind and the correct influence of line blocking on the bound-bound
transitions (cf.~Section~\ref{sec:nltemodel}, item~(ii)), since it
preserves the exact frequential position of the lines.

\subsubsection{The opacity sampling method}
\label{sec:opasamp}

Following the idea of the {\em opacity sampling}, a representative set
of frequency points is distributed in a logarithmic wavelength scale
over the relevant spectral range, and the radiative transfer equation
is solved for each point. (For O-stars the actual range depends on
$\Teff$; for hot objects the lower value is at $\approx 90\,{\rm \AA}$
and for cooler objects the upper value is at $2000\,{\rm \AA}$; note
that accurate ionization calculations require extending the line
blocking calculations to the range shortward of the \HeII\ edge ---
cf.~Pauldrach et al.,~1994)

In this way the exact solution is reached by increasing the number of
frequency points. A smooth transition is obtained when the number of
frequency points is increased up to the number --- $10^7$ ---
which is required to resolve the thermal width of the UV lines. It is
obvious however that convergence can be achieved already with
significantly less points (see below). Furthermore, special blocking
effects on selected bound-bound transitions can be investigated more
thoroughly by spreading additional frequency points around the line
transition of interest.

In the following subsection we will investigate how many sampling
points are required in order to represent the physical situation in a
correct way.

\paragraph{The influence of line blocking on the photoionization
integrals.}

\begin{figure}
\centerline{\includegraphics[width=8.5cm,height=7.5cm]{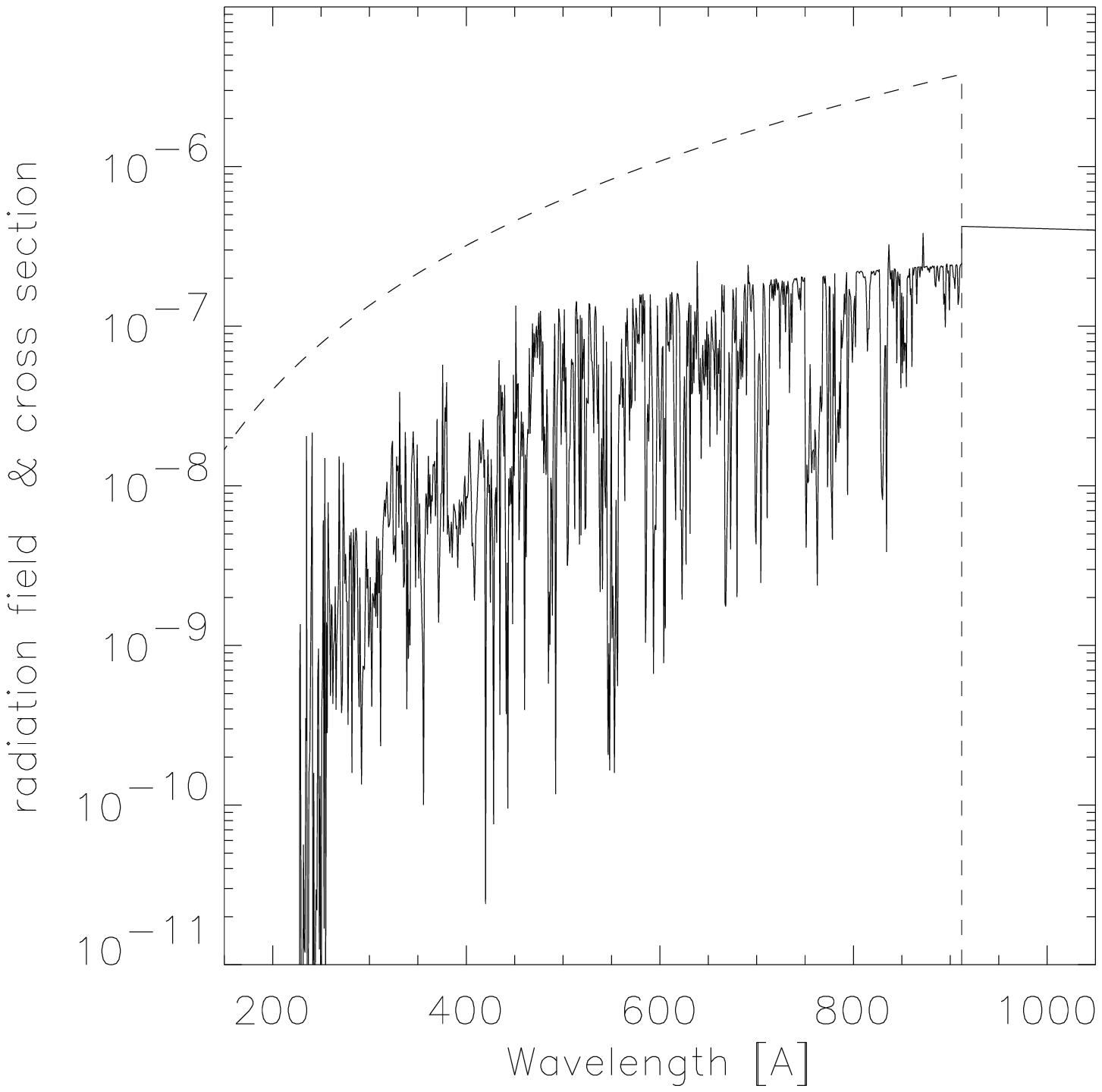}}
\vspace{-3mm}
\mycaption{Mean radiation field $J_{\nu}$ together with the
  photoionization cross section $\sigma_{ik}$ of the
  ground state of H (in arbitrary units).
\label{crossection}}
\centerline{\includegraphics[width=9cm,height=7.5cm]{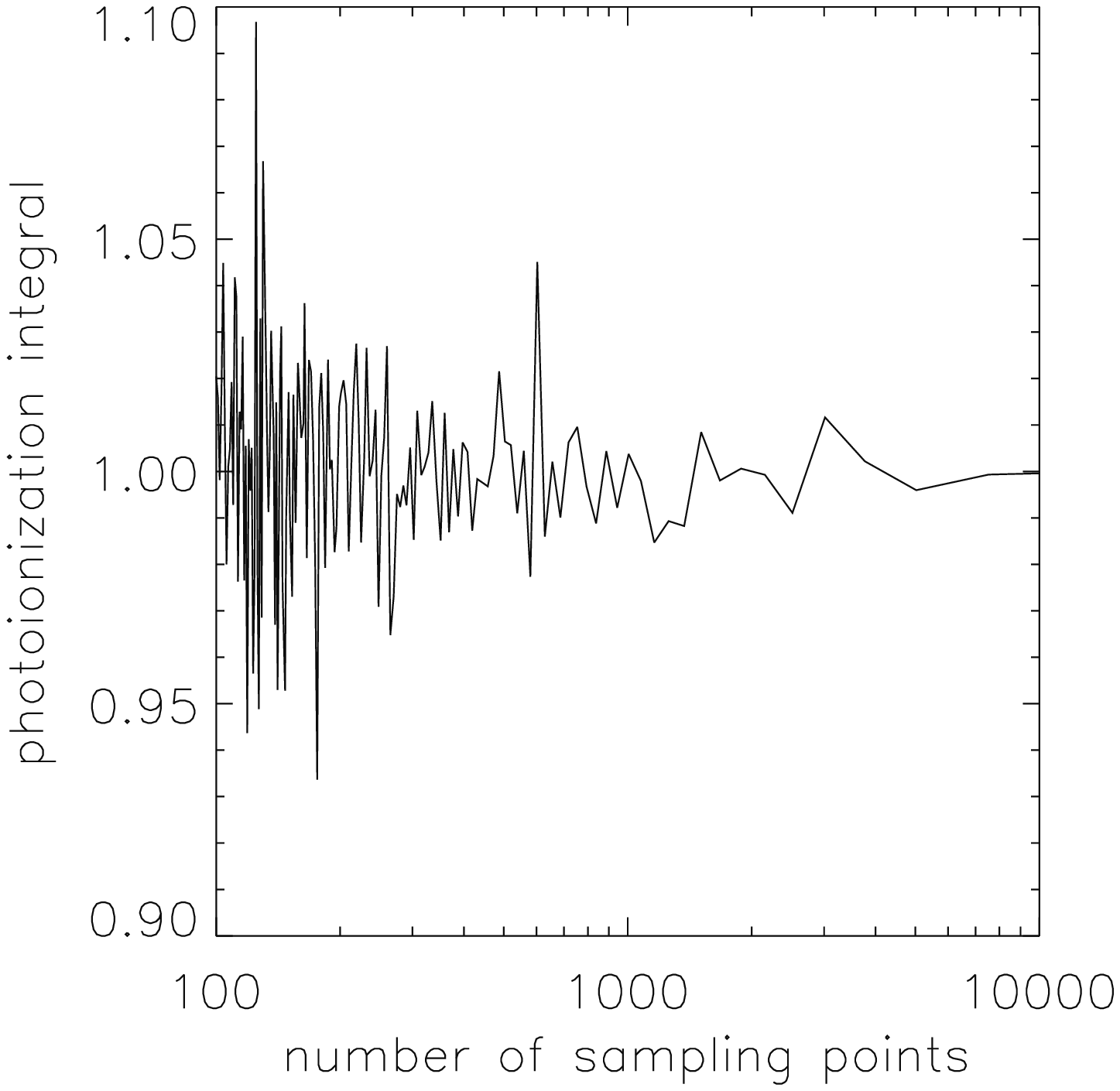}}
\vspace{-3mm}
\mycaption{Accuracy of the normalized photoionization
  integral $R_{ik}$ of the groundstate of H in dependence of an
  increasing number of sampling points within the Lyman
  continuum.
\label{photoisp}}
\end{figure}

The most important effect of line blocking on the emergent spectrum is
the influence on the ionization structure via the photoionization
integrals
\begin{equation}
\label{eq:RKi}
R_{ik} = {4 \pi} \int \frac{J_{\nu}}{h\nu} \sigma_{ik}(\nu)\,\d\nu .
\end{equation}
This can be verified from Fig.~\ref{crossection} where it is shown
that the mean radiation field $J_{\nu}$ changes rapidly over the
frequency interval covered by a typical smooth bound-free cross
section $\sigma_{ik}$ --- several $100\,{\rm {\AA}}$ are affected.
(Note that dielectronic resonances which may occur in addition are not
shown here.)

It is obvious from Fig.~\ref{crossection} that the photoionization rates
are sensitive functions of the blocking influence on $J_{\nu}$, and hence,
on the number of sampling points in the relevant frequency range.

In order to determine the number of sampling points required for an
accurate description we performed empirical tests by evaluating
typical photoionization integrals in dependence of an increasing
numbers of sampling points. Fig.~\ref{photoisp} shows as an example
the normalized photoionization integral of the ground state of
hydrogen. The result is that 1,000 sampling points within the Lyman
continuum guarantee a sufficient accuracy of about 1 to 2 percent. By
means of a separate investigation Sellmaier~(1996) showed the given
number of sampling points to be reasonable, since it reproduces the
actual line-strength distribution function quite well.

\paragraph{The treatment of the lineshift.}

The total opacity at a certain sampling frequency $\nu$ is given by
adding the line opacity $\chi_{\rm lb}$ to the continuum opacity
$\chi_{\rm c}$
\begin{equation}
  \chi_{\nu} = \chi_{\rm lb}(\nu) + \chi_{\rm c}(\nu),
\end{equation}
where $\chi_{\rm lb}$ is the sum over all (integrated) single line
opacities $\ol{\chi}_{\rm l}$ multiplied by the line profile function
$\varphi_{\rm l}(\nu)$
\begin{equation}
\label{eq:oplb}
   \chi_{\rm lb}(\nu) =
   \sum_{\rm lines} \ol{\chi}_{\rm l} \varphi_{\rm l}(\nu).
\end{equation}
Here $\ol{\chi}_{\rm l}$ is
\begin{equation}
 \label{eq:chil}
 \chil = \frac{h\nu_0}{4\pi}\left(n_i B_{ij}-n_j B_{ji}\right),
\end{equation}
and the analogous expression for the emissivity is
\begin{equation}
 \label{eq:etal}
 \etal = \frac{h\nu_0}{4\pi} n_j A_{ji}.
\end{equation}
($B_{ij}$, $B_{ji}$, und $A_{ji}$ are the Einstein coefficients of the
line transition at the frequency $\nu_0$, and $h$ is Planck's constant.)

\begin{figure}
\centerline{\includegraphics[width=9cm,height=8.0cm]{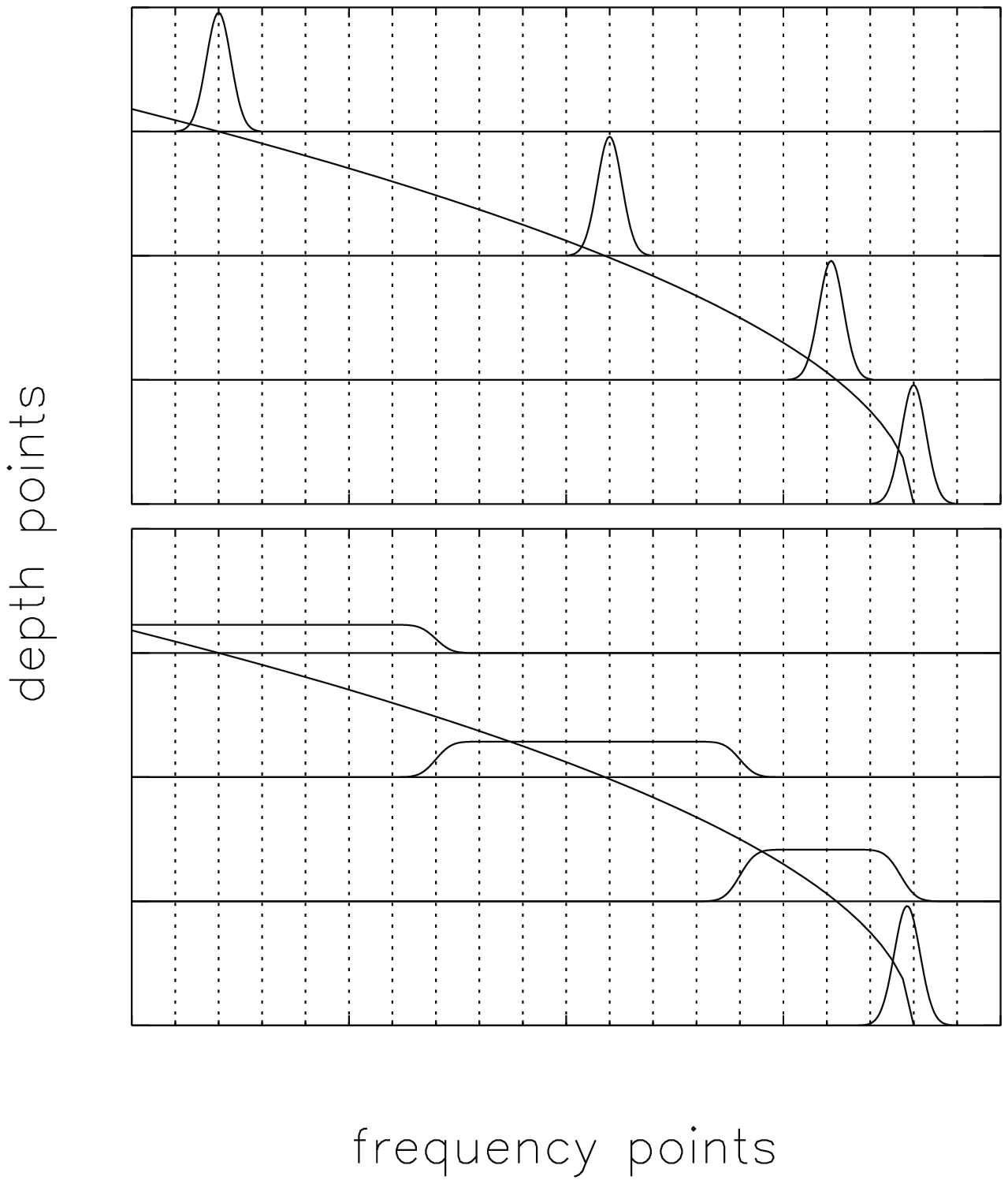}}
\mycaption{
  {\bf upper panel:} simply shifting the line profile along $\nu_{\rm
  CMF}$ (represented by the curve) at each radius grid point (standard
  opacity sampling) causes the line to be missed at most frequency
  points;
  {\bf lower panel:} this problem is solved by assuming a boxcar
  profile for each depth point with a width corresponding to the
  difference in Doppler shift between to successive radius points
  (``Doppler-spread opacity sampling'').}
\label{boxprofile}
\vspace{5mm}
\centerline{\includegraphics[width=\columnwidth]{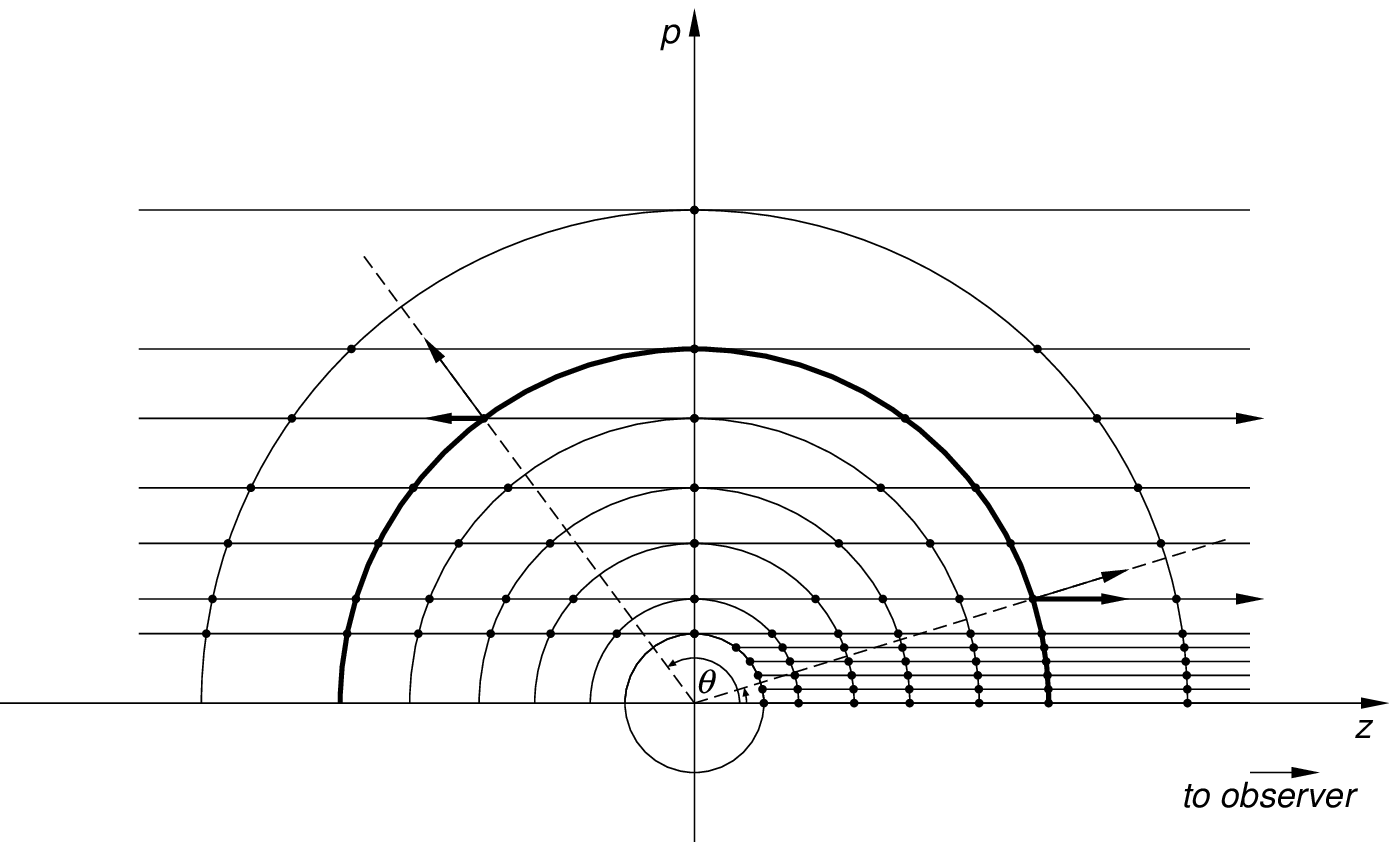}}
\mycaption{
  $(p,z)$-geometry for the spherically symmetric radiative transfer.
  For any given depth point, a different Doppler shift must in principle
  be considered for every $p$-ray, since the projected velocity
  varies with $\cos\theta$.
  However, as no analogy to the boxcar method exists in this case,
  in our opacity sampling method we take
  the Doppler shift of the central ray as being
  representative for all other rays (see text).}
\label{fig:pzgeo}
\end{figure}

In the static part of the atmosphere a line's opacity covers with its
(thermal and microturbulent) Doppler profile $\varphi_{\rm D}$ only a
very small interval around the transition frequency $\nu_0$
(illustrated in Fig.~\ref{boxprofile} on the right hand side of both
figures; note that with regard to our sampling grid about 40 percent
of the available lines are treated in this part). The effect of these
lines on the radiation field is nevertheless considerable
(cf.~Fig.~\ref{crossection}), if the lines are strong enough to become
saturated.

In the expanding atmospheres of hot stars the effect of line blocking
is enhanced considerably in the supersonic region due to the nonlinear
character of the radiative transfer. A velocity field $v(r)$ enables
the line to block the radiation also at other frequencies
$\nu = \nu_0 (1+v(r)/c)$, i.\,e., the Doppler shift increases the
frequency interval which can be blocked by a single line to a factor
of $\approx 100$. On the other hand, the velocity field reduces the
spatial area where a photon can be absorbed by a line. If a line is
optically thick, however, the effect of blocking will ultimately be
increased compared to a static photosphere.

The lineshift due to the velocity field is applied to the individual
line opacities before the summation in eq.~\ref{eq:oplb} is carried
out at each sampling and depth point (otherwise the effect of the
lineshift would be underestimated with respect to the ratio of line
width to sampling distance --- see below). However, in our approach
this is done by applying the Doppler shift of the radial ray to all
$p$-rays (see Fig.~\ref{fig:pzgeo}), ignoring the angular dependence of
the Doppler shift (see below). Apart from the intrinsic character of
the sampling method this is the most restrictive approximation in our
first iteration cycle. The frequency shift is thus treated in a
statistical sense, its main effect --- increase of the frequential
range of line absorption --- is nevertheless included properly and
that is what has to be iterated in this first cycle.

From the upper panel of Fig.~\ref{boxprofile} it is obvious that if
the line opacity is simply shifted along the comoving frame frequency
($\nu_{\rm CMF}$) to every radius point successively, many frequency
points will miss the line, since the radius grid is too coarse to
treat large lineshifts in the observer's frame. This behaviour is
corrected by convolving the intrinsic Doppler profile of the line with
a boxcar profile $\varphi_{\Delta v}$ representing the velocity range
around each radius point (Fig.~\ref{boxprofile}, lower panel).

The boxcar profile is the mean profile obtained by considering the
velocity shifts $\Delta v$ of the two corresponding intermesh points
($\nu_1$, $\nu_2$) on both sides of the regarded radius grid point in
the way that the gaps in the frequency grid are closed. This can be
expressed in terms of the Heaviside function $\theta$:
\begin{equation}
\label{eq:phiv}
  \varphi_{{\Delta} v}(\nu) =
  \frac{\theta(\nu_2 - \nu) - \theta(\nu_1-\nu)}{2(\nu_2-\nu_1)}
\end{equation}
$\nu_1$ and $\nu_2$ are the observer's frame frequencies belonging to
the velocities of two successive radius points ($r_1$ and $r_2$),
i.\,e., $\nu_{1,2} = \nu_0 (1+v(r_{1,2})/c)$. Assuming thermal Doppler
broadening for the intrinsic line profile,
\begin{equation}
\label{eq:phid}
  \varphi_{\rm D}(\nu) = \frac{ e^{-x^2} }{ \sqrt{\pi} \dnud }
  \quad\mbox{with}\quad
  x = \frac{\nu-\nu_0}{\dnud},
\end{equation}
where $\dnud$ is the thermal Doppler width, the convolution
$(\varphi_{\rm D} \otimes \varphi_{\Delta v})(\nu)$ results in the final
profile function
\begin{equation}
\label{eq:Phierf}
  \phi(\nu) = (\varphi_{\rm D} \otimes \varphi_{\Delta v})(\nu)
            =  \frac{ {\rm erf}(x_2-x) - {\rm erf}(x_1-x) }
                    { 2 (x_2-x_1) \dnud}.
\end{equation}
This profile can be used for the entire sub- and supersonic region.
For $\Delta v < v_{\rm therm}$ it gives, as a lower limit, the {\it
ordinary opacity sampling}, and for sufficiently high velocity
gradients ($\Delta v > v_{\rm therm}$) the integration over a radius
interval represents the {\em Sobolev optical depth} ($\tau_{\rm
Sob}(r)$) of a local resonance zone for a radial ray
\begin{eqnarray}
\label{eq:dtau}
 {\Delta} \tau
 & = & \int_{r_1}^{r_2} \ol{\chi}_{\rm l} \cdot \phi({\nu})
                            \,{\rm d}r \nonumber \\
 & \approx & \ol{\chi}_{\rm l} \cdot \frac{1-(-1)}{2(x_2-x_1)\cdot \dnud}
             \cdot (r_2-r_1) \nonumber \\
 & = &       \ol{\chi}_{\rm l} \cdot \frac{r_2-r_1}{v_2-v_1} \cdot
             \frac{c}{\nu_0} \nonumber \\
 & \approx & \ol{\chi}_{\rm l} \cdot \frac{c}{\nu_0} \cdot
             \left(\frac{{\rm d}v}{{\rm d}r}\right)^{-1} \nonumber \\
 & = &       \tau_{\rm Sob}(r).
\end{eqnarray}
Note that at sufficiently high velocity gradients all lines are
included in the radiative transfer if the sampling grid is fine enough
(see also \cite{} Sellmaier (1996)). In this case our {\it
Doppler-spread opacity sampling method} therefore becomes an {\it
exact solution}.

\begin{figure}
\centerline{\includegraphics[height=8.7cm,angle=90]{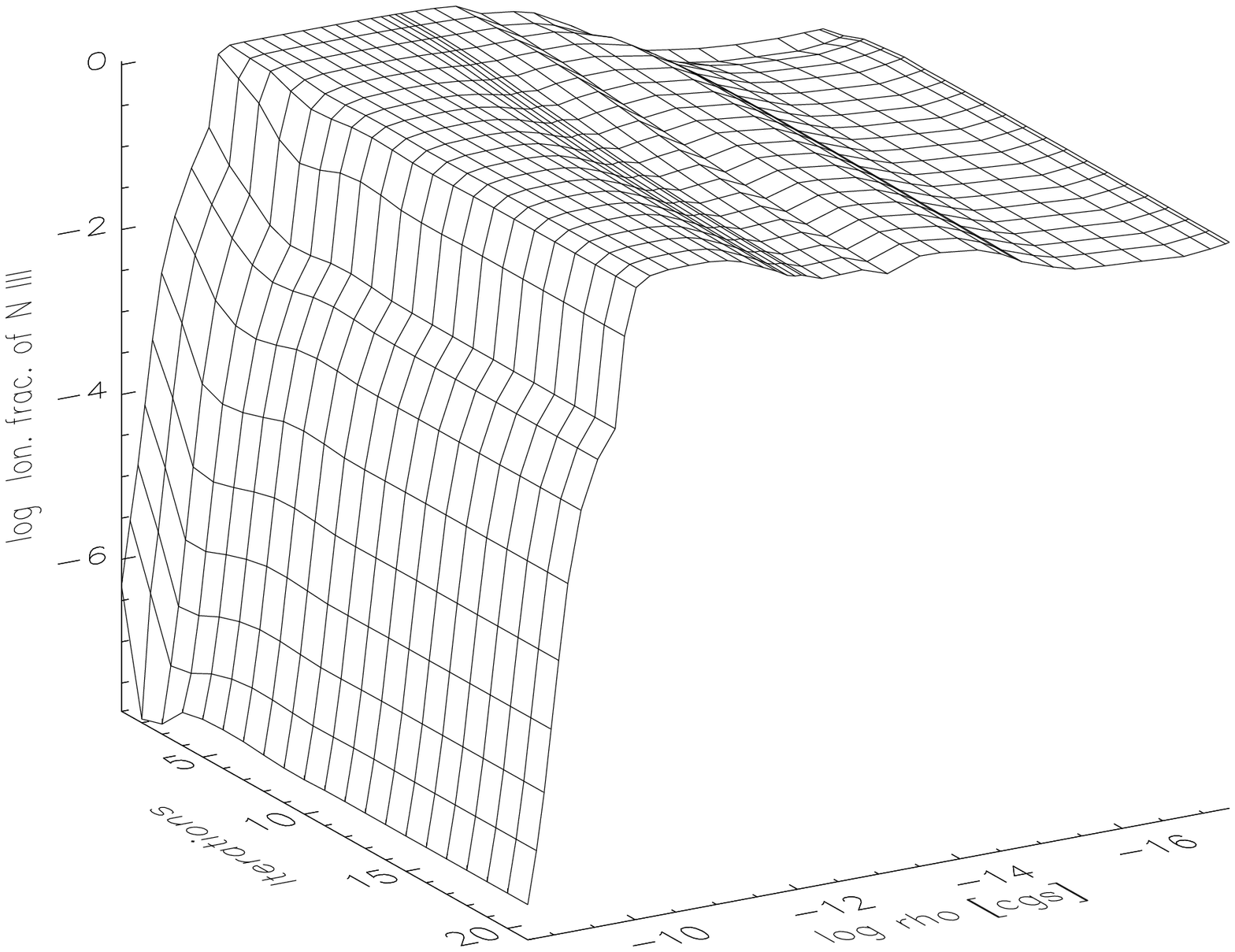}}
\centerline{\includegraphics[height=8.7cm,angle=90]{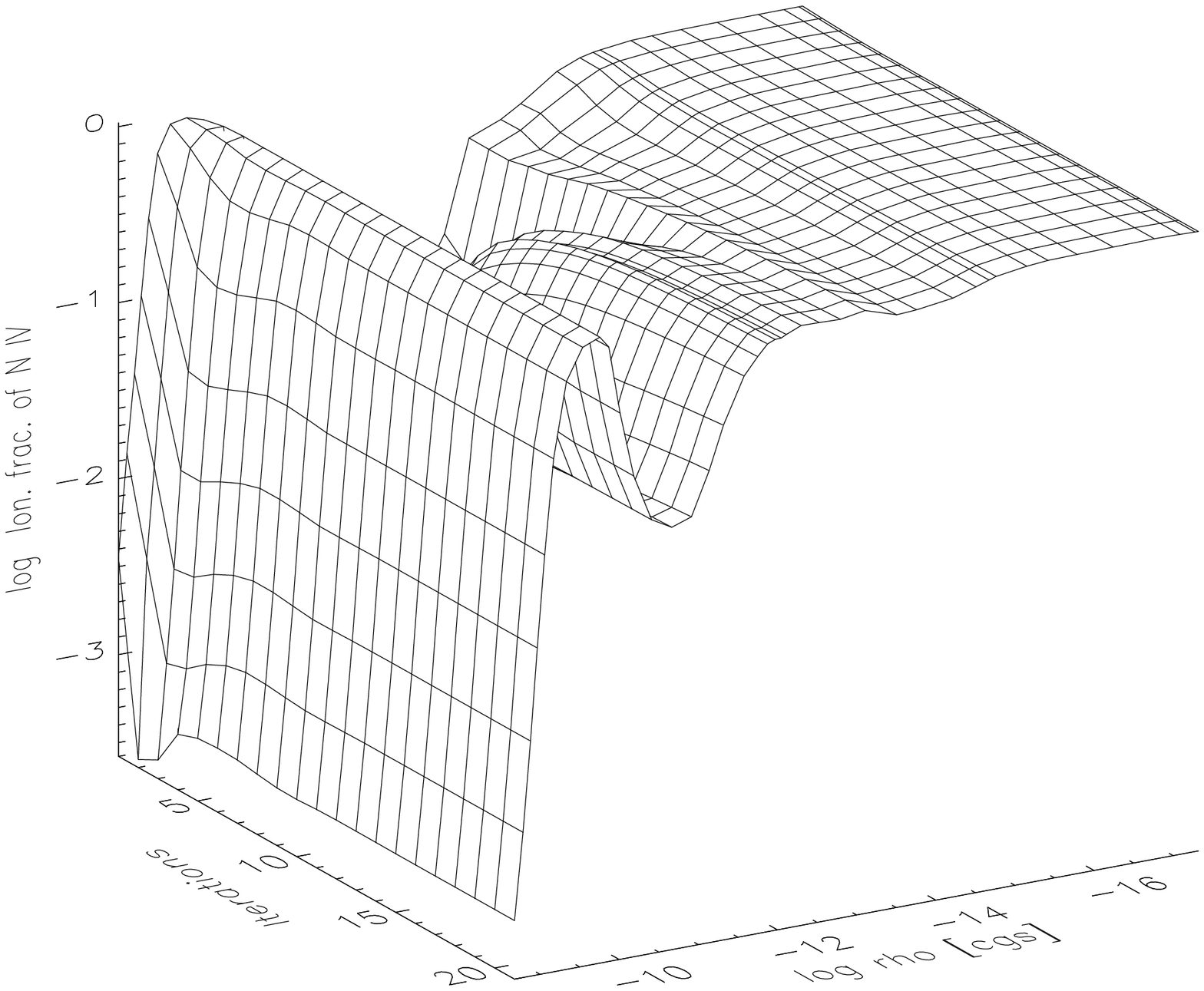}}
\centerline{\includegraphics[height=8.7cm,angle=90]{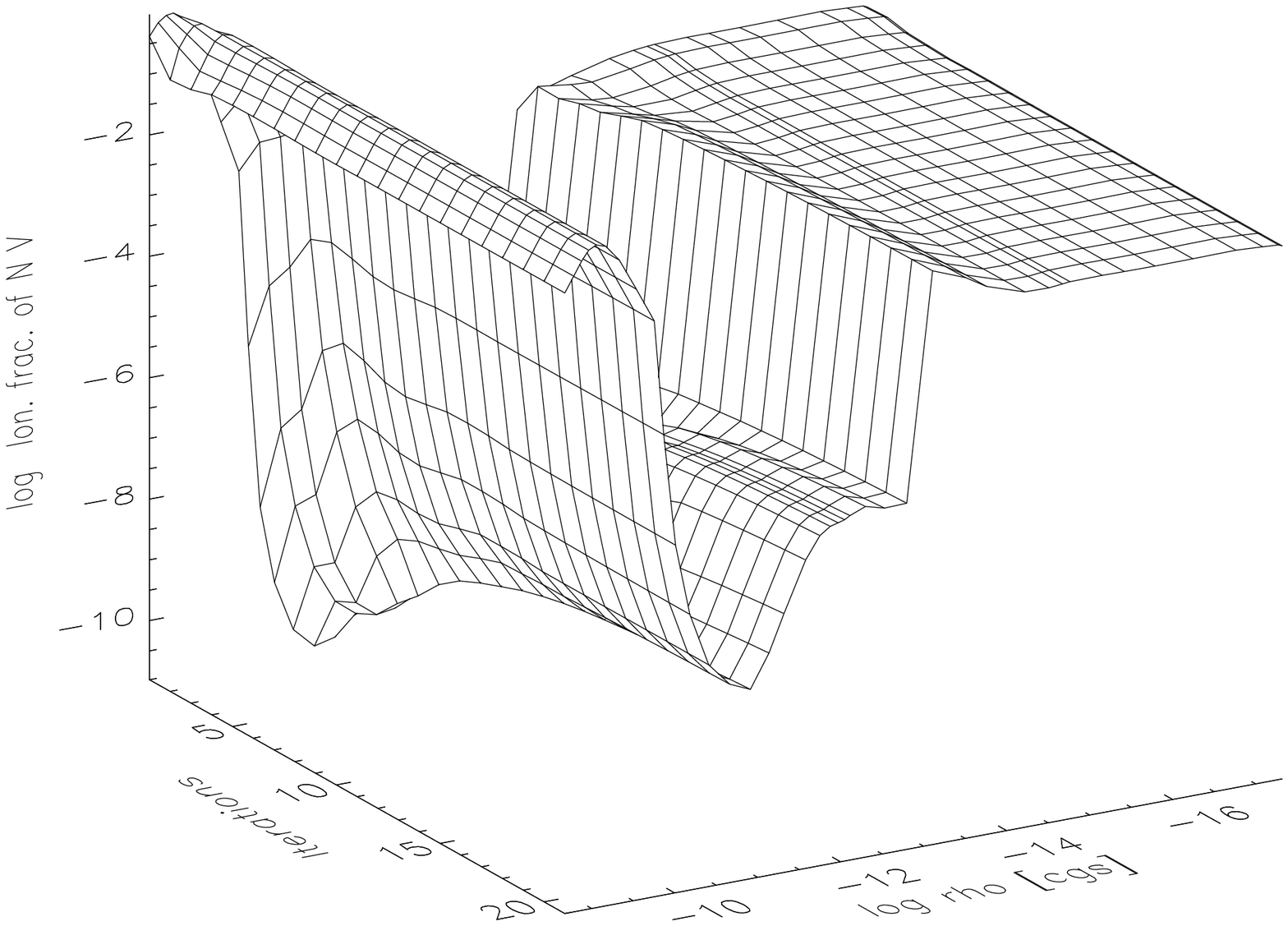}}
\mycaption{Logarithm of the
ionization fractions of
  \NIII\ ({\it upper panel}),
  \NIV\  ({\it middle panel}), and
  \NV\   ({\it lower panel})
  versus density and iteration block number
  for an O supergiant model
  ($\Teff=29\,000\,{\rm K}$, $\logg=3.0$, $\Rrs=27.0$).
  The region between two successive contour lines (one iteration
  block) corresponds to 30 iterations.}
\label{fig:converg1}
\end{figure}

In principle, one would have to account for the angular variation of
the Doppler shift in a similar manner (see Fig.~\ref{fig:pzgeo}), but as
no analogy to the boxcar profile method exists in this case, as
mentioned above we simply apply the (correctly calculated, radially
dependent) opacities ($\chi_{\rm lb}(\nu,r)$) of the central $p$-ray
to the other $p$-rays, regarding these opacities as being
representative. A welcome result of this simplification is the
fastness of the method, a very important consideration in this
iteration cycle.

Concerning our {\em WM-basic} program package running on a normal
scalar processor, however, the method is still not fast enough (a
model calculation would require an amount of computing time of about
20 hours).  The reason is that the Rybicki-method which is used in
this step for the solution of the second-order form of the equation of
transfer (cf.~Mihalas,~1978) requires more than 80\% of the computing
time of a model calculation.
(Note that the Rybicki-method is applied in each iteration just once
per frequency point; in order to improve the accuracy, the radiative
quantities are then further iterated internally by using the moments
equation of transfer (cf.~Mihalas,~1978). Because of strong changes in
the opacities and emissivities within the NLTE iteration cycle it is
necessary to start with the Rybicki-method nevertheless.) We have
therefore rethought the solution concept of the Rybicki-scheme and
developed a method which is 10 times faster on a vector processor and
3 to 5 times faster on a scalar processor --- the actual factor
depends on the quality of the level-2 {\sc blas} functions available
with professional compiler programs and which do most of the work in
our method (see \ref{app:rybick}).

In order to illustrate the behaviour of convergence of our method~I,
the ionization fractions of N\,{\sc iii}, {\sc iv}, and {\sc v} are
shown versus density and the iteration block number in
Fig.~\ref{fig:converg1} for the first 600 iterations as an example. As
displayed, the model convergences within 400 iterations --- the
remaining iterations are required to warrant the luminosity
conservation (see section~\ref{sec:lineblanket}). The steep increase of \NV\ in the
wind part results from the EUV and X-ray radiation produced by
shock-heated matter (see section~\ref{sec:shocks}).

We finally note that first results obtained with a preliminary version
of this procedure have already been published. Sellmaier et al.~(1996)
showed that their NLTE line-blocked O-star wind models solve the
longstanding \NeIII\ problem of \HII-regions for the first time, and
Hummel et al.~(1997) carried out NLTE line-blocked models for
classical novae.

\begin{figure}
\centerline{\includegraphics[height=\columnwidth,angle=-90]{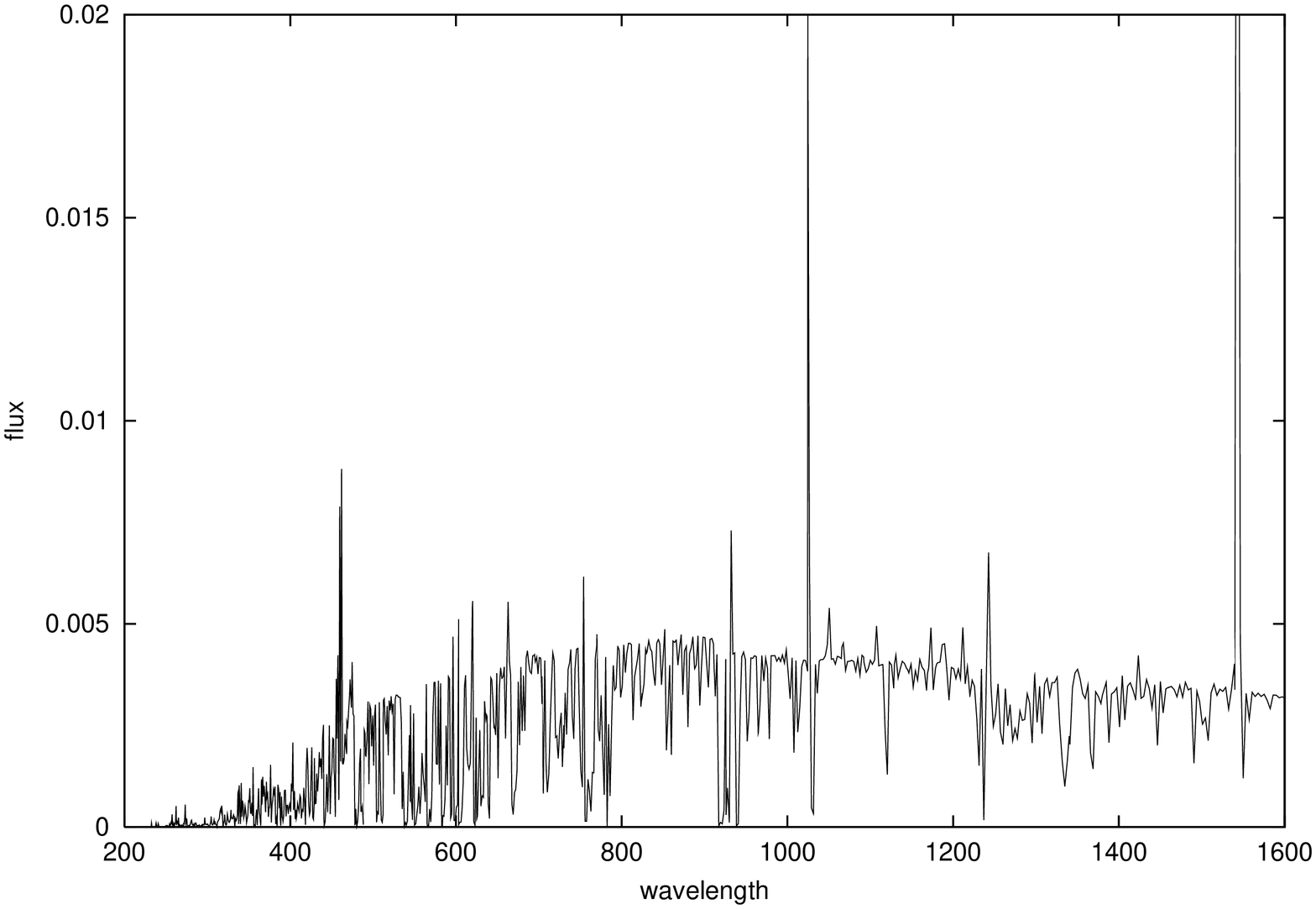}}
\centerline{\includegraphics[height=\columnwidth,angle=-90]{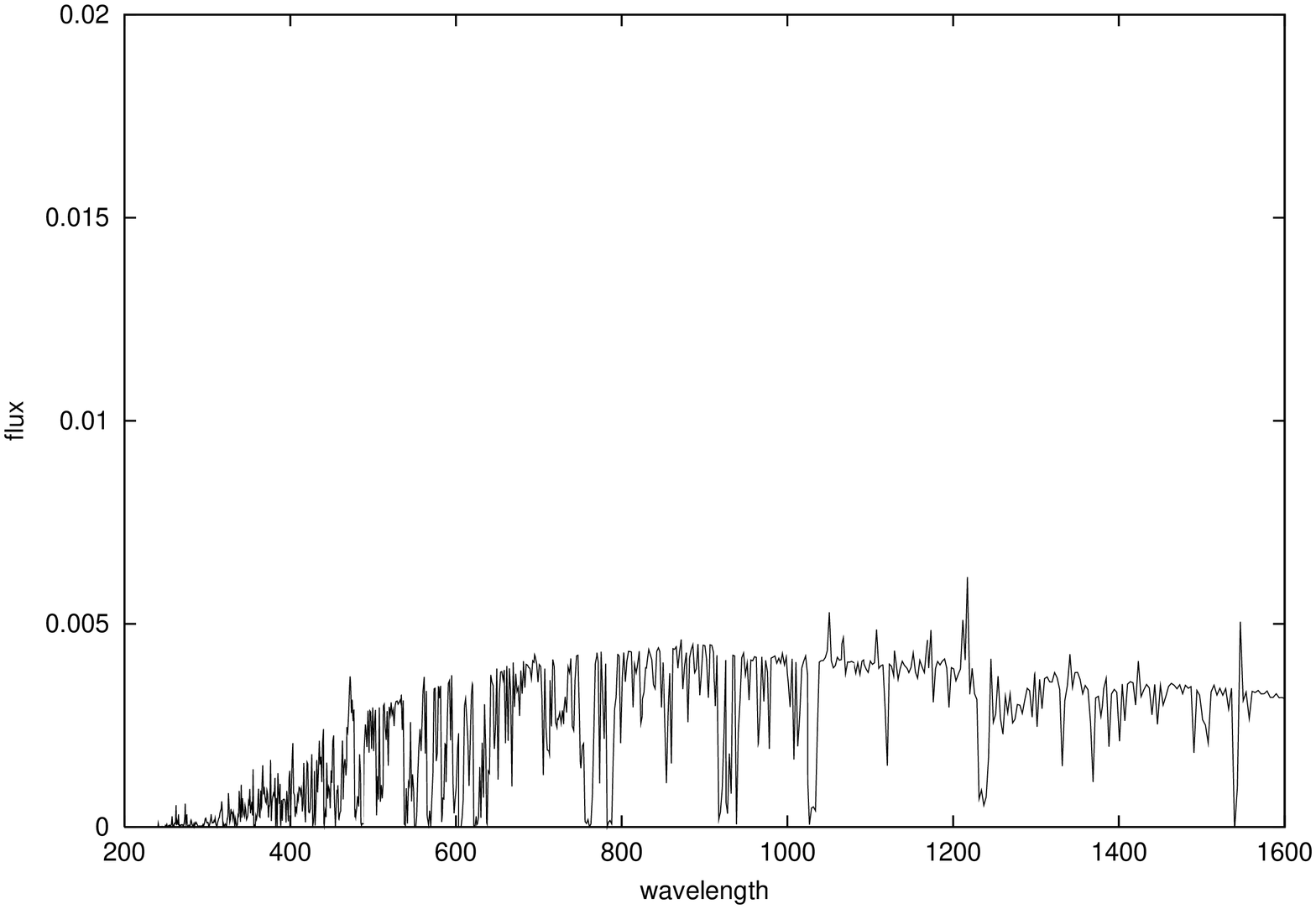}}
\mycaption{
Part of the synthetic EUV spectrum (200--1600\,\AA) of the \mbox{S-45}
supergiant model ($\Teff=45000\,{\rm K}$, $\logg=3.6$, $\Rrs=18.0$)
calculated with the {\it Doppler-spread opacity sampling method}. The
upper panel shows the spectrum resulting from the standard treatment
of the transfer equation which produces several strong artificial
emission lines. The lower panel shows the correct spectrum resulting
from our modified treatment.}
\label{fig:discret}
\end{figure}

\paragraph{Special problems.}


From first test calculations performed in the manner described we
recognized and solved two additional nontrivial problems:

\medskip
{\it The first problem} concerns the artificial effect of {\it
self-shadowing} (see above) which occurs because the incident
intensity used for the calculation of a bound-bound transition that
enters into the rate equations is already affected by the line
transition itself, since the opacity of the line has been used for
the computation of the radiative quantities in the previous iteration
step. If the lines contained in a frequency interval are of almost
similar strength, this is no problem, since the used intensity
$I_{\nu_n}(r)$ calculated at the sampling point represents a fair mean
value for the {\it true incident radiation} of the individual lines in
the interval. If however, a line has a strong opacity with a
dominating influence in the interval, the intensity taken at the
sampling point for the same bound-bound transition in the radiative
rates is much smaller than the {\it true incident radiation} for this
line, because the line has already influenced this value considerably.
In consequence the source function of this line is underestimated and
the radiative processes --- the scattering part is mostly affected ---
are not correctly described in the way that the line appears
systematically too weak.

The solution to this problem is rather simple: in calculating the
bound-bound rates of the dominating lines, we use an incident
intensity which is independent of the lines in the considered interval
(cf.~Pauldrach et al.,~1998).

\medskip

{\it The second problem} involves the discretization of the transfer
equation in its differential form, for computing the radiative quantities
(Feautrier method).
In the standard approach (see, for example, Mihalas,~1978)
the equation of transfer is written as a second-order differential
equation with the optical depth $\tau$ as the independent variable:
\begin{equation}
{\d^2 u \over \d\tau^2} = u - S,
\end{equation}
where $S$ is the source function and $u={1\over 2}(I^++I^-)$,
with $I^+$ and $I^-$ being the intensities in positive and
negative $\tau$ direction along the ray considered.

This differential equation is then converted to a set of difference
equations, one for each radius point $i$ on the ray,
\begin{eqnarray}
\left.{\d^2u\over\d\tau^2}\right|_{\tau_i}
&\approx&
{ \left.{\d u\over\d\tau}\right|_{\tau_{i+{1\over 2}}} -
  \left.{\d u\over\d\tau}\right|_{\tau_{i-{1\over 2}}} \over
  \tau_{i+{1\over 2}} - \tau_{i-{1\over 2}}}
\\&\approx&
{{\displaystyle u_{i+1}-u_i\over\displaystyle\tau_{i+1}-\tau_i} -
 {\displaystyle u_i-u_{i-1}\over\displaystyle\tau_i-\tau_{i-1}} \over
 {1\over 2}(\tau_{i+1}+\tau_i) - {1\over 2}(\tau_i+\tau_{i-1})},
\end{eqnarray}
resulting in a linear equation system
\begin{equation}
a_i u_{i-1} + b_i u_i + c_i u_{i+1} = S_i
\end{equation}
with coefficients
\begin{equation}
\begin{array}{rcl}
a_i&=&-\left({1\over2}(\tau_{i+1}-\tau_{i-1})(\tau_i-\tau_{i-1})\right)^{-1}
\smallskip\\
c_i&=&-\left({1\over2}(\tau_{i+1}-\tau_{i-1})(\tau_{i+1}-\tau_i)\right)^{-1}
\smallskip\\
b_i&=&1-a_i-c_i,
\end{array}
\end{equation}
(and appropriate boundary conditions). This linear equation system
has a tridiagonal structure and can be solved economically by standard
linear-algebra means.\footnote{
In practice, a Rybicki-type scheme (cf.~Mihalas, 1978; and
Appendix~\ref{app:rybick}, this paper) is used for solving the
equation systems for all $p$-rays simultaneously, since the source
function contains a scattering term (see eq.~\ref{eq:appatrf}) which
redistributes the intensity at each radius shell over all rays
intersecting that shell.}
Note that the equations contain only {\em differences} in $\tau$,
which can easily be calculated from the opacities
and the underlying $z$-grid (cf.~Fig.~\ref{fig:pzgeo}) as
\begin{equation}
\tau_{i+1}-\tau_i = {\textstyle{1\over2}}(\chi_{i+1}+\chi_i)(z_{i+1}-z_i),
\end{equation}
with $\chi_i$ being the opacity at depth point $i$.

The equation systems are well-behaved if the opacities and source
functions vary only slowly with $z$. Caution must be taken if this
cannot be guaranteed, for example, {\em whenever a velocity field is
involved} at strong ionization edges or with the opacity sampling
method at strong lines, since the velocity field shifts the lines in
frequency, causing large variations of the opacity from depth point to
depth point for a given frequency.  In particular, a problematic
condition occurs if a point with a larger-than-average source function
$S_i$ and low opacity $\chi_i$ borders a point with a high opacity
$\chi_{i+1}$ (and low or average source function $S_{i+1}$). In
reality, this large source function should have little impact, since
it occurs in a region of low opacity, and thus the emissivity is
small. However, the structure of the equations is such that the
emission is computed to be on the order of
\begin{equation}
\begin{array}{rcl}
\Delta I & \approx & \overline S \cdot \Delta\tau
\smallskip\\ &\approx &
\textstyle{1\over2}(S_{i+1}+S_i)\cdot{1\over2}(\chi_{i+1}+\chi_i)(z_{i+1}-z_i),
\end{array}
\end{equation}
where, if the other quantities are comparatively small (in accordance
with our assumptions), the term $S_i \chi_{i+1}$ dominates,\footnote{
The physical reason for the failure of the system is that the source
function only has meaning relative to its corresponding opacity.
Multiplying the source function from one point with the opacity at
another point is complete nonsense.}
leading to artificially enhanced emission.  In
Figure~\ref{fig:discret} (upper panel) we show the exaggerated
emission of the strongest spectral lines in the emergent flux of a
stellar model computed using this standard discretization, leading to
false results.  Even a simple example can serve to illustrate this
effect, as demonstrated in Appendix~B.

However, with a subtle modification of the equation system
coefficients the method can nevertheless be salvaged.  The subtle
point involves writing the transfer equation as an equation not in
$\tau$, but in $z$ for derivation of the coefficients, since only
this formulation treats correctly the $z$-dependence of $\chi$:
\begin{equation}
{1\over\chi}{\d\over\d z}\left({1\over\chi}{\d u\over\d z}\right)=u-S.
\end{equation}
(Note that the grid should still be spaced so as to cover $\tau$
more-or-less uniformly.)
Again approximating the differential equation with a system of
differences we obtain
\begin{eqnarray}
\lefteqn{ {1\over\chi_i}\left.\left({\d\over\d z}
\left({1\over\chi}{\d u\over\d z}\right)\right)\right|_i \approx }\nonumber\\
&\qquad\approx&
{1\over\chi_i}
{ \left.\left({1\over\chi}{\d u\over\d z}\right)\right|_{i+{1\over2}} -
  \left.\left({1\over\chi}{\d u\over\d z}\right)\right|_{i-{1\over2}}
\over z_{i+{1\over2}}-z_{i-{1\over2}} } \\
&\qquad\approx&
{1\over\chi_i}
{ {\displaystyle 1\over\displaystyle\overline{\chi}_{i+1,i}}
  {\displaystyle u_{i+1}-u_i\over\displaystyle z_{i+1}-z_i}-
  {\displaystyle 1\over\displaystyle\overline{\chi}_{i,i-1}}
  {\displaystyle u_i-u_{i-1}\over\displaystyle z_i-z_{i-1}}
  \over {1\over 2}(z_{i+1}+z_i) - {1\over 2}(z_i+z_{i-1}) },
\end{eqnarray}
so that
\begin{equation}
\label{eq:trfcoeff}
\begin{array}{rcl}
a_i&=&-\left({1\over2}\chi_i(z_{i+1}-z_{i-1}) \cdot
\overline{\chi}_{i,i-1}(z_i-z_{i-1})\right)^{-1}
\smallskip\\
c_i&=&-\left({1\over2}\chi_i(z_{i+1}-z_{i-1}) \cdot
\overline{\chi}_{i+1,i}(z_{i+1}-z_i)\right)^{-1}
\smallskip\\
b_i&=&1-a_i-c_i.
\end{array}
\end{equation}
Even though these coefficients seem not too different from those of
the standard method, their impact on the computed radiation field is
significant, as witnessed by the drastic improvement in the emergent
flux shown in the lower panel of Figure~\ref{fig:discret}.  The
crucial difference in the coefficients is that the first factor in $a$
and $c$ now contains only the {\em local} opacity.  (We naturally make
the corresponding changes in the coefficients of the moments equation
as well.)

\begin{figure*}
\centerline{\includegraphics[height=\textwidth,angle=-90]{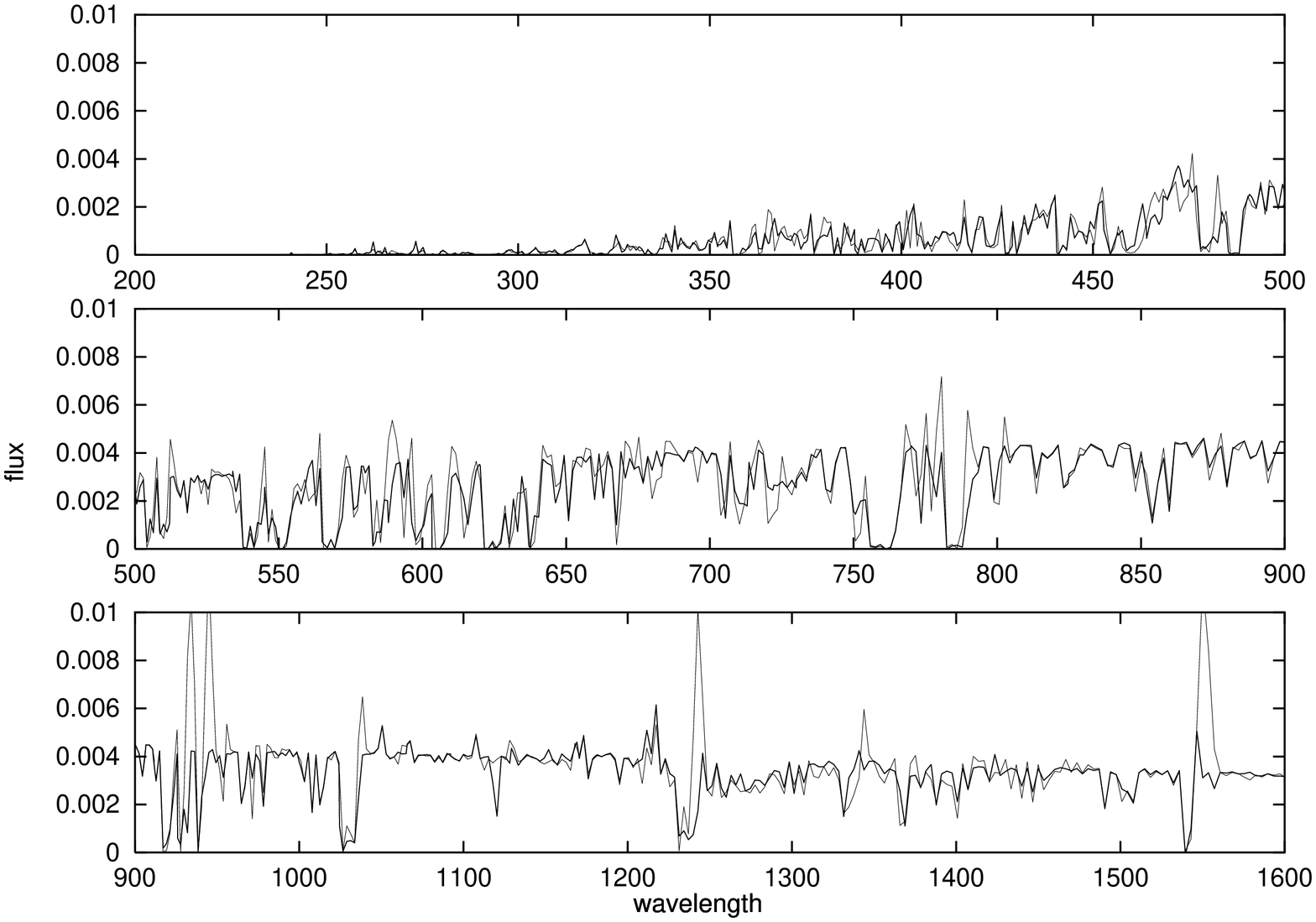}}
\mycaption{
Comparison of the synthetic EUV spectrum (200--1600\,\AA) of the S-45
supergiant model ($\Teff=45000\,{\rm K}$, $\logg=3.6$, $\Rrs=18.0$)
calculated with the {\it Doppler-spread opacity sampling method}
(thick line) and the detailed method (thin line) on the same frequency
grid.  Due to simplifications in our implementation the sampling
method cannot produce P~Cygni emission; nevertheless it provides an
extremely good basis for the final iterations using the detailed
method.}
\label{fig:sampform}
\end{figure*}

Test calculations have shown that for the second factor in the
coefficients the geometric mean
(an arithmetic mean on a logarithmic scale)
\begin{equation}
\overline{\chi}_{i+1,i} = \sqrt{\chi_{i+1}\cdot\chi_i}
\end{equation}
gives good results, as demonstrated in Figure~\ref{fig:sampform},
where the spectrum of a model computed with the opacity sampling
method is compared to that of our detailed radiative line transfer,
described in the next section.
Considering the relative coarseness of the opacity sampling method,
and the fact that the detailed line transfer suffers none of the
approximations of the sampling method, the agreement is indeed
remarkable.
Note again that through our {\em single-p-ray approximation} for the
sampling opacities (see above), our method~I (opacity sampling)
{\em cannot} produce P~Cygni profiles, since the P~Cygni
emission is a direct result of the different Doppler shifts of a
particular spectral line along different rays.

\subsection{The detailed radiative line transfer}

The detailed radiative line transfer (method~II) removes the two most
significant simplifications of our opacity sampling method (method~I),
i.\,e., it accounts for:

\arabiclist
\item Correct treatment of the angular variation of the
opacities,
\item Spatially resolved line profiles\footnote{
Note that this will not by itself solve the problem of
self-shad\-owing, since that is an intrinsic property of any method
using an ``incident radiation'' in solving for the bound-bound
radiative rates with a continuum already affected by the transition
being considered. In the iteration cycle using method~II we therefore
also have to apply our correction for self-shadowing.}
(implying correct treatment of multi-line effects).
\end{list}

Whereas in method~I the former is completely ignored, the lack of
spatial resolution was already compensated for to a large extent
through the use of our Doppler-spread sampling.  (Multi-line
interaction is partly included in our method~I, but without regard for
the sign of the Doppler shift (using just that of the central ray),
and without regard for the order of the lines along the ray within a
radius interval, as the Doppler-spread sampling effectively ``maps''
the lines to the nearest radius point.)

\begin{figure*}
\centerline{\includegraphics[width=\textwidth]{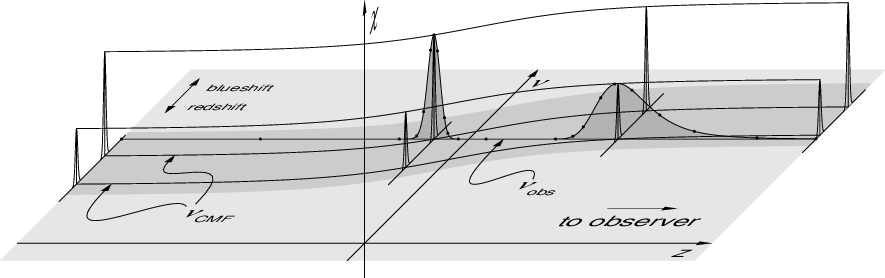}}
\mycaption{
Diagram (not to scale) illustrating the basic relationship of
the rest-frame frequencies of spectral lines ($\nu_{\rm CMF}$)
to observer's frame frequency ($\nu_{\rm obs}$) for one
particular (non-core) $p$-ray in the spherically symmetric geometry
(cf.~Figure~\ref{fig:pzgeo}).
Shown are two spectral lines which get shifted across the observer's
frame frequency by the velocity field in the wind.
The dots represent the stepping points of the adaptive microgrid used
in solving the transfer equation in the detailed radiative line transfer.}
\label{fig:formal}
\end{figure*}

With all major approximations removed, the biggest shortcoming that
remains in method~II is that only Doppler broadening is considered for
the lines, as Stark broadening has not yet been implemented.  However,
this is of no relevance for the UV spectra, as it concerns only a few
lines of Hydrogen and Helium in the optical frequency range. It will,
however, be important for our future planned analysis of the optical H
and He lines.  (Stark broadening is not considered in the sampling
method either, but here this is of minor significance, as all other
approximations are much more serious.) Method~II would remain a
sampling in frequency if the frequential resolution were chosen so
coarse that the whole Doppler profile of a line would fit between two
frequency points; only if the velocity field carries a spectral line
over a point in the frequency grid will that line be considered for
radiative transfer (at the correct radial position). Nevertheless
(with the exception of missing Stark broadening), {\em at all points
of the frequency grid the radiative transfer is solved for correctly}.

In contrast to our method~I, where the symmetry and our assumption of
only radially (not angular) dependent Doppler shifts
allowed solving the transfer equation for only one quadrant,\footnote{
The 2nd-order differential representation of the transfer equation
accounts for both the left- and right-propagating radiation
simultaneously, the unknowns being the symmetric averages of the two.}
a correct treatment of the both red {\em and} blue Doppler-shifted
line opacities (see Figure~\ref{fig:pzgeo}) requires a solution in two
quadrants\footnote{ A one-quadrant solution is also possible, but
requires both a red- and a blue-shifted opacity for each $(p,z)$-point
and separate treatment of the left- and right-directed radiation, thus
being equivalent in computational effort to the two-quadrant solution
that solves for radiation going in only one direction.} (corresponding
to, from the observer's viewpoint, the front and back hemispheres; the
rotational symmetry along the line-of-sight is taken care of through
the angular integration weights).

\begin{figure*}
\centerline{\includegraphics[height=\textwidth,angle=-90]{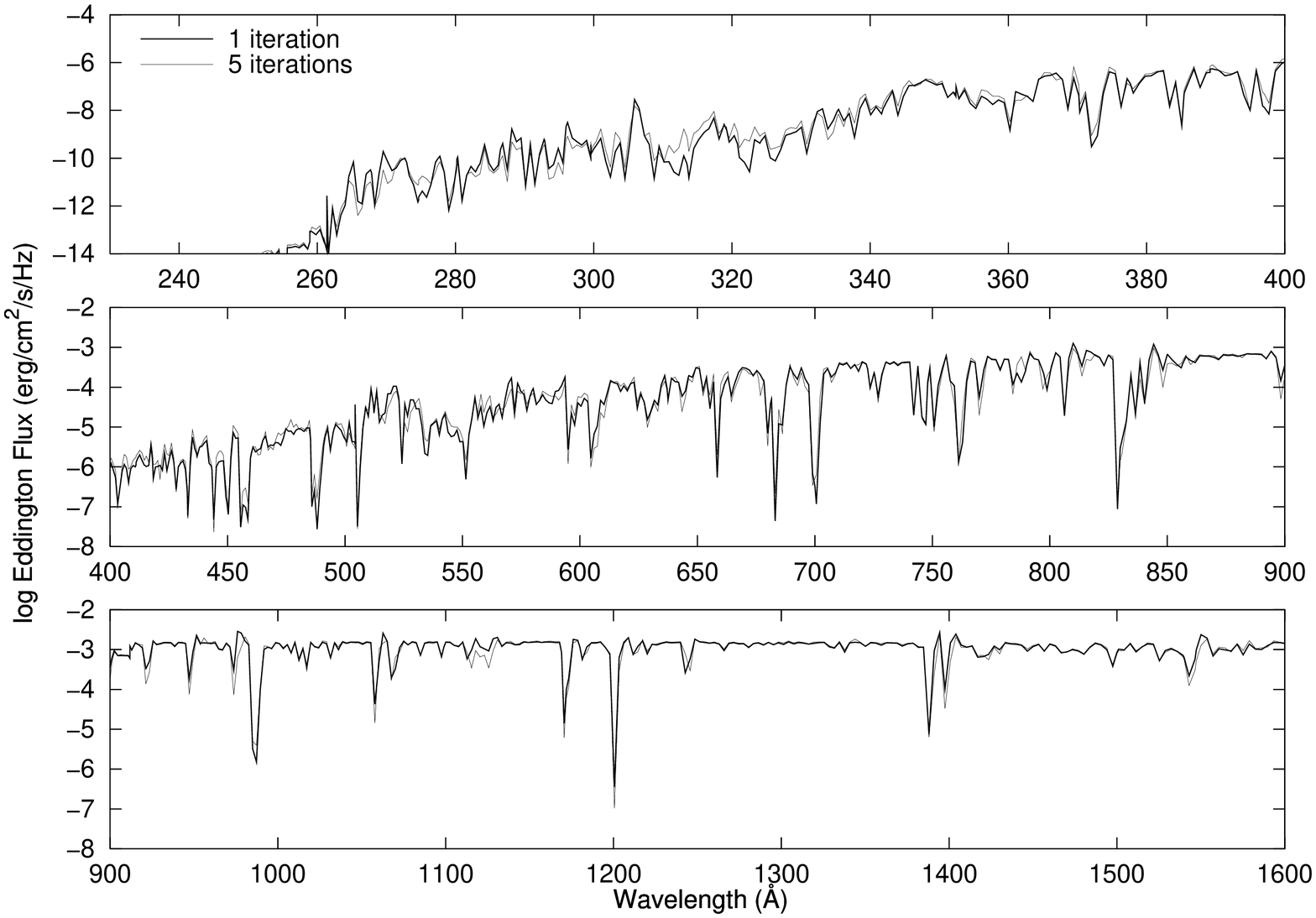}}
\mycaption{
Flux of our S-29 supergiant model
($\Teff=29\,000\,{\rm K}$, $\logg=3.0$, $\Rrs=27.0$)
after 1 and 5 iteration blocks of method~II.}
\label{fig:convergII}
\end{figure*}

\begin{figure*}
\centerline{\includegraphics[height=\textwidth,angle=-90]{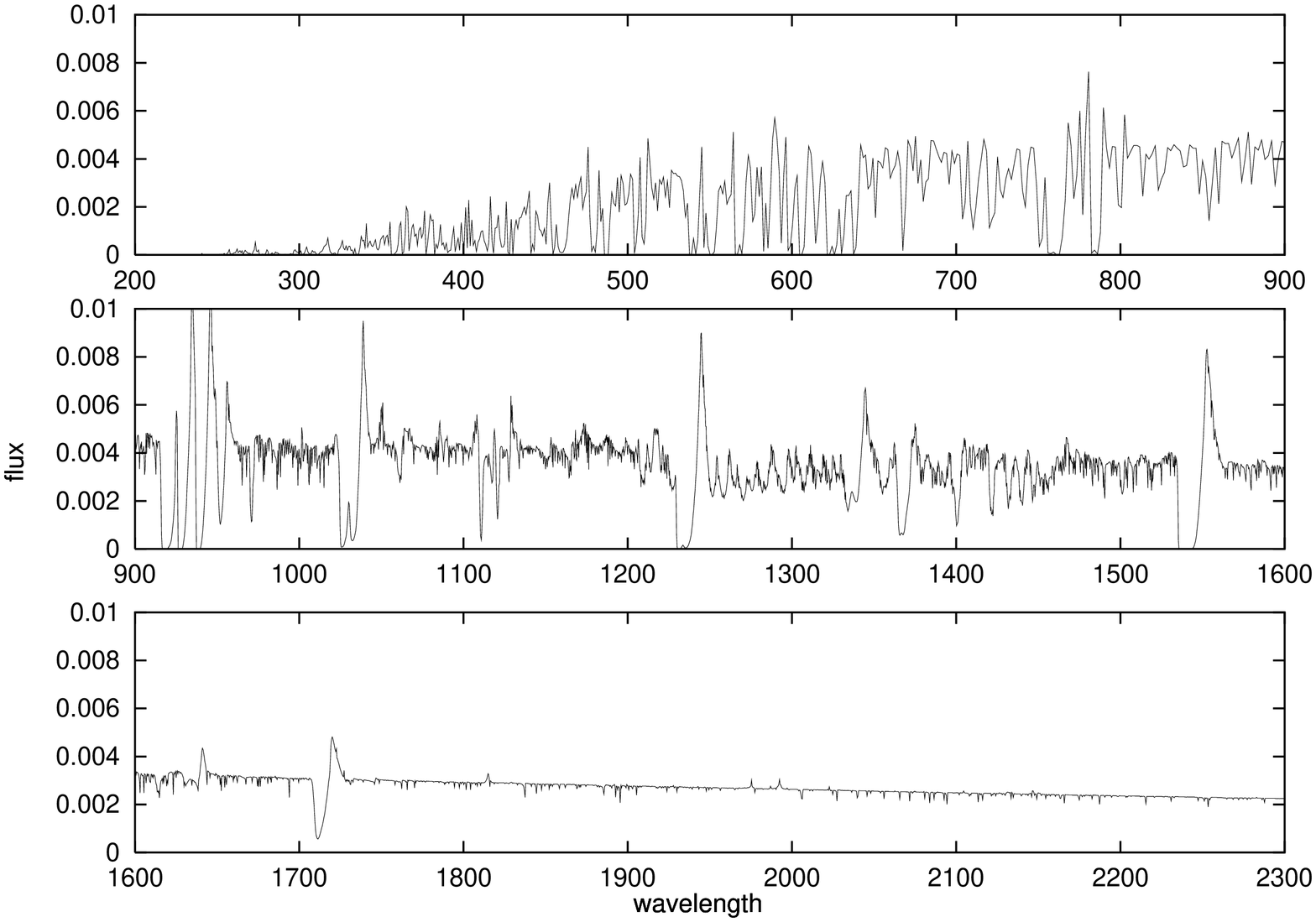}}
\mycaption{
The final --- completely converged --- spectrum
from 200 to 2300\,\AA\ of the S-45 supergiant model.
The observable UV region redwards of the Lyman edge has been computed with
a high resolution (for comparison with observed spectra).}
\label{fig:flux}
\end{figure*}

The method employed is an adaptation of the one described by Puls and
Pauldrach (1990), using an integral formulation of the transfer
equation and an adaptive stepping technique which ensures that the
optical depth in each step (``microgrid'') does not exceed
$\Delta\tau=0.3$, so that the radiation transfer in each
micro-interval can be approximated to high accuracy by an analytical
formula assuming a linear run of opacity and emissivity between the
micro-interval endpoints:
\begin{equation}
I(\tau_0)
=\int_{\tau_0}^{\tau_n} S(\tau)e^{-(\tau-\tau_0)}\,\d\tau
+I(\tau_n)e^{-(\tau_n-\tau_0)},
\end{equation}
where the integral is performed as a weighted sum on the microgrid
\begin{eqnarray}
\lefteqn{\int_{\tau_0}^{\tau_n} S(\tau)e^{-(\tau-\tau_0)}\,\d\tau=}
\nonumber\\
&=&\sum_{i=0}^{n-1}\left(e^{-(\tau_i-\tau_0)}
\int_{\tau_i}^{\tau_{i+1}} S(\tau)e^{-(\tau-\tau_i)}\,\d\tau\right),
\end{eqnarray}
each ``microintegral'' being evaluated as
\begin{equation}
\int_{\tau_i}^{\tau_{i+1}} S(\tau)e^{-(\tau-\tau_i)}\,\d\tau
=w_i^{(a)} S(\tau_i) + w_i^{(b)} S(\tau_{i+1})
\end{equation}
with weights
\begin{equation}
w_i^{(a)} = 1-{1-e^{-\Delta\tau_i}\over\Delta\tau_i}, \quad
w_i^{(b)} = {1-e^{-\Delta\tau_i}\over\Delta\tau_i}-e^{-\Delta\tau_i}
\end{equation}
where $\Delta\tau_i = \tau_{i+1}-\tau_i$.
Note that in constructing the opacities and emissivities, all line
profile functions $\varphi_{\rm l}$ (cf.~eq.~\ref{eq:oplb}) are
evaluated correctly for the current microgrid-$(z,p)$-coordinate on
the ray.  Only the slowly-varying occupation numbers (or equivalently,
the integrated, frequency-independent line opacities
$\overline\chi_{\rm l}$ and emissivities $\overline\eta_{\rm l}$) and
the velocity field are interpolated between the regular radius grid
points.

Figure~\ref{fig:formal} depicts schematically the relationship between
the Doppler-shifted frequencies of spectral lines (which are constant
in the comoving frame) and the observer's frame frequency for which
the radiative transfer is being calculated.  The figure also
illustrates the {\em line overlap} in accelerating, expanding
atmospheres: lines clearly separated in the comoving frame (slices
parallel to the $(\nu,\chi)$-plane) overlap in the observer's frame
(slice parallel to the $(z,\chi)$-plane at $\nu_{\rm obs}$) due to
large Doppler shifts many times the intrinsic (thermal and
microturbulent) linewidth.  The areas shaded in dark gray correspond
to the spatially resolved Sobolev resonance zones of the two lines for
this particular observer's frame frequency and $p$-ray.  Note that the
dimensions are not to scale, i.\,e., the intrinsic width of the lines,
and consequently the thickness in $z$ of the resonance zones, has been
greatly exaggerated in relation to the total velocity shift.

\begin{figure*}
\centerline{\includegraphics[width=12.5cm,angle=90]{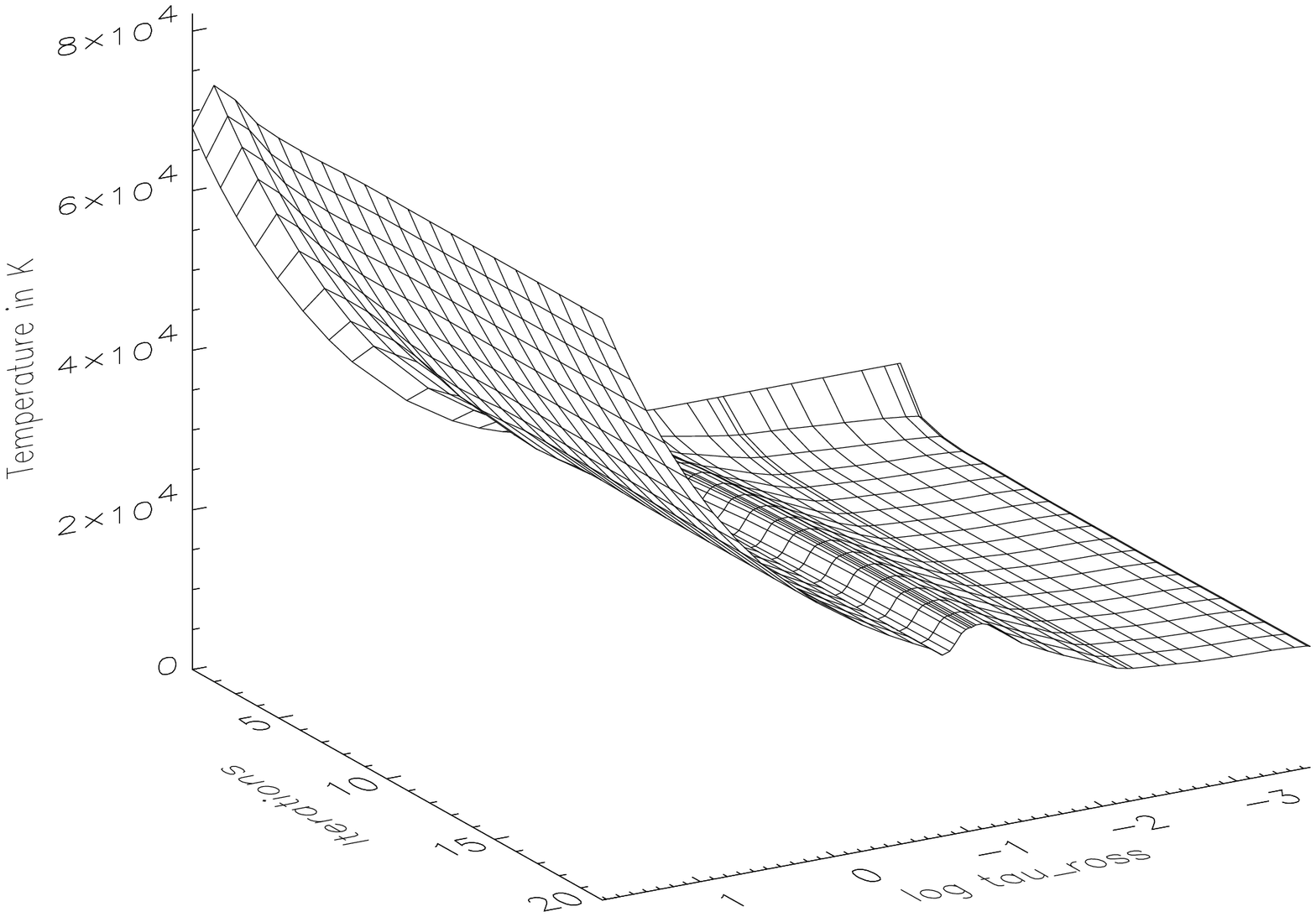}}
\mycaption{
The temperature structure versus the Rosseland optical depth and
the iteration block number for the same O supergiant model as in
Fig.~\ref{fig:converg1} ($\Teff=29\,000\,{\rm K}$, $\logg=3.0$,
$\Rrs=27.0$).}
\label{fig:convergt1}
\end{figure*}

All lines whose maximum Doppler shift $\Delta\nu=\pm\nu_0 \vinf/c$
puts them in range of the observer's frame frequency for which the
radiative transfer is being calculated are considered for that
frequency point.  In Figure~\ref{fig:formal}, these correspond to
those lines whose rest frequencies lie in the gray band in the
$(\nu,z)$-plane at $z=0$.

After the occupation numbers have converged in the iteration cycle
using method~I, one iteration block with method~II is usually
sufficient for full convergence of the model, as demonstrated by
Figure~\ref{fig:convergII}, where the emergent spectrum of our S-30
model after 1 iteration block of method~II is compared to the
spectrum resulting from 5 iteration blocks.

A high-resolution spectrum is computed for the purpose of comparison
with observations (wavelength range usually from 900 to 1600\,\AA)
after full convergence of the model.  This spectrum is generated with
exactly the same procedure as used for the detailed line blocking
calculations, and would thus correspond to a further iteration block of
method~II. The high-resolution spectrum is then merged with the
lower-resolution blocking flux ($\lambda < 900\,{\rm \AA}$) for the
final flux output (Figure~\ref{fig:flux}).

\subsection{Line blanketing}
\label{sec:lineblanket}

Line absorption and emission also has an important effect on the
atmospheric {\em temperature structure}. The corresponding influence
on the radiation balance is usually referred to as {\em line
blanketing}. The objective now is to calculate an atmospheric
temperature stratification which conserves the radiative flux and
which treats the impact of the line opacities and emissivities
properly. In principle there are three methods for calculating electron
temperatures in model atmospheres. The commonly used one is based on
the condition of radiative equilibrium. The second one uses a flux
correction procedure, and the third one is based on the thermal
balance of heating and cooling rates. As the first method has some
disadvantages (see below), we use the second and the third method
(\cite{}cf.~Pauldrach et al.,~1998). In deeper layers ($\tauross>0.1$)
where true absorptive processes dominate we use the flux correction
procedure, and the thermal balance is used in the outer part of the
expanding atmosphere ($\tauross<0.1$), where scattering processes
start to dominate.

\paragraph{The flux correction procedure.}

\begin{figure*}
\centerline{\includegraphics[width=12.5cm,angle=90]{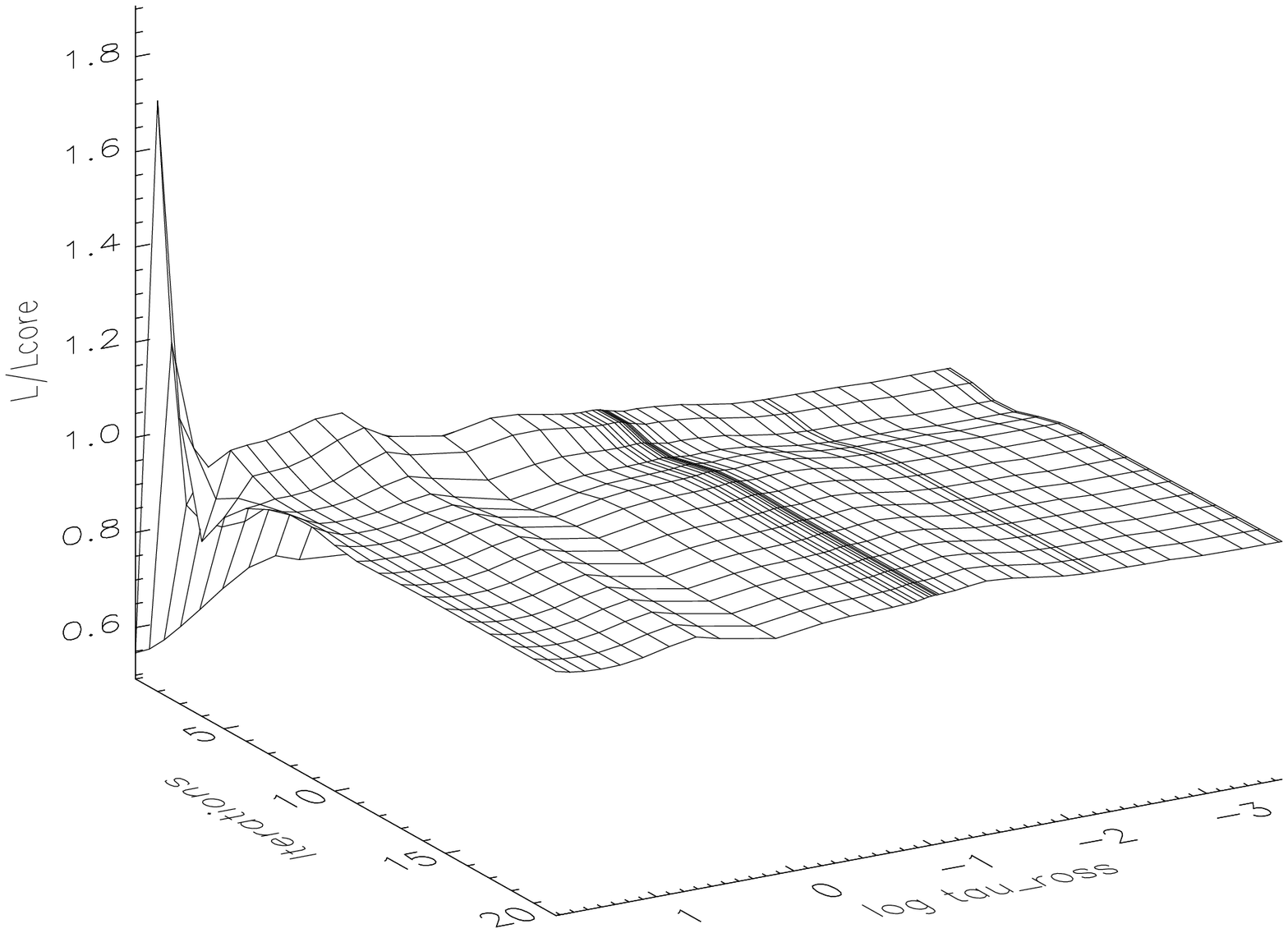}}
\mycaption{
The flux conservation versus the Rosseland optical depth and the
iteration block number for the same O supergiant model as in
Fig.~\ref{fig:converg1} ($\Teff=29\,000\,{\rm K}$, $\logg=3.0$,
$\Rrs=27.0$). The accuracy of the flux conservation is on the 1\%~level
for the final iteration blocks.}
\label{fig:convergt2}
\end{figure*}

The idea of this method is straightforward: The local temperature has
to be adjusted in such a way that the radiative flux is conserved.
This requires, however, that the temperature is the dominant parameter
on which the flux depends, and that the effect of a change in
temperature on the flux is known. The first condition is certainly the
case for $\tauross>0.1$. With regard to the second condition, a law
for the temperature structure is required which is controlled by some
global parameters that can be adjusted in a proper way in order to
conserve the flux. The ``Hopf function'', usually applied for the grey
case, has an appropriate functional dependence which has been adapted
to the spherical NLTE case by \cite{}Santolaya-Rey et al.~(1997)
recently in a general way. The main characteristic of the Hopf method
is that the Rosseland optical depth is the decisive parameter on which
the temperature stratification depends.  Thus, in deeper layers the
temperature structure can be calculated efficiently by using this new
concept of the {\it NLTE Hopf function}:
\begin{equation}
\label{eq:nhopf}
T^4(r) = \Teff^4 {3 \over 4} {\taur \over \tauross}
\left( \qn(\taur)+\tauross \right)
\end{equation}
where $\taur$ is the radial optical depth in the spherical case,
\begin{equation}
\d\taur = \chiross(r) \left( \frac{R_\ast}{r} \right)^2 \d r,
\end{equation}
and
\begin{equation}
\qn(\taur) \simeq \qinf + (\q0-\qinf) \exp(-\gamma \taur)
\end{equation}
is the spherical NLTE Hopf function, where the parameters $\q0$,
$\qinf$, and $\gamma$ are fitted to a predefined run of the
$\qn(\tauross)$ stratification (cf.~\cite{}Santolaya-Rey et al.,~1997).
Test calculations performed with fixed parameters ($\q0$, $\qinf$, and
$\gamma$) and without metal lines lead to almost identical results for
the temperature structures obtained by our code and the completely
independently developed code of
Santolaya-Rey et~al.\ (cf.~\cite{}Pauldrach et al.,~1998). The
reliability of the method has been further proven by the resulting
flux conservation which turns out to be on the 1\%~level.

In the next step, {\it line blocking} has to be {\it treated
consistently}.  Although the line processes involved are complex, they
always increase the Rosseland optical depth ($\tauross$). In the
deeper layers ($\tauross>0.1$) this leads directly to an enhancement
of the temperature law (backwarming). Using the method of the {\it
NLTE Hopf functions} we thus have to increase the parameter $\q0$
first by using the flux deviation at $\tauross \approx 3$. This
parameter is updated in the corresponding iteration cycle until the
flux is conserved at this depth on the 0.5\%~level. Afterwards the
same is done with the parameter $\qinf$ at an optical depth of
$\tauross \approx 0.1$. In case the flux deviation at $\tauross
\approx 3$ becomes larger than 0.5\% the parameter $\q0$ is iterated
again with a higher priority. As a last step in this procedure the
parameter $\gamma$ is adjusted in order to conserve the flux at an
optical depth of $\tauross \approx 1$.

The resulting temperature structure and the corresponding flux
deviation for an O~supergiant model ($\Teff=29\,000\,{\rm K}$,
$\logg=3.0$, $\Rrs=27.0$) are shown in Fig.~\ref{fig:convergt1} and
Fig.~\ref{fig:convergt2}, respectively.  As can be inferred from the
figure, the flux is conserved for this model with an accuracy of a few
percent. (We note that from test calculations where line blocking was
treated, but parameters of the NLTE Hopf function were held fixed,
thus effectively ignoring blanketing effects, we found a flux
deviation which already starts in the inner part ($\tauross<50$) and
reaches a value of up to 50\% at $\tauross\approx 0.1$.) This clearly
shows the importance of blanketing and backwarming effects and the
need to include them. As these test calculations have also shown that
absorptive line opacities dominate the total opacity down to an
optical depth of $\tauross>0.1$, the temperature structure is
influenced by backwarming effects in the entire atmosphere ---
cf.~eq.~\ref{eq:nhopf}.

\paragraph{The thermal balance.}

In the outer part of the expanding atmosphere ($\tauross<0.1$), where
scattering processes start to dominate, the effects of the line
influence on the temperature structure are more difficult to treat. Of
the two possible treatments, calculating for radiative equilibrium or
for thermal balance, we have chosen the latter one, as the convergence
of the radiative equilibrium method turned out to be problematic since
the $\tauross$-values are small in this part, and, hence, most
frequency ranges are optically thin.  (This has recently also been
proven by \cite{}Kubat et al.~(1999), where the corresponding
equations of the method are also presented.) In calculating the
heating and cooling rates (\cite{}Hummer and Seaton,~1963), all
processes that affect the electron temperature have to be included ---
bound-free transitions (ionization and recombination), free-free
transitions, and inelastic collisions with ions. For the required
iterative procedure we make use of a linearized Newton-Raphson method
to extrapolate a temperature that balances the heating and cooling
rates.

Fig.~\ref{fig:convergt1} displays for the model above the resulting
temperature structure vs.\ the number of iterations and shows a pronounced
bump and a successive decrease of the temperature in the outer atmospheric
part. (Note that the mismatch of the heating and cooling rates which
immediately goes to 0\% in the outer part ($\tauross<0.07$) where it
is applied for correcting the temperature structure has already been
presented by \cite{}Pauldrach et al.~(1998).)


\subsection{Revised inclusion of EUV and X-ray radiation}
\label{sec:shocks}

The {\it EUV and X-ray radiation} produced by cooling zones which
originate from the simulation of {\it shock heated matter} arising
from the non-stationary, unstable behaviour of radiation driven winds
(see \cite{}Lucy and Solomon~(1970), who found that radiation driven
winds are inherently unstable, and \cite{}Lucy and White~(1980) and
Lucy~(1982), who explained the X-rays by radiative losses of
post-shock regions where the shocks are pushed by the non-stationary
features) is, together with K-shell absorption, included in our
radiative transfer. The primary effect of the EUV and X-ray radiation
is its influence on the ionization equilibrium with regard to high
ionization stages like \NV\ and \OVI\ (cf.~the problem of
``superionization'', the detection of the resonance lines of \OVI,
\NV, \SVI\ in stellar wind spectra (cf.~\cite{}Snow and Morton, 1976);
in a first step this problem was investigated theoretically by
\cite{}Cassinelli and Olson,~1979) where the contribution of enhanced
direct photoionization due to the EUV shock radiation is as important
as the effects of Auger-ionization caused by the soft X-ray radiation
(cf.~\cite{}Pauldrach,~1987; Pauldrach et al.,~1994 and~1994a). In
order to treat this mechanism accurately it is obviously important to
describe the radiation from the shock instabilities in the stellar
wind flow properly. Note that in most cases a small fraction of this
radiation leaves the stellar wind to be observed as soft X-rays with
$L_x/L_{\rm bol}\approx 10^{-7}$ (cf.~\cite{}Chlebowski et al.,~1989).
Thus, the reliability of the shock description can be further
demonstrated by a comparison to X-ray observations, by ROSAT for
instance.

In principle, a correct calculation of the creation and development
of the shocks is required for the solution of the problem. This means
that a detailed theoretical investigation of time-dependent radiation
hydrodynamics has to be performed (for exemplary calculations see
\cite{}Owocki, Castor, and Rybicki~(1988) and Feldmeier~(1995)).
However, these calculations favour the picture of a stationary
``cool wind'' with embedded randomly distributed shocks where the
shock distance is much larger than the shock cooling length in the
accelerating part of the wind. They also indicate that only a small
amount of high velocity material appears with a filling factor not
much larger than $f\approx 10^{-2}$, and jump velocities of about
$u_{\rm s} = 300\dots700\,{\rm km/s}$ which give immediate post-shock
temperatures of approximately $T_{\rm s} = 1\times 10^{6}$ to $8\times
10^{6}\,{\rm K}$.  We also note that the reliability of these results
was already demonstrated by a comparison to ROSAT-observations
(cf.~\cite{}Feldmeier et al.,~1997).

On the basis of these results we had developed an empirical
approximative description of the EUV and X-ray radiation, where the
shock emission coefficient
\begin{equation}
\label{eq:p12}
\epsilon_\nu^{\rm s}(r)=\frac {f}{4\pi}n_{\rm p}n_{\rm e}
  \Lambda_\nu(T_{\rm s}(r)\,n_{\rm e})
\end{equation}
was incorporated in dependence of the volume emission coefficient
$\Lambda_\nu$ calculated by using the \cite{}Raymond and Smith~(1977)
code for the X-ray plasma, the velocity-dependent post shock
temperatures $T_{\rm s}$, and the filling factor $f$ which enter as
fit parameters --- these values are determined from a comparison of
the calculated and observed ROSAT ``spectrum''. With this description
the effects on the high ionization stages (\NV, \OVI) lead to
synthetic spectral lines which reproduce the observations almost
perfectly (cf.~Pauldrach et al.,~1994 and~1994a). However, with this
method we were not able to reproduce the ROSAT-observations with the
same model parameters simultaneously (see below). We therefore had to
determine the filling factor and the post-shock temperatures by a
separate and hence in view of our concept not consistent procedure
(cf.~\cite{}Hillier et al.,~1993).  In order to overcome this problem
refinements to our method are obviously required.

In the present treatment the outlined approximative description of the
EUV and X-ray radiation has been revised. The major improvement
consists of the consideration of {\em cooling zones} of the randomly
distributed shocks embedded in the stationary cool component of the
wind. Up to now we had assumed, for reasons of simplicity, that the
shock emission is mostly characterized by the immediate post-shock
temperature, i.\,e., we considered {\em non-stratified, isothermal}
shocks.  This, however, neglects the fact that shocks have a cooling
structure with a certain range of temperatures that contribute to the
EUV and X-ray spectrum. Our revision comprises two modifications to
the shock structure. The first one concerns the inner region of the
wind, where the cooling time can be regarded to be shorter than the
flow time.  Here the approximation of {\em radiative} shocks can be
applied for the cooling process (cf.~\cite{}Chevalier and
Imamura,~1982).  The second one concerns the outer region, where the
stationary terminal velocity is reached, the radiative acceleration is
negligible, and the flow time is therefore large. Here radiative
cooling of the shocks is of minor importance and the cooling process
can be approximated by {\em adiabatic expansion} (cf.~\cite{}Simon and
Axford~(1966), who investgated a pair of reverse and forward shocks
that propagate through an ambient medium under these circumstances).
For our purpose we followed directly the modified concept of
isothermal wind shocks presented recently by \cite{}Feldmeier et
al.~(1997).

Compared to eq.~\ref{eq:p12} we account for the density and
temperature stratification in the shock cooling layer by replacing the
values of the volume emission coefficient ($\Lambda_\nu(T_{\rm
s}(r)\,n_{\rm e})$) through adequate integrals over the cooling zones
denoted by $\hat \Lambda_\nu(T_{\rm s}(r))$. Thus, $\epsilon_\nu^{\rm
s}(r)$ is replaced by
\begin{equation}
\hat\epsilon_\nu^{\rm s}(r)=\frac {f}{4\pi}n_{\rm p}n_{\rm e}
\hat\Lambda_\nu(T_{\rm s}(r)),
\end{equation}
where
\begin{equation}
\label{eq:feld1}
\hat\Lambda_\nu(T_{\rm s}(r))=\pm \frac{1}{x_{\rm s}}
\int\limits_r^{r\pm x_{\rm s}}\hat f^2(r')\Lambda_\nu(T_{\rm s}(r')\cdot
\hat g(r'))\,\d r',
\end{equation}
and $r$ is the location of the shock front, $r'$ is the cooling length
coordinate with a maximum value of $x_{\rm s}$, the plus sign
corresponds to forward and the minus sign to reverse shocks, and $\hat f(r')$
and $\hat g(r')$ denote the normalized density and temperature structures with
respect to the shock front. The improvement of our treatment is now
obviously directly connected to the description of the latter
functions. In the present step we used the analytical approximations
presented by Feldmeier et al.~(1997), which are based on the two
limiting cases of radiative and adiabatic cooling layers behind shock
fronts (see above).


\begin{figure*}
\centerline{\includegraphics[height=14cm,angle=270]{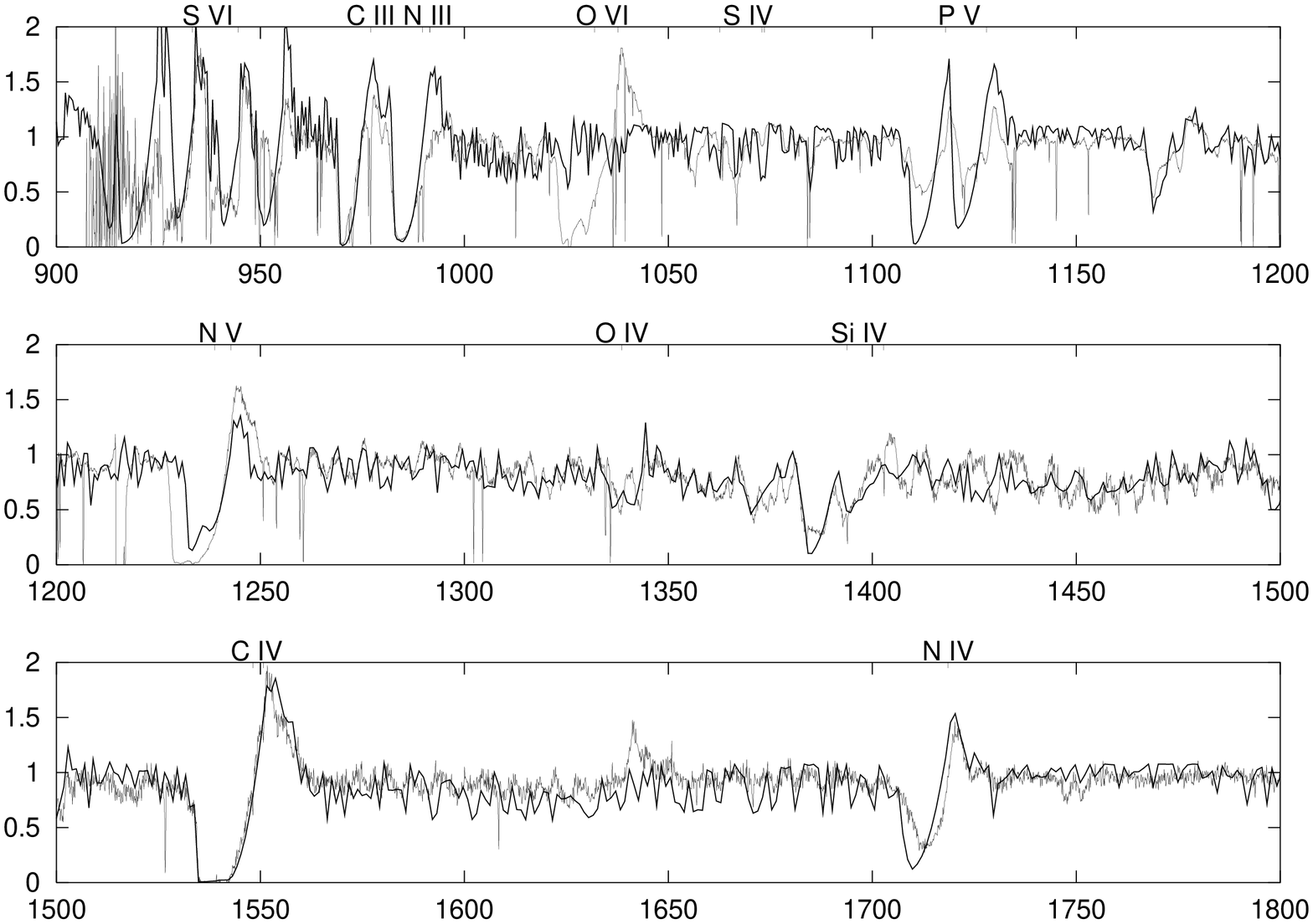}}
\mycaption{
Calculated and observed UV spectrum for the O4f-star $\zeta$~Puppis.
The calculated spectrum belongs to a model where the influence of
shock emission has been neglected. The high resolution observations
have been obtained with the IUE and Copernicus satellites. (Note that
the improved blocking and blanketing treatment has not been considered
for the model calculations of this object --- see text.)}
\label{fig:ovi_0}
\vspace{2mm}
\centerline{\includegraphics[height=14cm,angle=270]{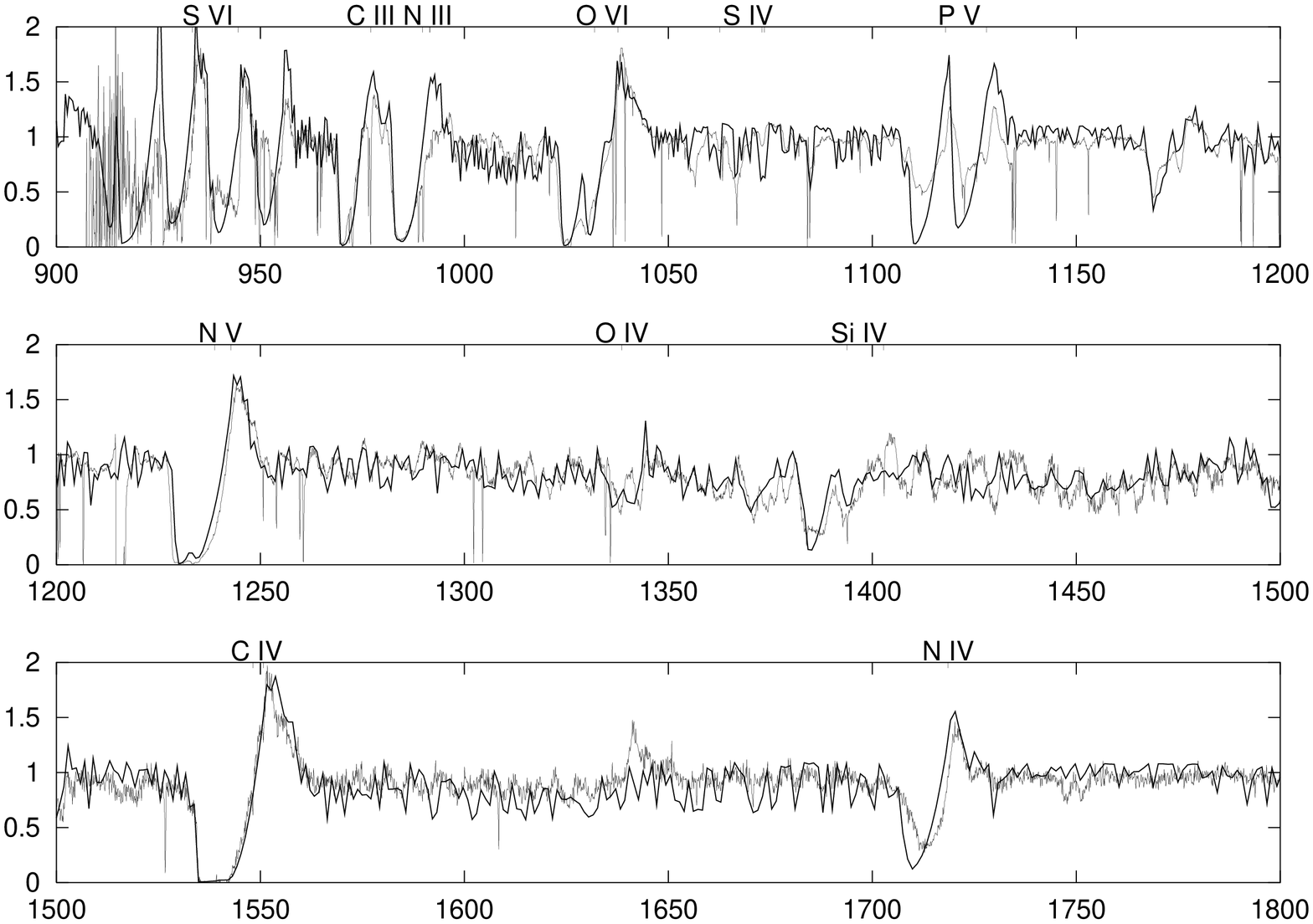}}
\mycaption{
Calculated and observed UV spectrum for the O4f-star $\zeta$~Puppis.
The calculated spectrum belongs to a model where the influence of
shock emission has been included --- former method (model 1).}
\label{fig:ovi_1}
\end{figure*}

\subsubsection{Test calculations}

In the following we present results of test calculations showing the
influence of our modified treatment of shock emission. For this
purpose we selected the O4f-star $\zeta$~Puppis as a test object and
ignored for the corresponding model calculations the improved blocking
and blanketing treatment discussed above. This restriction makes our
results directly comparable to those of Pauldrach et al.~(1994a), who
used the old, simplified treatment for the shock emission. The stellar
parameters of $\zeta$~Puppis, used as basic input for our models, have
been adopted from Pauldrach et al.~(1994) (see Table~\ref{tbl:zpuppara})
together with the abundances
\begin{center}
\begin{tabular}{r@{~$=$~}l@{\quad}r@{~$=$~}l}
$Y_{\rm He}$ & $1.20\,Y_{{\rm He},\odot}$ &
$Y_{\rm C}$  & $0.35\,Y_{{\rm C},\odot}$ \\[1mm]
$Y_{\rm N}$  & $8.00\,Y_{{\rm N},\odot}$ &
$Y_{\rm O}$  & $0.75\,Y_{{\rm O},\odot}$
\end{tabular}
\end{center}
($Y_{\rm X}:=n_{\rm X}/n_{\rm H}$,
where $Y_{{\rm X},\odot}$ denotes the solar abundance.)
For all other abundances solar values were used.

\begin{table}[b]
\begin{center}
\begin{tabular}{|c|c|c|c|c|c|c|}
\hline  & & & & & \\
$\log(\frac{L}{L_\odot})$ & $\frac{T_{\rm ef\/f}}{10^3{\rm K}}$&
$\log g$&$\frac{R_\star}{R_\odot}$&$\frac{v_\infty}{{\rm km/s}}$&
$\frac{\dot M}{10^{-6}M_\odot/yr}$\\
 & & & & & \\
\hline  & & & & & \\
6.006 & 42 & 3.625 & 19 & 2250 & 5.9\\
 & & & & & \\  \hline
\end{tabular}
\end{center}
\mycaption{
The stellar parameters of the O4f-star $\zeta$~Puppis.}
\label{tbl:zpuppara}
\begin{center}
\begin{tabular}{|c|c|c|}
\hline  & & \\ $\log (\frac{T_{\rm s}}{{\rm K}})$
&$\frac{f}{10^{-3}}$&$\log(\frac{N_{\rm H}}{{\rm cm}^{-2}})$\\   & & \\
\hline  & & \\  6.75 & 4.3 & 20.00\\  & & \\  \hline
\end{tabular}
\end{center}
\mycaption{
Parameters required for describing the cooling zones of the shocked gas.}
\label{tbl:feld}
\end{table}

Although the final objective of our treatment is the determination of
the maximum post-shock temperature $(T_{\rm s})$ and the filling
factor ($f$) from a comparison of the calculated and observed ROSAT
spectrum, we have also adopted these values for the present test
calculations from the similar fits performed by Feldmeier et
al.~(1997). The values are given in Table~\ref{tbl:feld} together with
the interstellar column density of hydrogen ($\log(N_{\rm H})$,
cf.~\cite{}Shull and van Steenberg, 1985).

\begin{figure}
\centerline{\includegraphics[height=\columnwidth,angle=270]{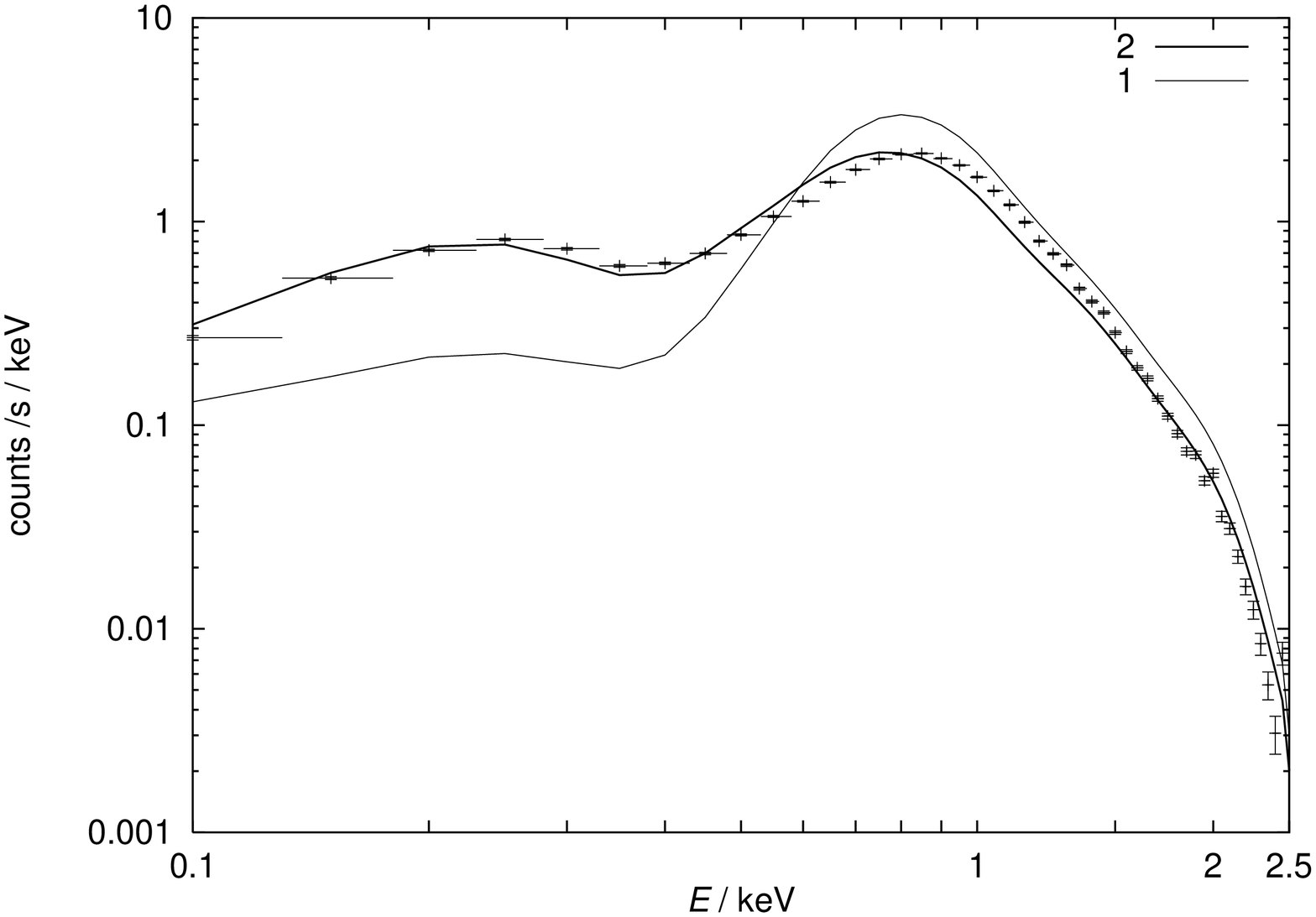}}
\mycaption{Comparison of the ROSAT-observations (error bars)
with the results of model~2 (thick line) and model~1 (thin line) for
$\zeta$~Puppis.}
\label{fig:rosat}
\vspace{3mm}
\centerline{\includegraphics[height=\columnwidth,angle=270]{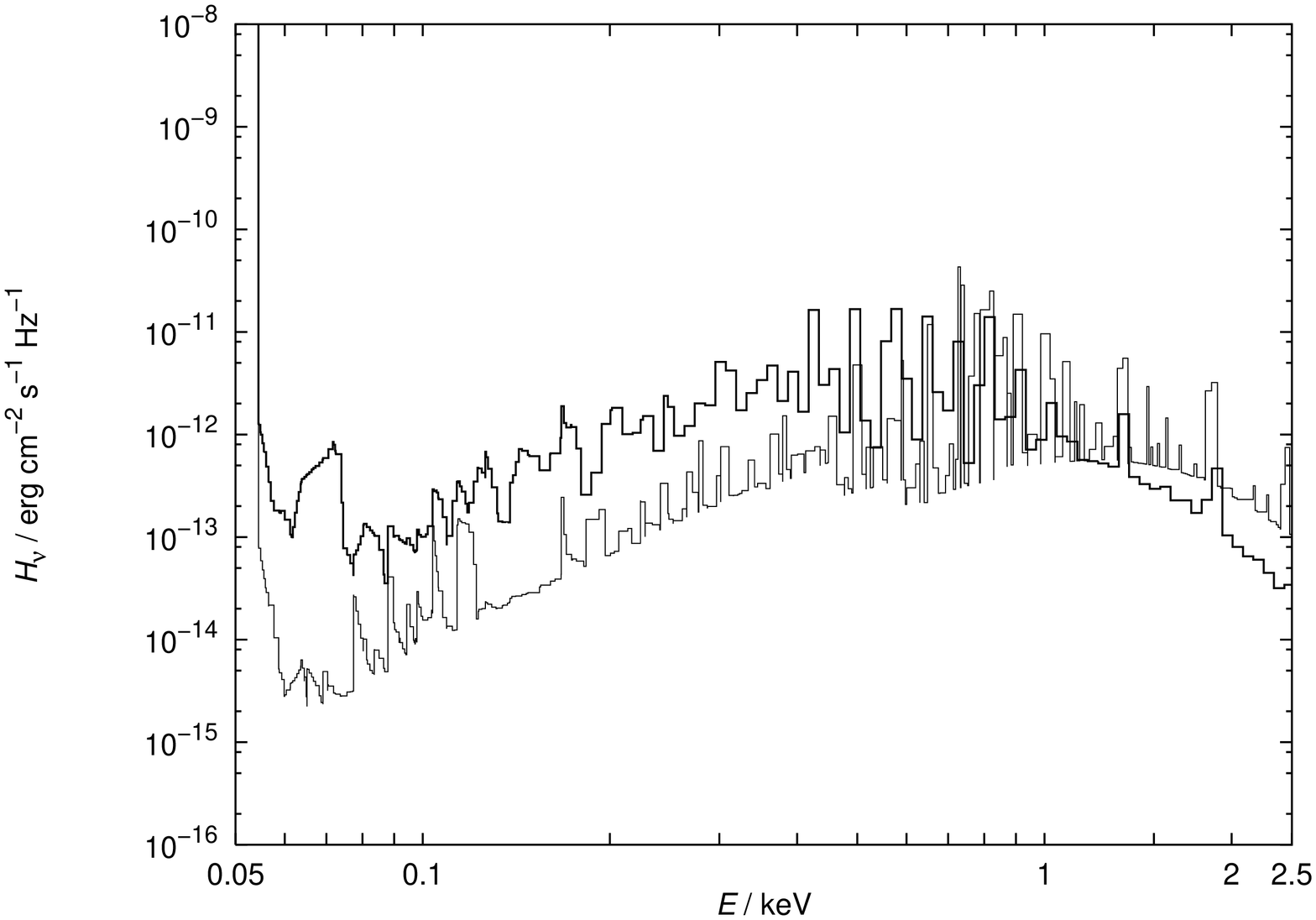}}
\mycaption{The Eddington flux in the EUV and X-ray band resulting from
model~2 (thick line) and model~1 (thin line) for $\zeta$~Puppis.
Note that the maximum shock temperatures are identical for both models.}
\label{fig:hnu}
\vspace{3mm}
\centerline{\includegraphics[height=\columnwidth,angle=270]{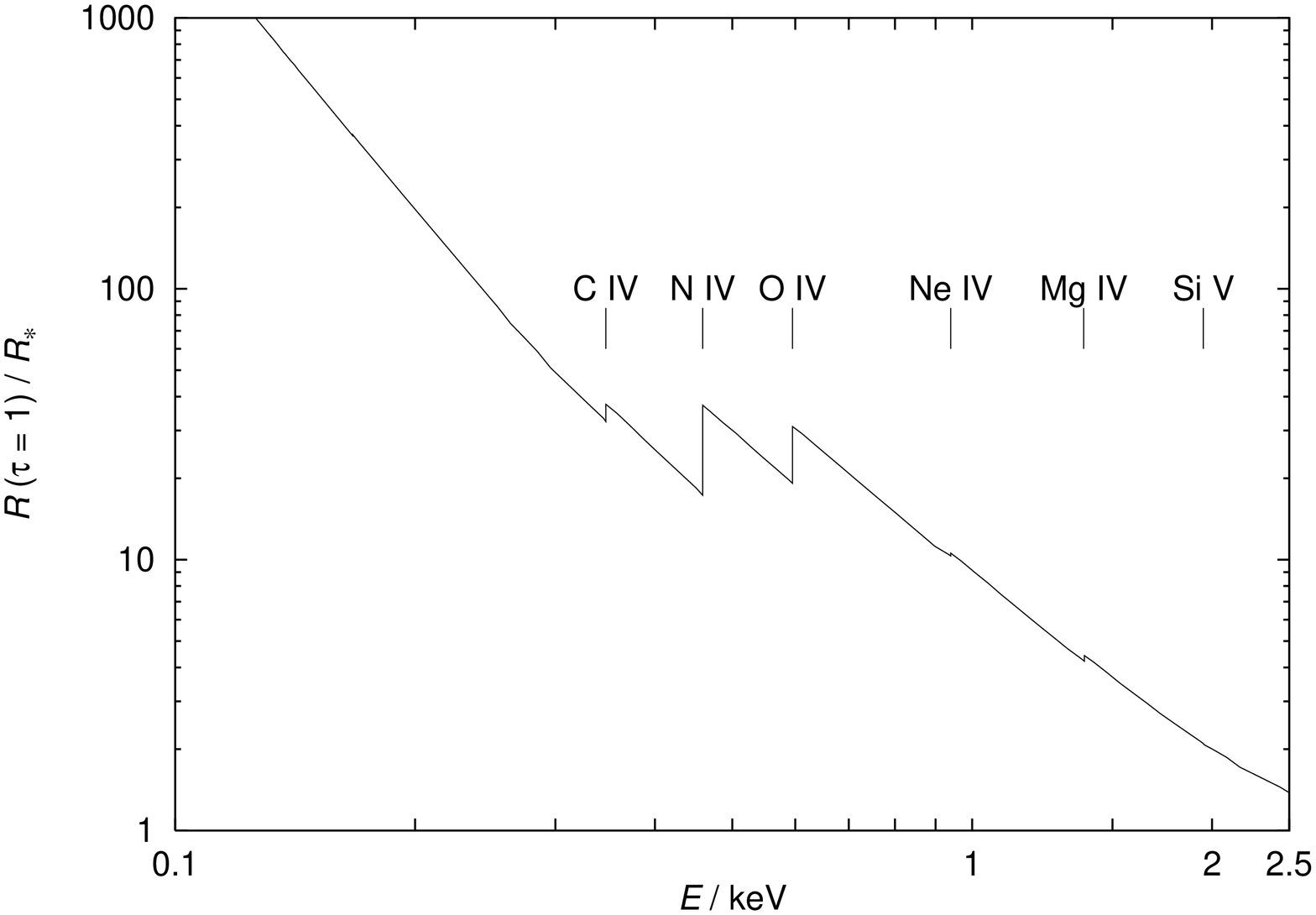}}
\mycaption{Spatial location of the optical depth unity in the
relevant energy band of ROSAT, displaying the origin of the observed flux.
Also shown is the influence of the K-shell opacities for model~2 of
$\zeta$~Puppis.}
\label{fig:kshell}
\end{figure}

We start with a spectrum synthesis calculation where EUV and
X-ray radiation by shock heated matter {\it is neglected}. The
comparison between the observed and the synthetic spectrum
(Fig.~\ref{fig:ovi_0}) shows clearly that the strong observed
resonance lines of \NV\ and \OVI\ are not reproduced by the model.
This striking discrepancy illustrates what is meant by the problem of
{\it ``superionization''}. In Fig.~\ref{fig:ovi_1} we demonstrate,
however, that this problem has already been solved by making use of
the EUV and X-ray radiation resulting from the treatment of isothermal
shocks (previous method --- model~1).  The observed resonance lines of
\NV\ and \OVI\ are reproduced quite well, apart from minor
differences. Thus it seems that the wind physics are correctly
described. That this is actually not the case can be inferred from
Fig.~\ref{fig:rosat} where the ROSAT PSPC spectrum (error bars) is
shown together with the result of model~1 (thin line). The deficiency
of the non-stratified isothermal shocks is obvious --- the model
yields much too little radiation in the soft X-ray part (shortward of
$0.7\,\keV$ the spectrum is more likely characterized by a cooler
shock component of $\log T_{\rm s} \approx 6.30$) and considerably too
much in the harder energy band.

\begin{figure*}
\centerline{\includegraphics[height=14cm,angle=270]{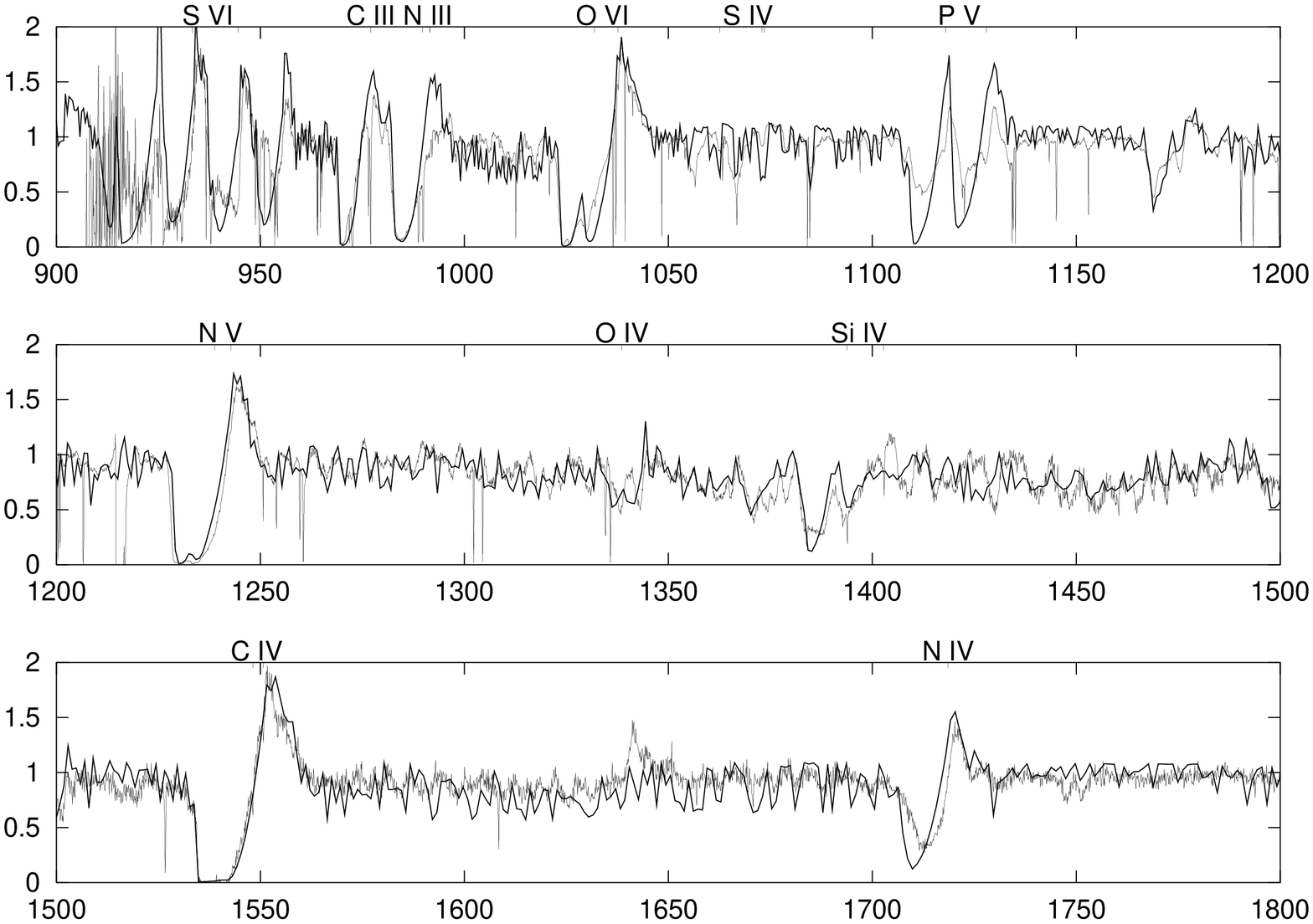}}
\mycaption{
Calculated and observed UV spectrum for the O4f-star $\zeta$~Puppis.
The calculated spectrum belongs to a model where the influence of
shock emission has been included in accordance with our improved
method (model~2).}
\label{fig:ovi_n}
\end{figure*}

Following the strategy outlined above we now investigate how far the
structured cooling zones behind the shocks can influence this negative
result. Fig.~\ref{fig:rosat}, which shows in addition the calculated X-ray
spectrum of our new model (model~2, thick line), illustrates the improvement.
Strikingly, the new calculations can quite well reproduce the ROSAT PSPC
spectrum and the comparison shown is at least of the same quality as that
obtained by Feldmeier et al.~(1997) with their best fit (see also
\cite{}Stock,~1998) --- note that the total X-ray luminosity of this model
is given by
\begin{equation}
L_{\rm x}=\int_{0.1\,\keV}^{2.5\,\keV}L_\nu\,\d\nu =
10^{-7.1} L_{\rm bol}.
\end{equation}
Actually, it is the fact that, compared to the non-stratified
isothermal shocks, the post-shock cooling zones with their temperature
stratifications radiate much more efficiently in the soft spectral
band which leads to the improved fit. This is portrayed in
Fig.~\ref{fig:hnu}.
Fig.~\ref{fig:kshell} shows the location of the optical depth unity in
the relevant energy band of ROSAT. Apart from displaying the influence
of the K-shell opacities, it becomes evident from this figure that the
wind is optically thick up to large radii, especially in the soft
X-ray band. This fact reduces the significance of the fit of the ROSAT
spectrum, because most of the observed X-ray radiation is obviously
emitted in the outermost part of the wind and thus only the properties
of the radiation produced in this region can be analyzed from the
observed spectrum. This, however, is not the case for the EUV and
X-ray radiation which populates the occupation numbers connected with
the resonance lines of \NV\ and \OVI, since due to their P-Cygni
structure these lines provide information about the complete wind
region and the properties of the influencing radiation produced in the
the whole wind region can therefore be analyzed by means of the
spectral line diagnostics.  Hence, for the significance of our
modified method it is therefore extremely convincing that the
synthetic UV-spectrum resulting from model~2 reproduces the observed
resonance lines of \NV\ and \OVI\ as well, as is shown in
Fig.~\ref{fig:ovi_n}.  That both model~1 and model~2 yield a good fit
of the P-Cygni lines shows, on the other hand, that distinguishing
between two different models from the profiles alone is not always
possible.  The fact that our new treatment accounting for the
structured cooling zones behind the shocks solves not only the problem
of {\it ``superionization''}, but reproduces for the first time
consistently the ROSAT PSPC spectrum {\em as well as} the resonance
lines of \NV\ and \OVI\ gives us confidence in our present approach.
(Note that the {\em new treatment} of the X-ray radiation is not yet
available in the download version of the code; it will be implemented
in an upcoming version (2.$x$).)

\def\0{\phantom{0}}
\def\x{\phantom{.}}
\begin{table*}
\mycaption{
The parameters of our basic model grid stars.
The Zanstra integrals given here are defined as
$\displaystyle Q_X=\int_{\nu_X}^\infty {H_\nu\over h\nu}\,\d\nu$,
where $h\nu_X$ is the ionization energy of ion $X$.}
\label{tbl:gridparams}
\begin{center}
\begin{tabular}{lcccccccc}
\hline
\hline
Model & $\Teff$ & $\logg$ & $R$ & $\vinf$ & $\Mdot$ &
 $\log Q_{\rm H}$ & $\log Q_{\rm He^+}$ & $H_\nu(5480\,\AA)$ \\
& (K) & (cgs) & ($R_\odot$) & (km/s) & ($10^{-6} M_\odot/{\rm yr}$) &
 & & (10$^{-3}$ erg/s/cm$^2$/Hz) \\
\hline
\hline
\multicolumn{9}{c}{Dwarfs} \\
\hline
D-30   & 30000 & 3.85 & 12 & 1800 & \00.008    & 21.42 & \08.42 & 0.3702 \\
D-35   & 35000 & 3.80 & 11 & 2100 & \00.05\0   & 22.65 &  11.41 & 0.4771 \\
D-40   & 40000 & 3.75 & 10 & 2400 & \00.24\0   & 23.15 &  17.63 & 0.5859 \\
D-45   & 45000 & 3.90 & 12 & 3000 & \01.3\0\0  & 23.45 &  18.99 & 0.6817 \\
D-50   & 50000 & 4.00 & 12 & 3200 & \05.6\0\0  & 23.69 &  20.28 & 0.7743 \\
D-55   & 55000 & 4.10 & 15 & 3300 & 20\x\0\0\0 & 23.89 &  20.17 & 0.8881 \\
\hline
\multicolumn{9}{c}{Supergiants} \\
\hline
S-30   & 30000 & 3.00 & 27 & 1500 & \05.0  & 22.32 & \06.39 & 0.4229 \\
S-35   & 35000 & 3.30 & 21 & 1900 & \08.0  & 22.88 & \09.70 & 0.4935 \\
S-40   & 40000 & 3.60 & 19 & 2200 & 10\x\0 & 23.19 &  11.24 & 0.5998 \\
S-45   & 45000 & 3.80 & 20 & 2500 & 15\x\0 & 23.48 &  11.84 & 0.7160 \\
S-50   & 50000 & 3.90 & 20 & 3200 & 24\x\0 & 23.71 &  18.34 & 0.8204 \\
\hline
\end{tabular}
\end{center}
\end{table*}

\section{Results}

In the following we apply our improved code for expanding atmospheres
to a basic model grid of O-stars. The objectives of these calculations
are to present ionizing fluxes which can be used for the quantitative
analysis of emission line spectra of \HII-regions and Planetary
Nebulae, and to prove our method and demonstrate its reliability by
means of synthetic UV spectra which are qualitatively compared to
corresponding observations.  (Note that for the standard model
calculations the EUV and X-ray shock radiation is not included ---
using our {\em WM-basic} program package this should always be the
first step. For succeeding models in an advanced stage we have used
solely our previous method based on isothermal shocks
(cf.~Section~\ref{sec:shocks}), since this is the method which is
presently available for WM-basic and thus the models presented in the
following can be reproduced by this offered tool.)

Finally, one of the grid models is chosen for a detailed comparison
between observed and calculated synthetic spectra, where the primary
objective has been to develop diagnostic tools for the verification of
stellar parameters, and the determination of abundances and stellar
wind properties entirely from the UV spectra.
This has been carried out for a cooler O9.5~Ia supergiant,
$\alpha$~Cam --- a cooler object has been chosen since several
aspects tend to make these generally more problematic, such as the
ionization balance (more stages are affected) and the optical
thickness of the continuum in the wind part.

\subsection{The basic model grid}

In this section we present the ionizing fluxes and synthetic spectra
of a basic model grid of O-stars of solar metallicity, comprising
dwarfs and supergiants with effective temperatures ranging from 30,000
to 50,000\,K.  The model parameters, summarized in
Table~\ref{tbl:gridparams}, were chosen in accordance with the range
of values deduced from observations as tabulated by \cite{}Puls et
al.~(1996).

In Fig.~\ref{fig:fluxes} we show for each model the primary result,
the ionizing emergent flux together with the corresponding continuum.
(We note that this is the continuum obtained from opacities and
emissivities resulting from the full line blocking and blanketing
calculation, which is quite different from the one that would be
obtained if only continuum opacities were used in the iteration
cycle.) It can be verified from the figure that the influence of the
line opacities, i.\,e., the difference between the continuum and the
total flux, increases from dwarfs to supergiants and from cooler to
hotter effective temperatures. Both points are not surprising, because
they are directly coupled to the mass loss rate ($\Mdot$) which
increases exactly in the same manner (cf.~Table~\ref{tbl:gridparams}).
Due to the increasing $\Mdot$ the optical depth of the lines also
increases in the wind part and in consequence the line blocking effect
is more pronounced. This behaviour, however, saturates for objects
with effective temperatures larger than $\Teff=45\,000\,{\rm K}$,
since in this case higher main ionization stages are encountered
(e.\,g., \FeV\ and \FeVI) which are shown to have less bound-bound
transitions (cf.~\cite{}Pauldrach,~1987). Thus, as can be verified
from Fig.~\ref{fig:fluxes} the effect of line blocking is strongest
for supergiants of intermediate $\Teff$. In Table~\ref{tbl:gridparams}
we present the numerical values of the integrals of ionizing photons
emitted per second for H ($\log Q_{\rm H}$) and \HeII\ ($\log Q_{\rm
He^+}$), as well as the flux at the reference wavelength
$\lambda=5480\,{\rm\AA}$, which can be used directly to calculate
Zanstra ratios and Str\"omgren radii.

The next step is to demonstrate the reliability of the calculated
emergent fluxes.  As the wavelength region shortwards of the Lyman
edge usually cannot be observed and thus a direct comparison of the
fluxes with observations is not possible, an indirect method to test
their accuracy is needed.  In principle, two such methods exist.  The
first one is to test the ionizing fluxes by means of their influence
on the emission lines of gaseous nebulae, i.\,e., using the ionizing
fluxes as input for nebular models and comparing the calculated
emission line strengths to observed ones. However, as a first step
this procedure is questionable, since the diagnostics of gaseous
nebulae is still not free from uncertainties --- dust clumps, complex
geo\-metric structure, etc.\ --- and therefore, if discrepancies are
encountered, it is difficult to decide which of the assumptions is
responsible for the disagreement. (As an example we mention the
\NeIII-problem discussed comprehensively by \cite{}Rubin et al.~(1991)
and Sellmaier et al.~(1996).) Rather, nebular modeling and diagnostics
should be able to build upon the reliability of the ionizing fluxes,
and thus the quantitative accuracy of the fluxes needs to be tested
independently of their use in nebular emission line analysis.

The second --- and in the light of the difficulties discussed above,
the only trustworthy --- method is quite analogous, but instead of an
external nebula involves {\em the atmosphere of the star itself}.  The
rationale is that the emergent flux is but the outer value of a
radiation field {\em calculated selfconsistently throughout the entire
wind}, which influences the ionization balance at {\em all depths}.
This ionization balance can be traced reliably through the strength
and structure of the wind lines formed everywhere in the atmosphere.
Hence it is a natural and important step to test the quality of the
ionizing fluxes by virtue of their direct product: {\it the UV spectra
of O~stars}.

\begin{figure*}
\vspace{10mm}
\centerline{
\includegraphics[height=\columnwidth,angle=-90]{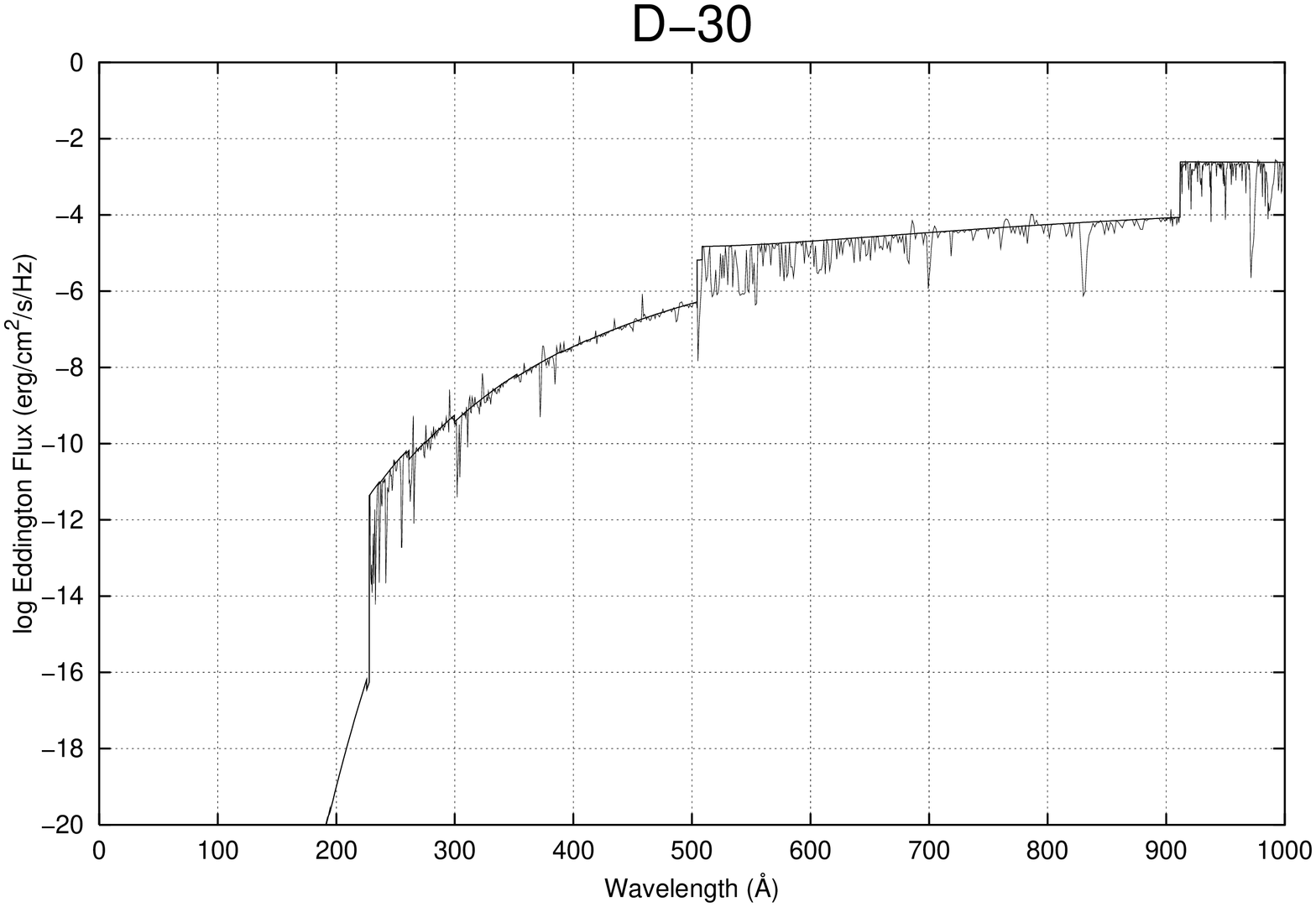} \hfill
\includegraphics[height=\columnwidth,angle=-90]{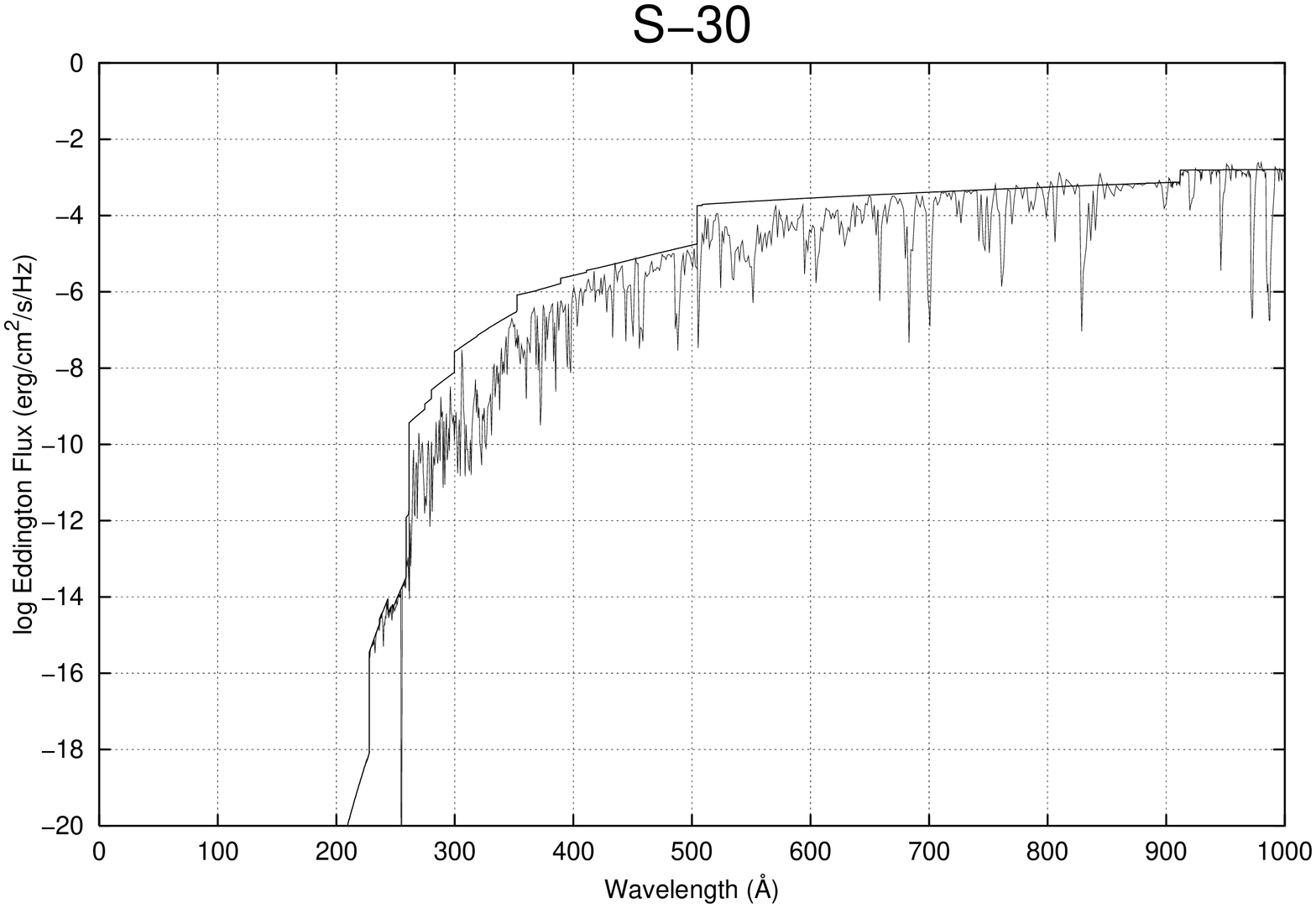}}
\vspace{10mm}
\centerline{
\includegraphics[height=\columnwidth,angle=-90]{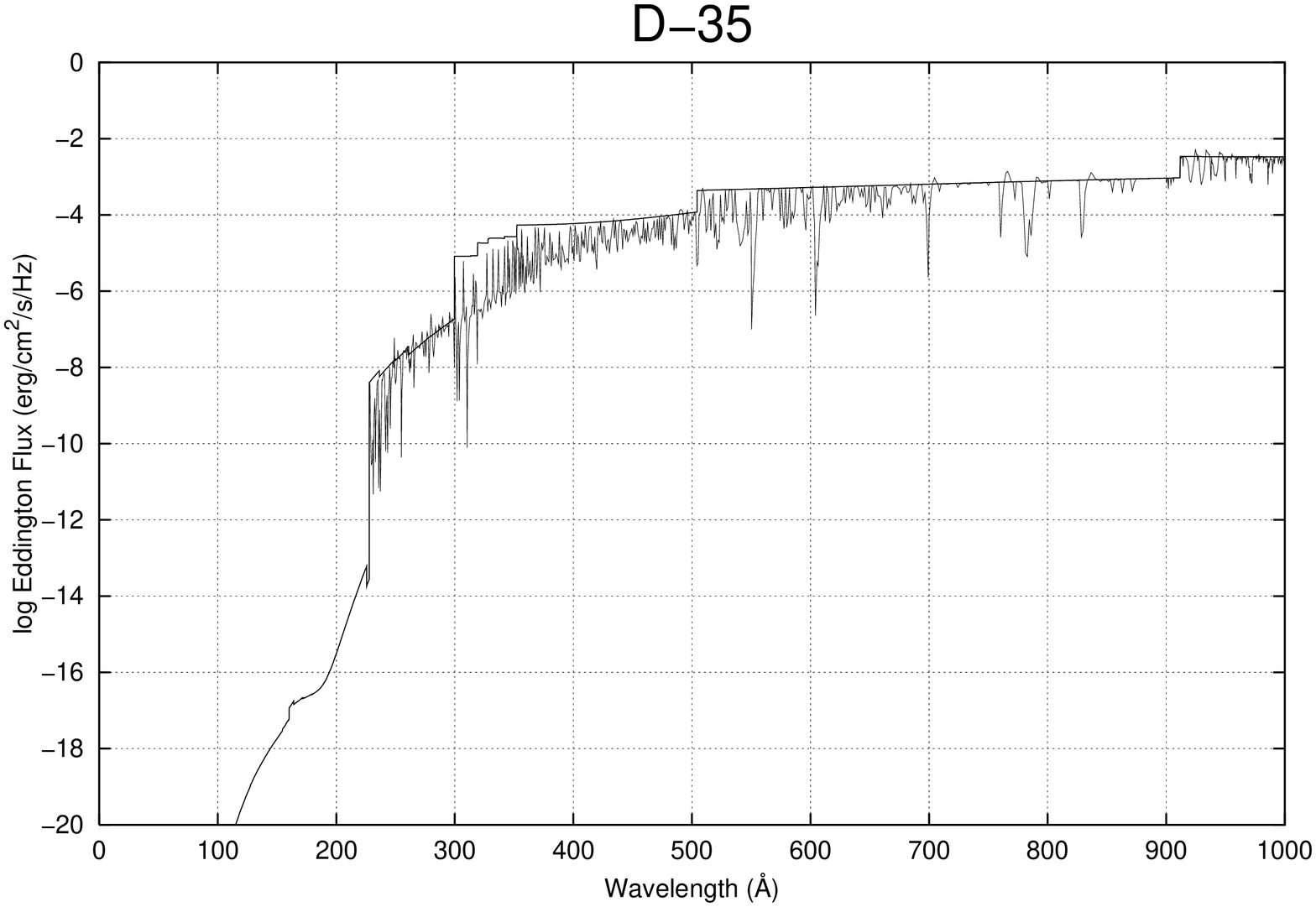} \hfill
\includegraphics[height=\columnwidth,angle=-90]{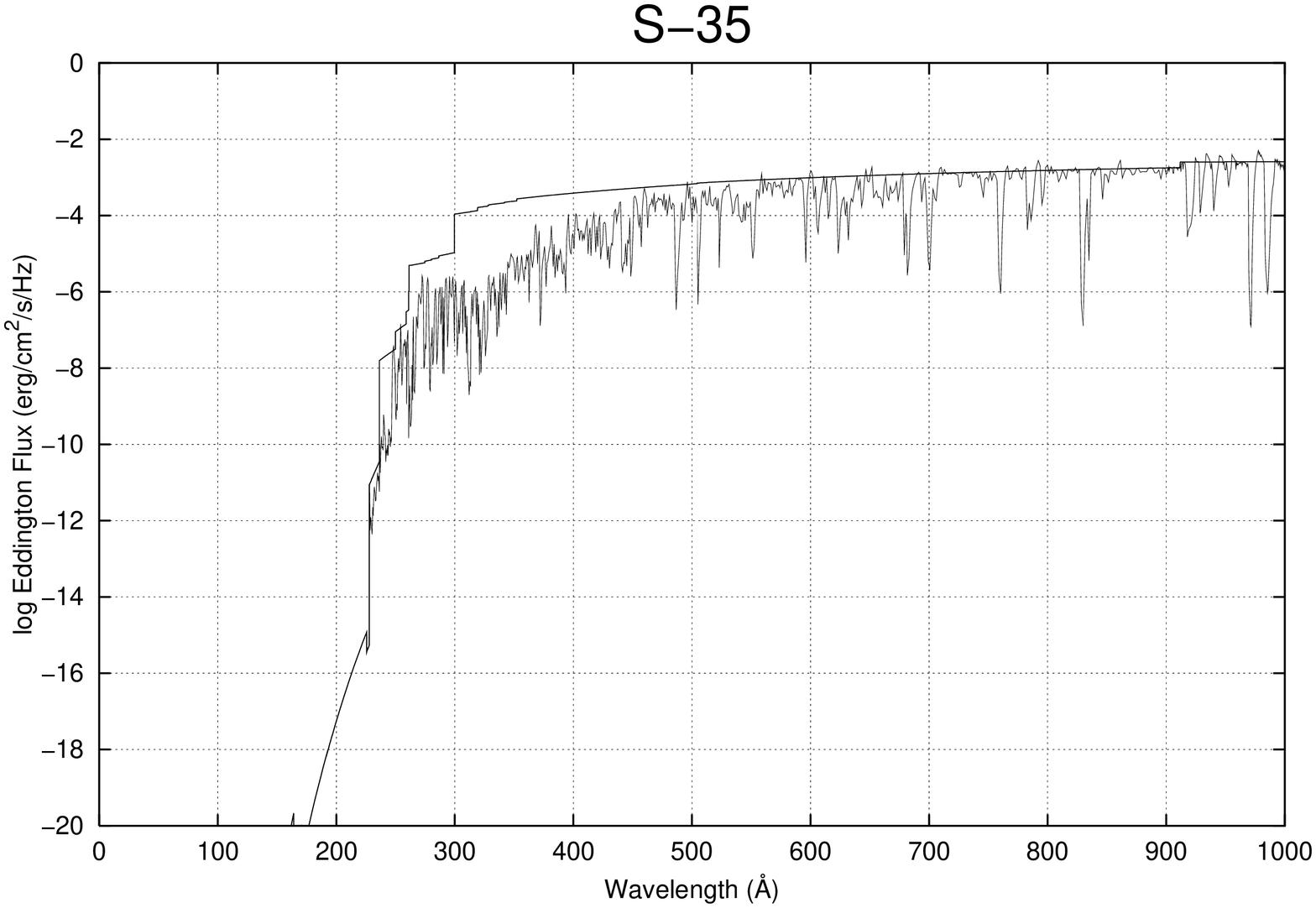}}
\vspace{10mm}
\centerline{
\includegraphics[height=\columnwidth,angle=-90]{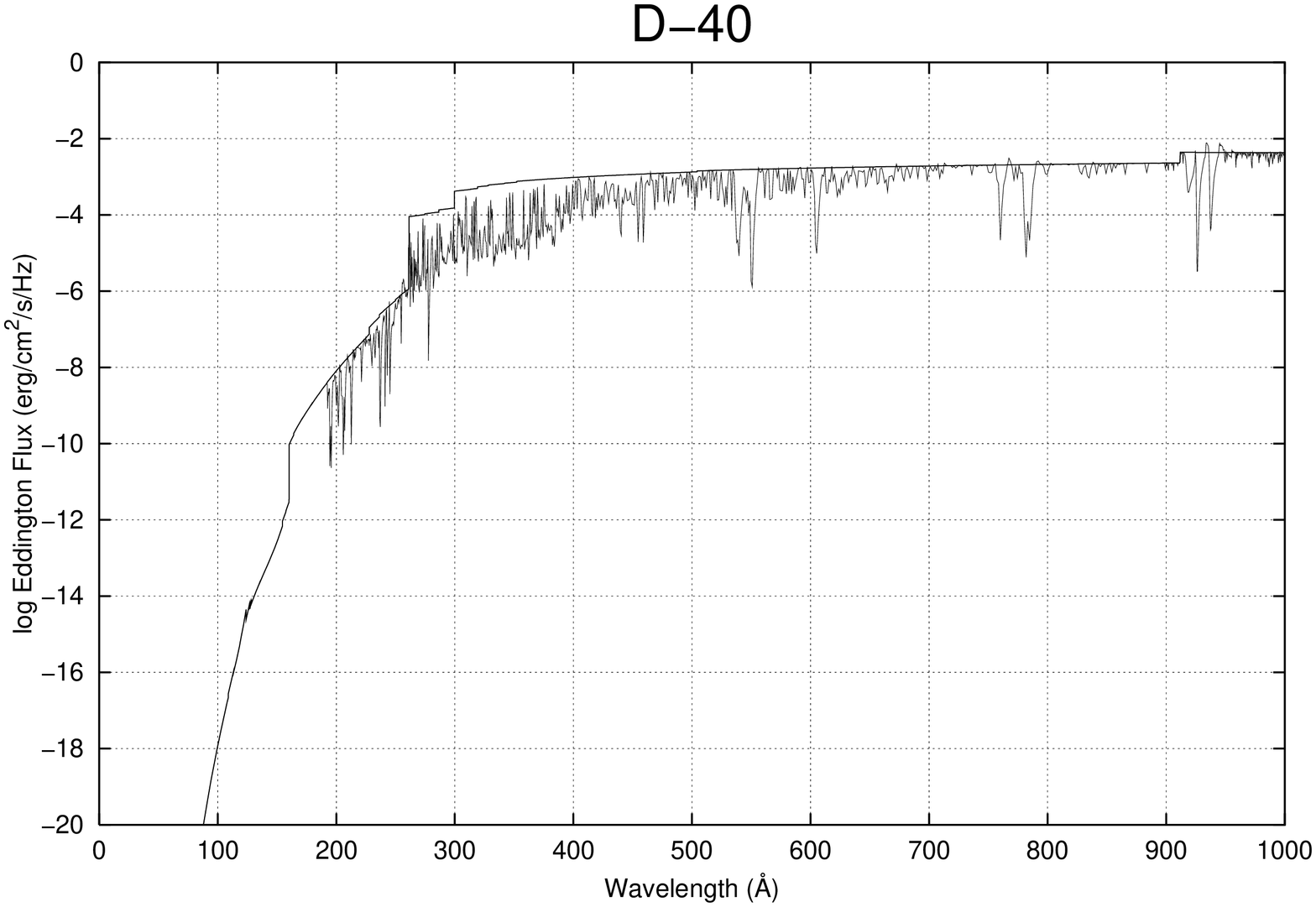} \hfill
\includegraphics[height=\columnwidth,angle=-90]{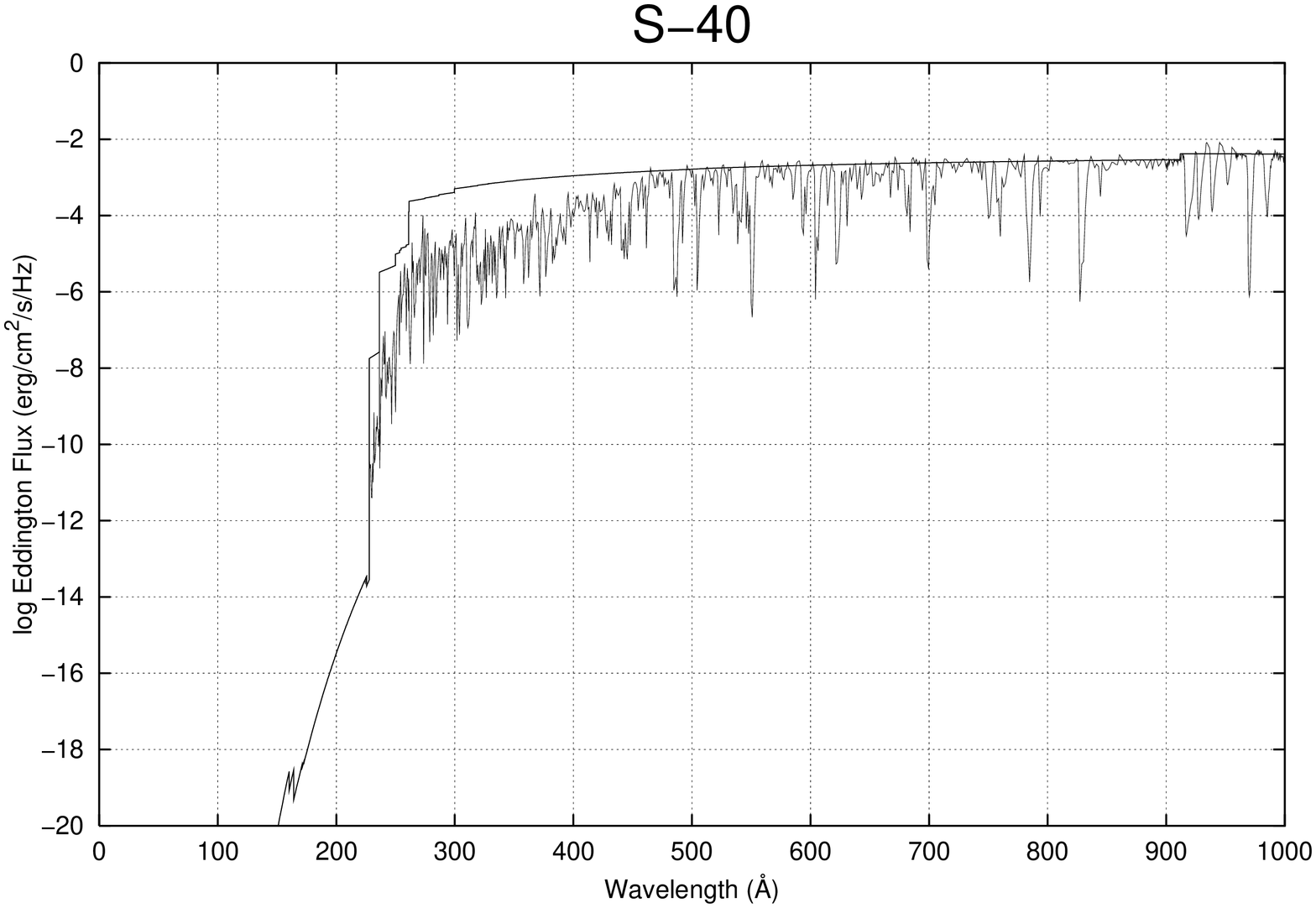}}
\vspace{5mm}
\mycaption{
Calculated ionizing fluxes (Eddington flux in cgs units) versus
wavelength of the model grid stars; dwarfs on the left, supergiants on
the right.}
\label{fig:fluxes}
\end{figure*}

\begin{figure*}
\vspace{10mm}
\centerline{
\includegraphics[height=\columnwidth,angle=-90]{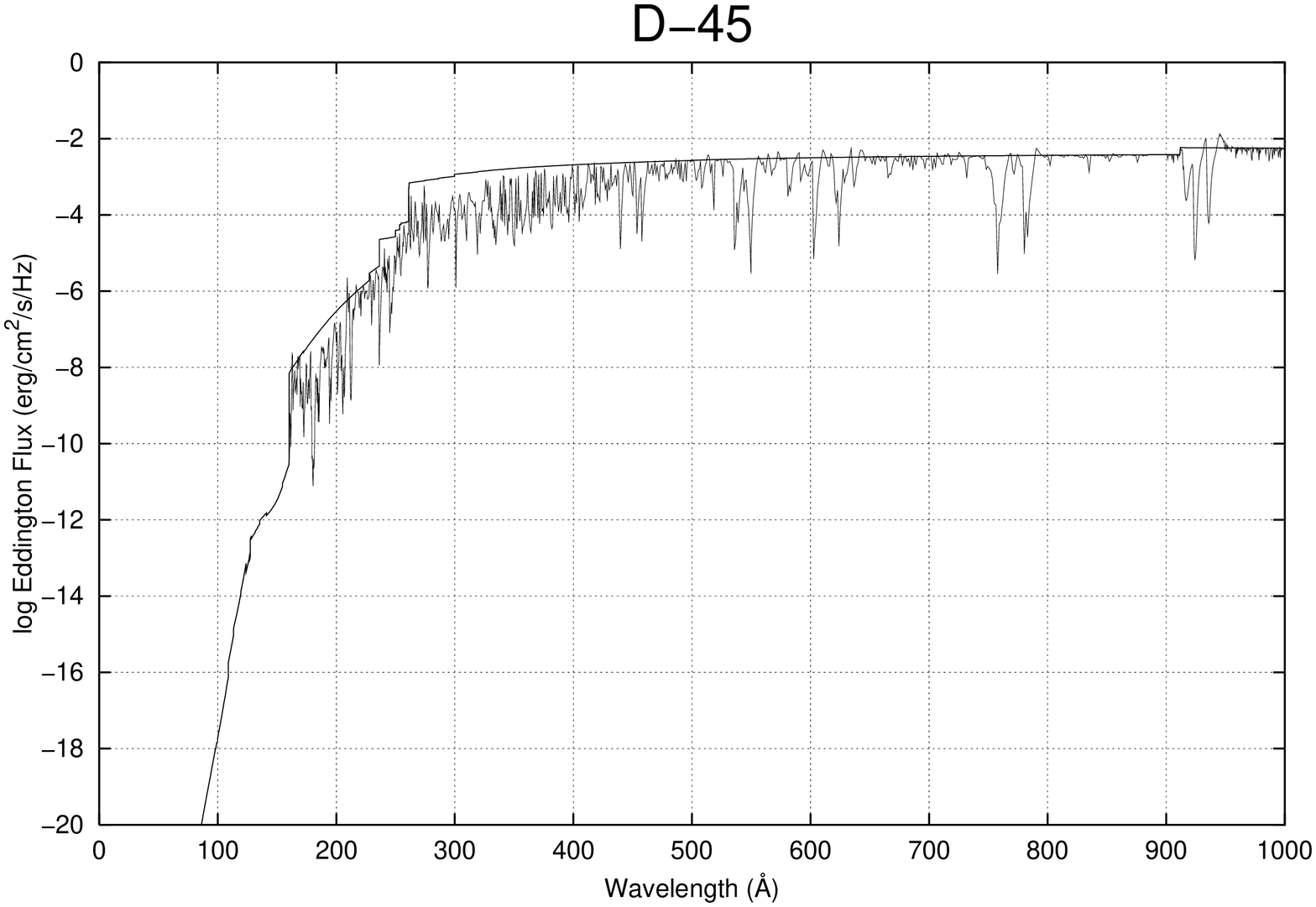} \hfill
\includegraphics[height=\columnwidth,angle=-90]{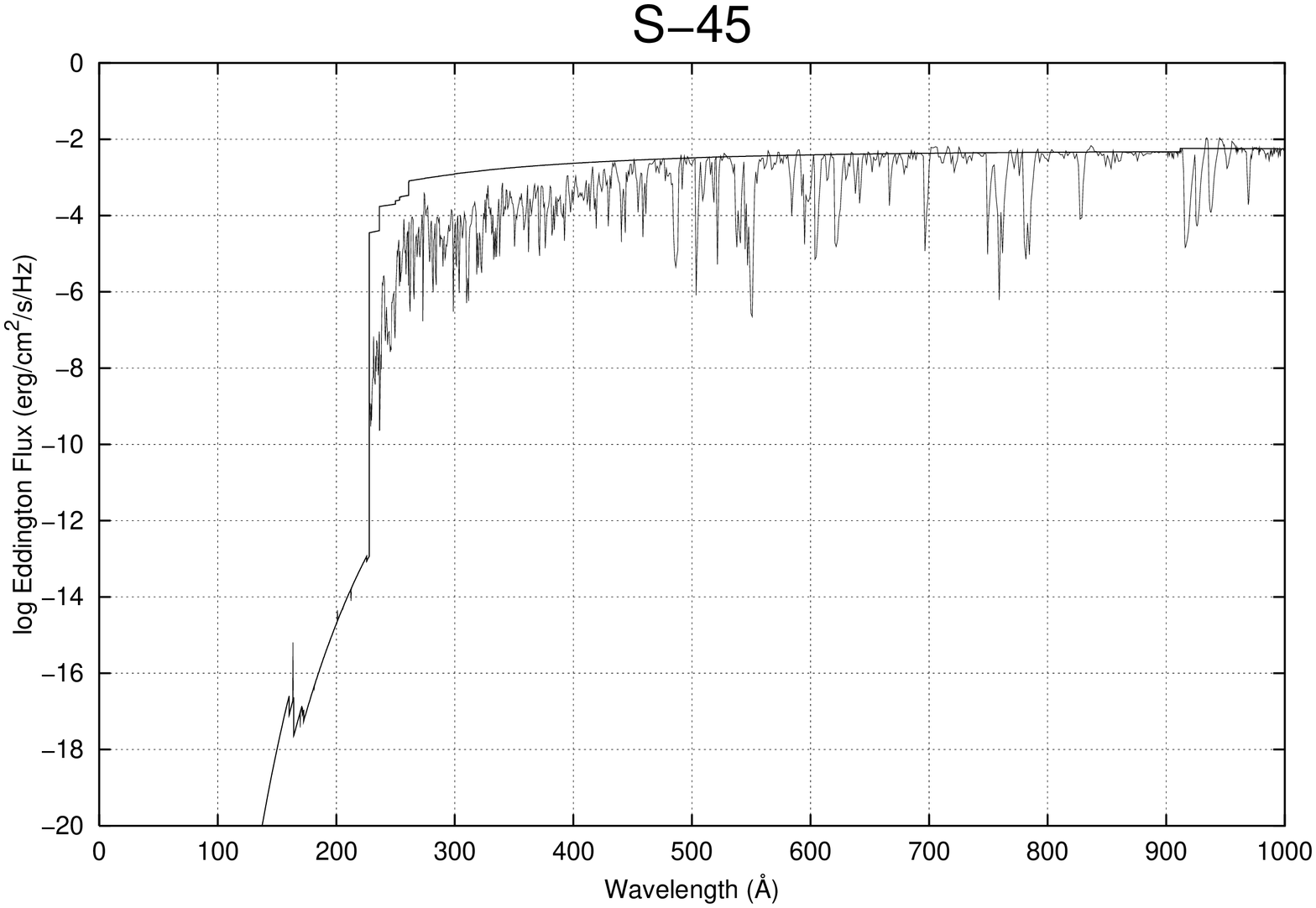}}
\vspace{10mm}
\centerline{
\includegraphics[height=\columnwidth,angle=-90]{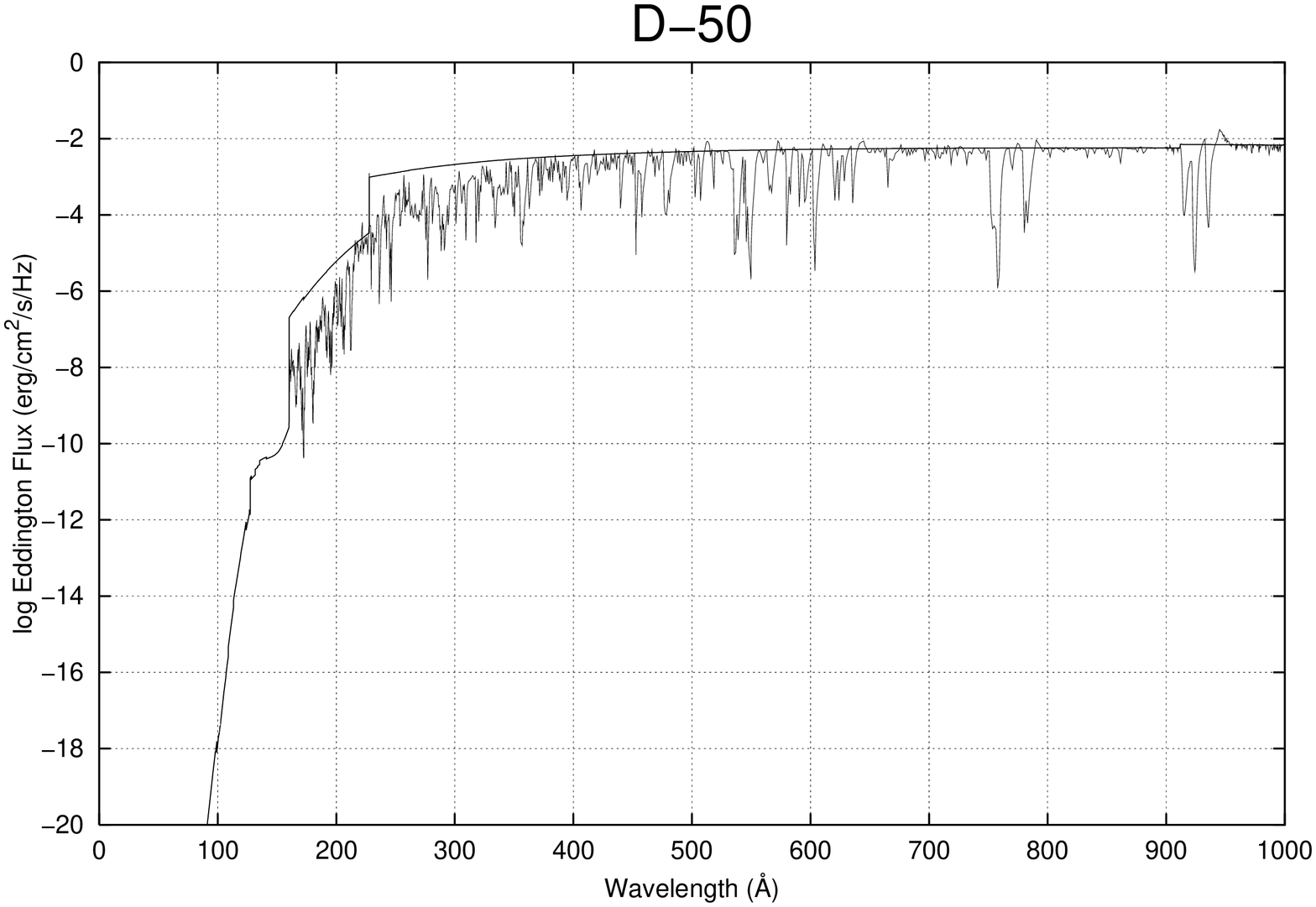} \hfill
\includegraphics[height=\columnwidth,angle=-90]{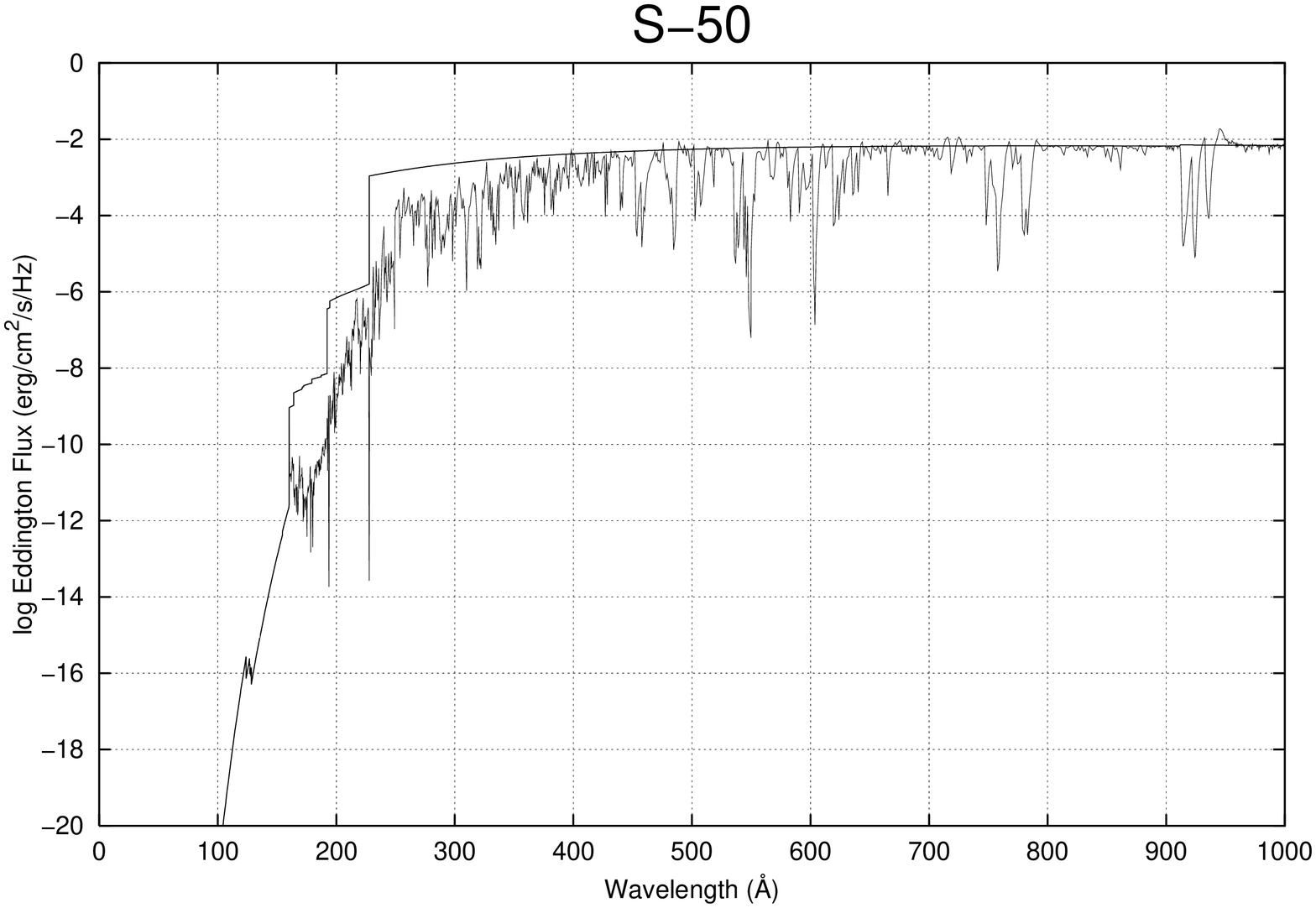}}
\vspace{10mm}
\centerline{
\includegraphics[height=\columnwidth,angle=-90]{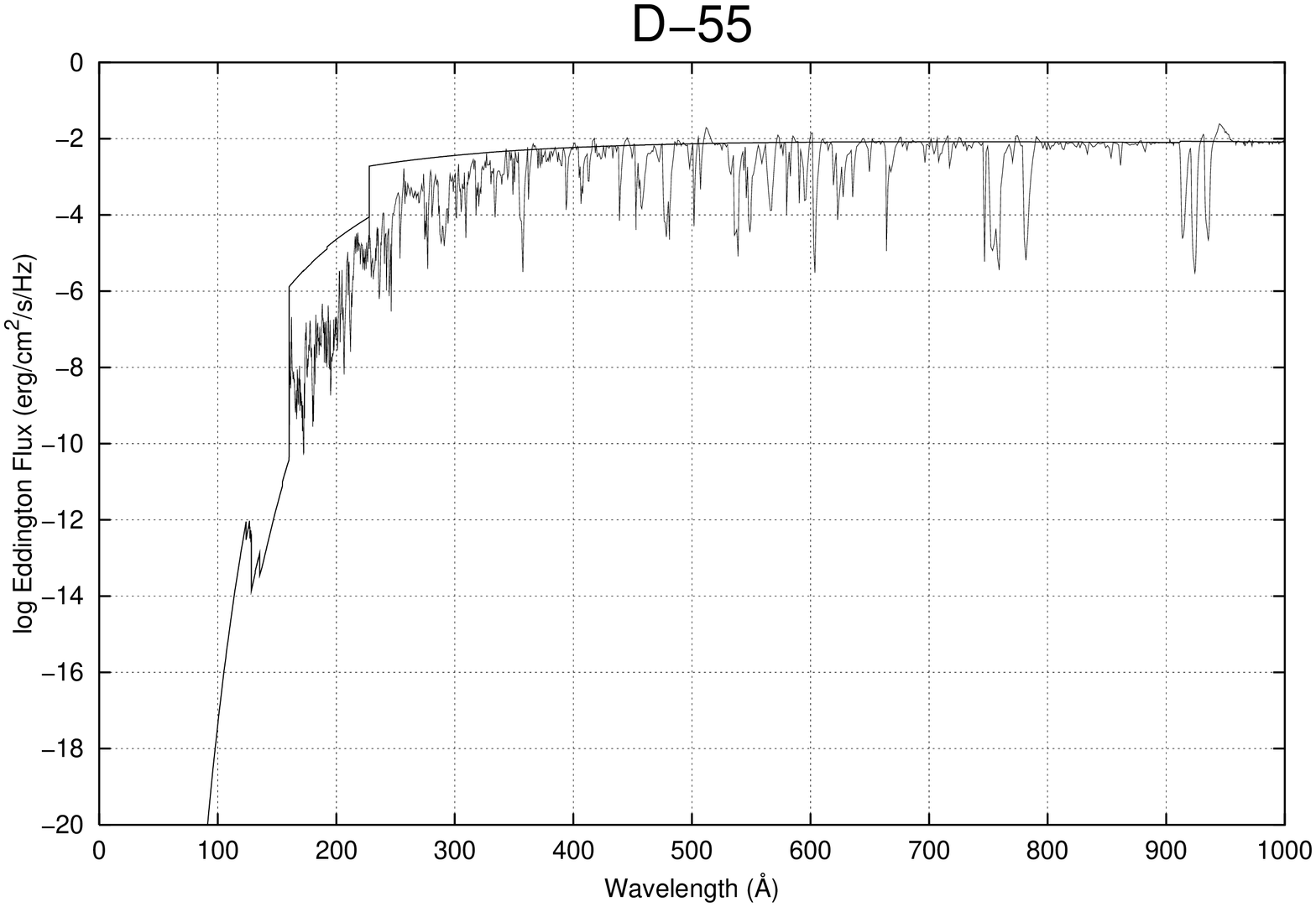} \hfill}
\addtocounter{figure}{-1}
\vspace{5mm}
\mycaption{ (continued)~
Calculated ionizing fluxes (Eddington flux in cgs units) versus
wavelength of the model grid stars; dwarfs on the left, supergiants on
the right.}
\end{figure*}

\subsection{Qualitative comparison with observations}

\def\y{\phantom{$\le$}}
\begin{table*}
\mycaption{
Model grid stars and real-world examples with similar spectral types.
The parameters of the observed example stars are from Puls et al.~(1996).}
\label{tbl:gridexamples}
\begin{center}
\begin{tabular}{lccccc}
\hline
\hline
\hfil Model~/\hfil & $\Teff$ & $\logg$ & $R$ & $\vinf$ & $\Mdot$ \\
\hfil Example\hfil & (K) & (cgs) & ($R_\odot$) & (km/s) & ($10^{-6} M_\odot/{\rm yr}$) \\
\hline
\hline
\multicolumn{6}{c}{Dwarfs} \\
\hline
D-30                    & 30000 & 3.85 & 12\x\0 & 1800 & \y   0.008   \\
HD 149757 ($\zeta$ Oph) & 32500 & 3.85 & 12.9   & 1550 & $\le$0.03\0  \\[1ex]
D-40                    & 40000 & 3.75 & 10\x\0 & 2400 & \y   0.24\0  \\
HD 217068               & 40000 & 3.75 & 10.3   & 2550 & $\le$0.2\0\0 \\[1ex]
D-50                    & 50000 & 4.00 & 12\x\0 & 3200 & \y   5.6\0\0 \\
HD 93250                & 50500 & 4.00 & 18\x\0 & 3250 & \y   4.9\0\0 \\
\hline
\multicolumn{6}{c}{Supergiants} \\
\hline
S-30                    & 30000 & 3.00 & 27\x\0 & 1500 & \05.0   \\
HD 30614 ($\alpha$ Cam) & 30000 & 3.00 & 29\x\0 & 1550 & \05.2   \\[1ex]
S-40                    & 40000 & 3.60 & 19\x\0 & 2200 &  10\x\0 \\
HD 66811 ($\zeta$ Pup)  & 42000 & 3.60 & 19\x\0 & 2250 & \05.9   \\[1ex]
S-50                    & 50000 & 3.90 & 20\x\0 & 3200 &  24\x\0 \\
HD 93129A               & 50500 & 3.95 & 20\x\0 & 3200 &  22\x\0 \\
\hline
\end{tabular}
\end{center}
\end{table*}

The test is performed by means of synthetic UV spectra which are
qualitatively compared to observed IUE spectra. For this comparison we
have chosen, for each model of a subset of our grid, a real object
from the list of Puls et al.~(1996) whose supposed stellar and wind
parameters come very close to those of the model.  The parameters of
the model stars and the selected real objects are summarized in
Table~\ref{tbl:gridexamples}.  (The influence of shock radiation on
the models has been neglected at this qualitative step.)

\begin{figure*}
\vspace{10mm}
\centerline{
\includegraphics[height=\columnwidth,angle=-90]{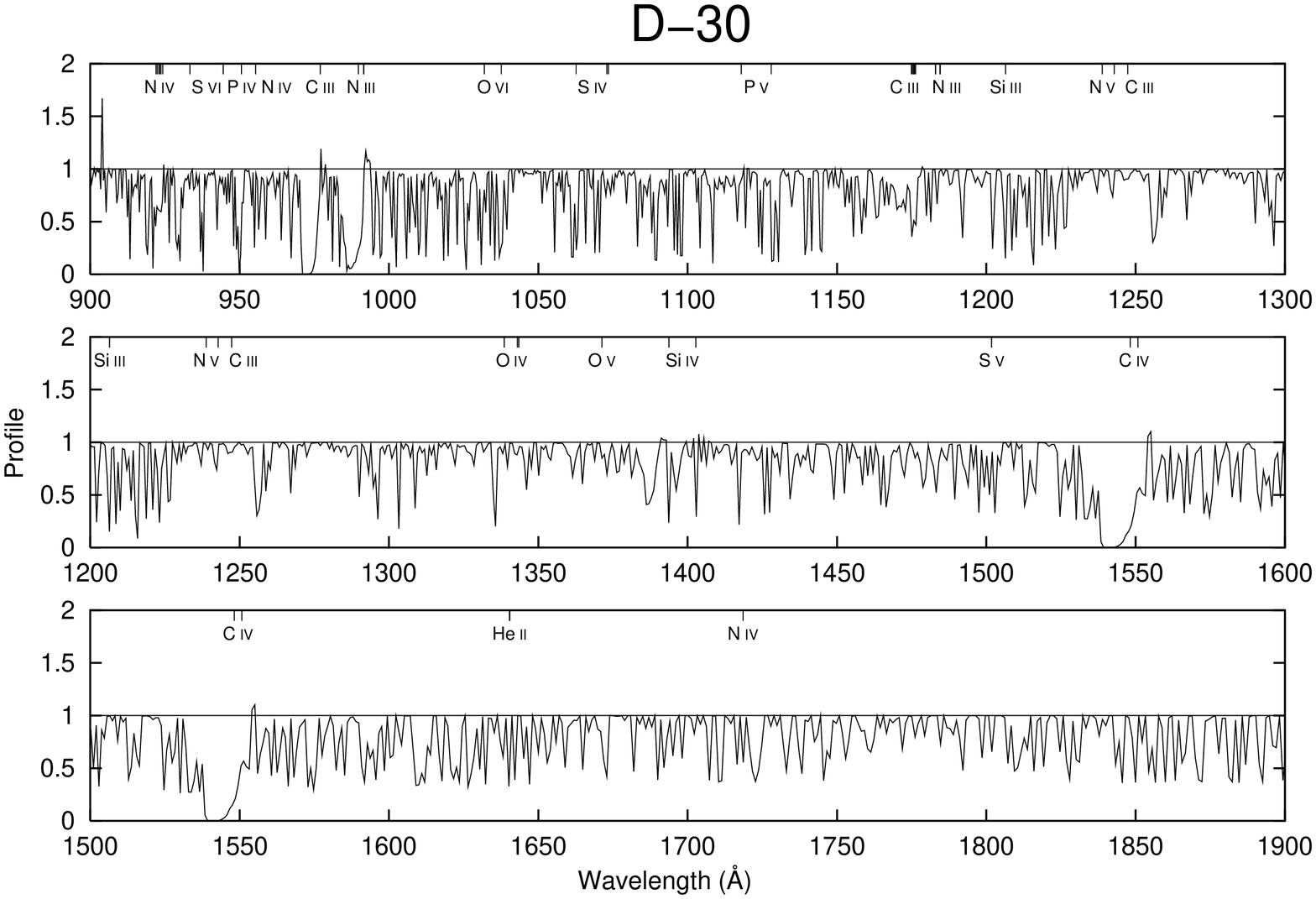} \hfill
\includegraphics[height=\columnwidth,angle=-90]{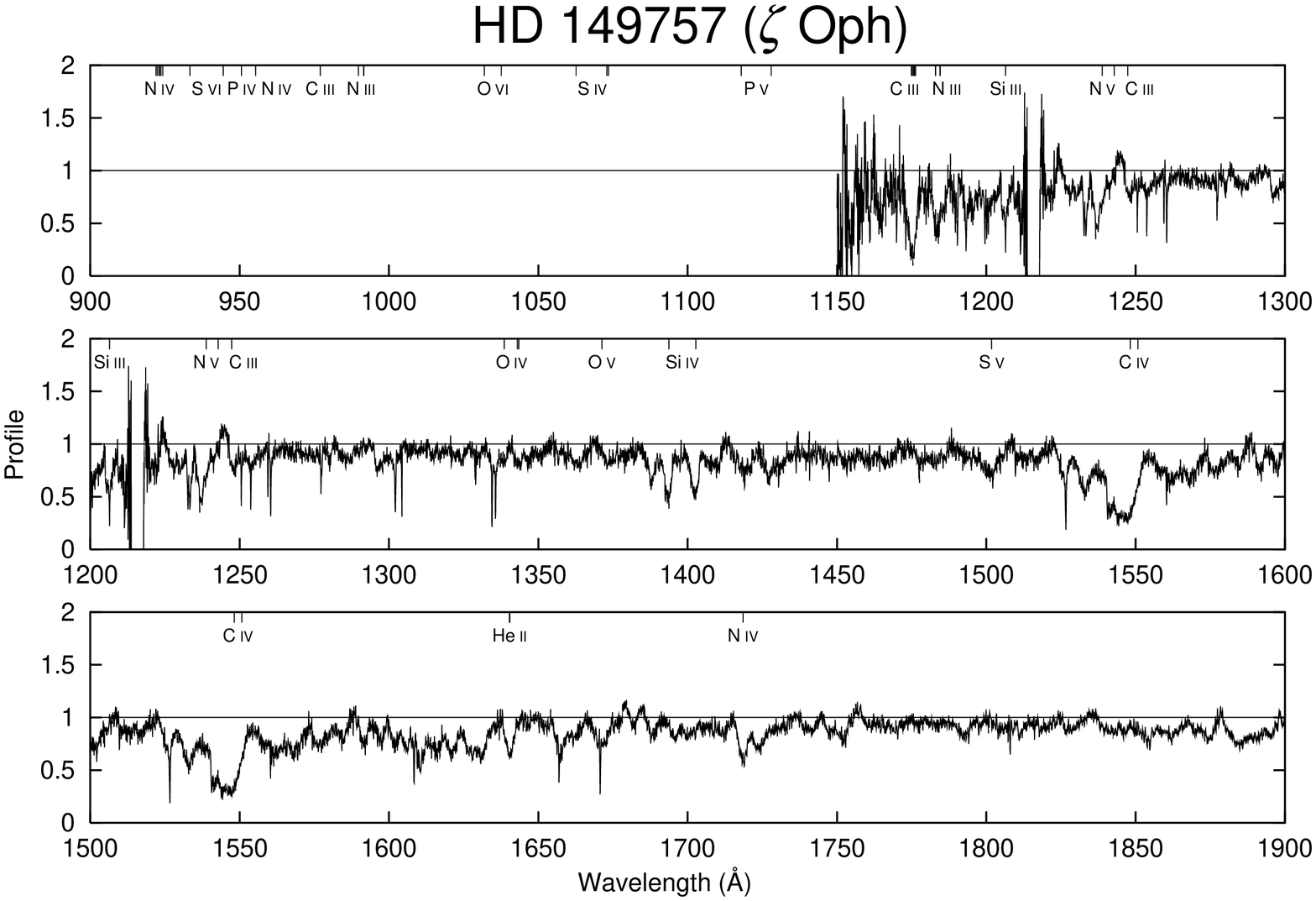}}
\vspace{10mm}
\centerline{
\includegraphics[height=\columnwidth,angle=-90]{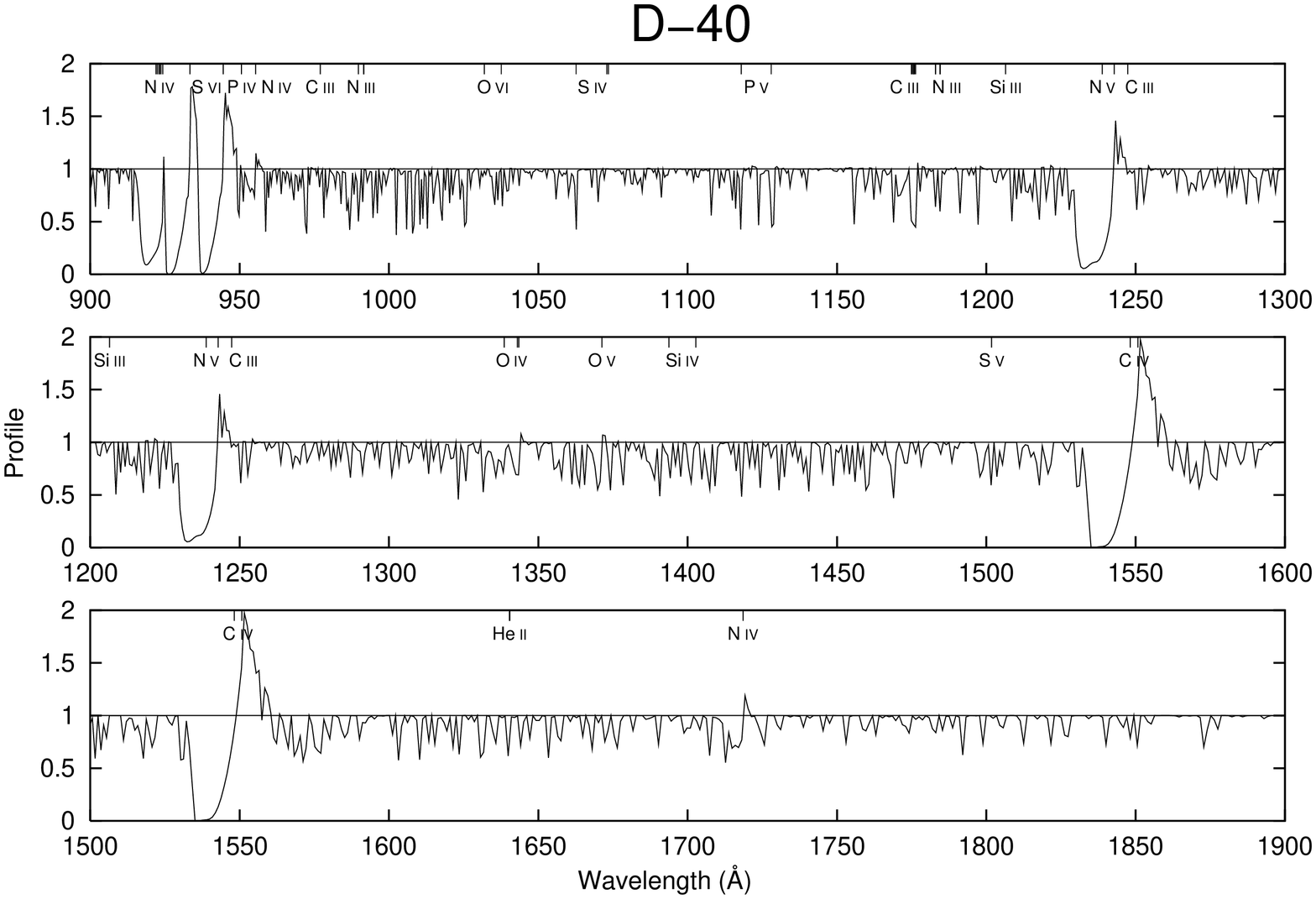} \hfill
\includegraphics[height=\columnwidth,angle=-90]{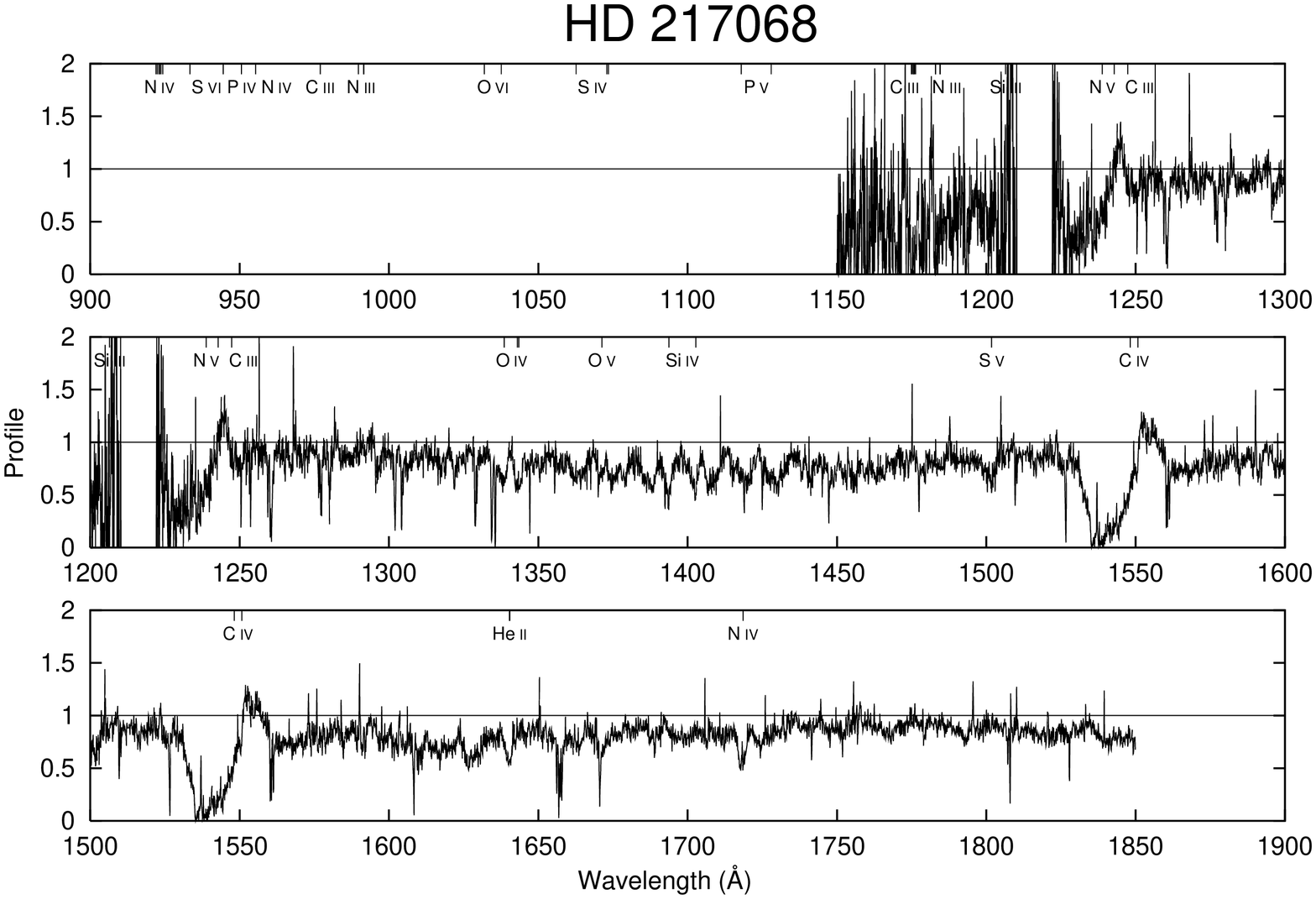}}
\vspace{10mm}
\centerline{
\includegraphics[height=\columnwidth,angle=-90]{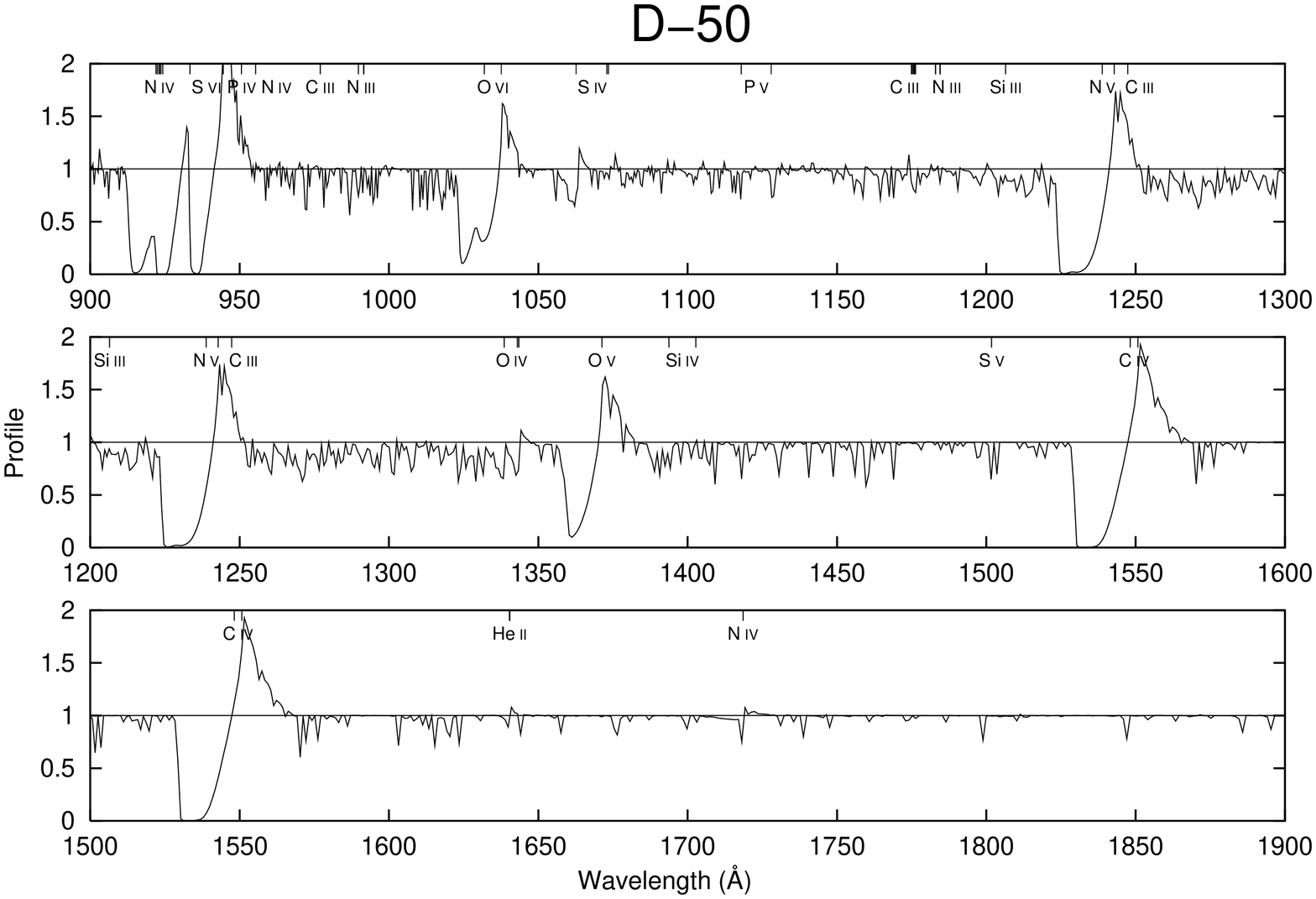} \hfill
\includegraphics[height=\columnwidth,angle=-90]{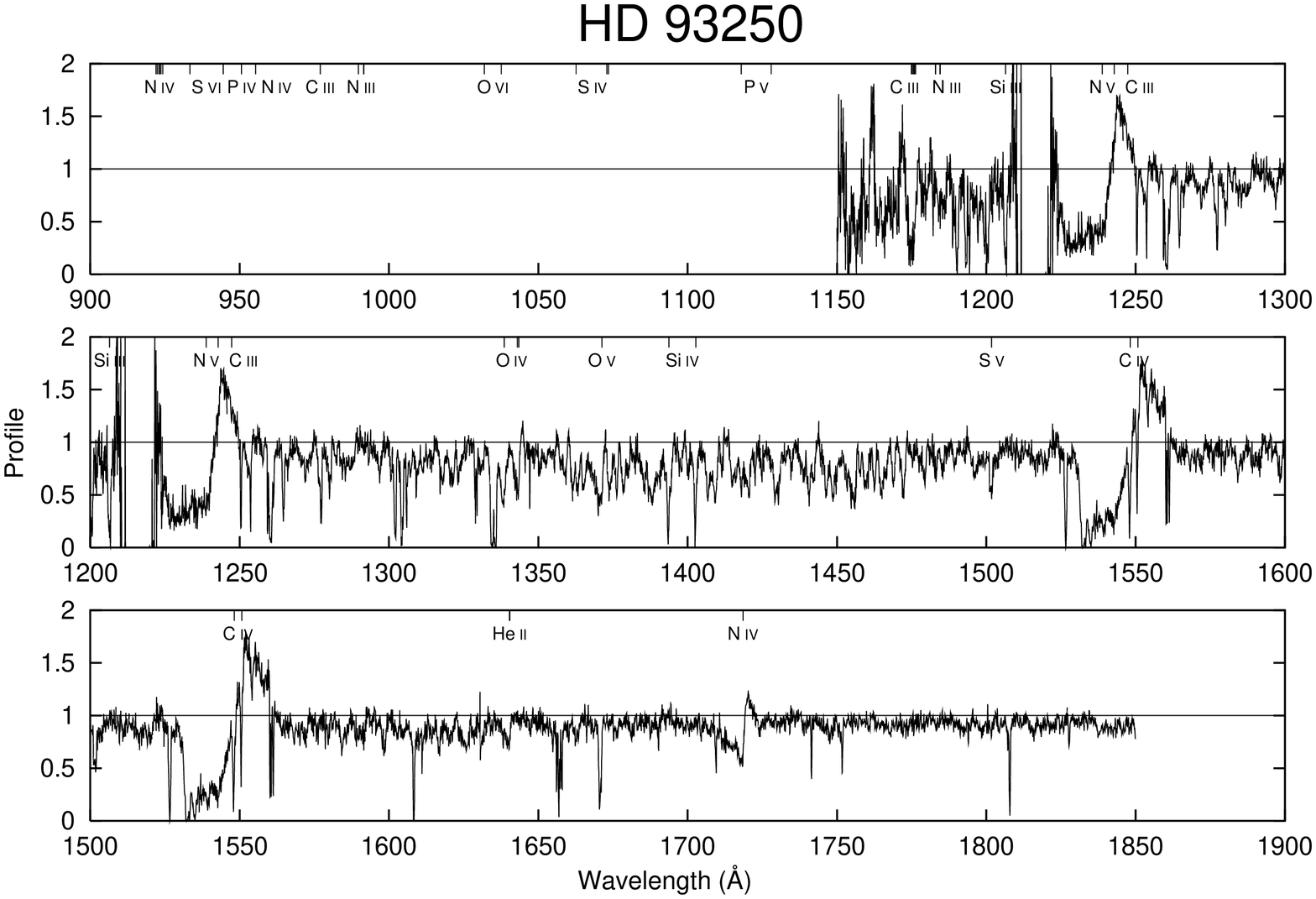}}
\vspace{5mm}
\mycaption{
Calculated UV spectra of the model grid dwarfs (left) compared to
observed IUE spectra (right) of stars of similar spectral type.}
\label{fig:spD}
\end{figure*}

First we investigate the spectra of the dwarf models. As can be
inspected from Fig.~\ref{fig:spD} the comparison of the models D-30
and D-40 with their counterparts HD~149757 and HD~217068 show in
principle an overall agreement, whereas the D-50 model, compared with
its counterpart HD~93250, shows a severe discrepancy concerning the
\OV\ subordinate line at 1371\,\AA\ (the calculated line is much too
strong) and a less pronounced discrepancy of the \NIV\ subordinate
line at 1718\,\AA\ (the calculated line is somewhat too weak). Hence
we have to realize that either the wind physics is not completely
described, or the stellar or the wind parameters of this model are too
different from those of HD~93250.

\begin{figure*}
\vspace{10mm}
\centerline{
\includegraphics[height=\columnwidth,angle=-90]{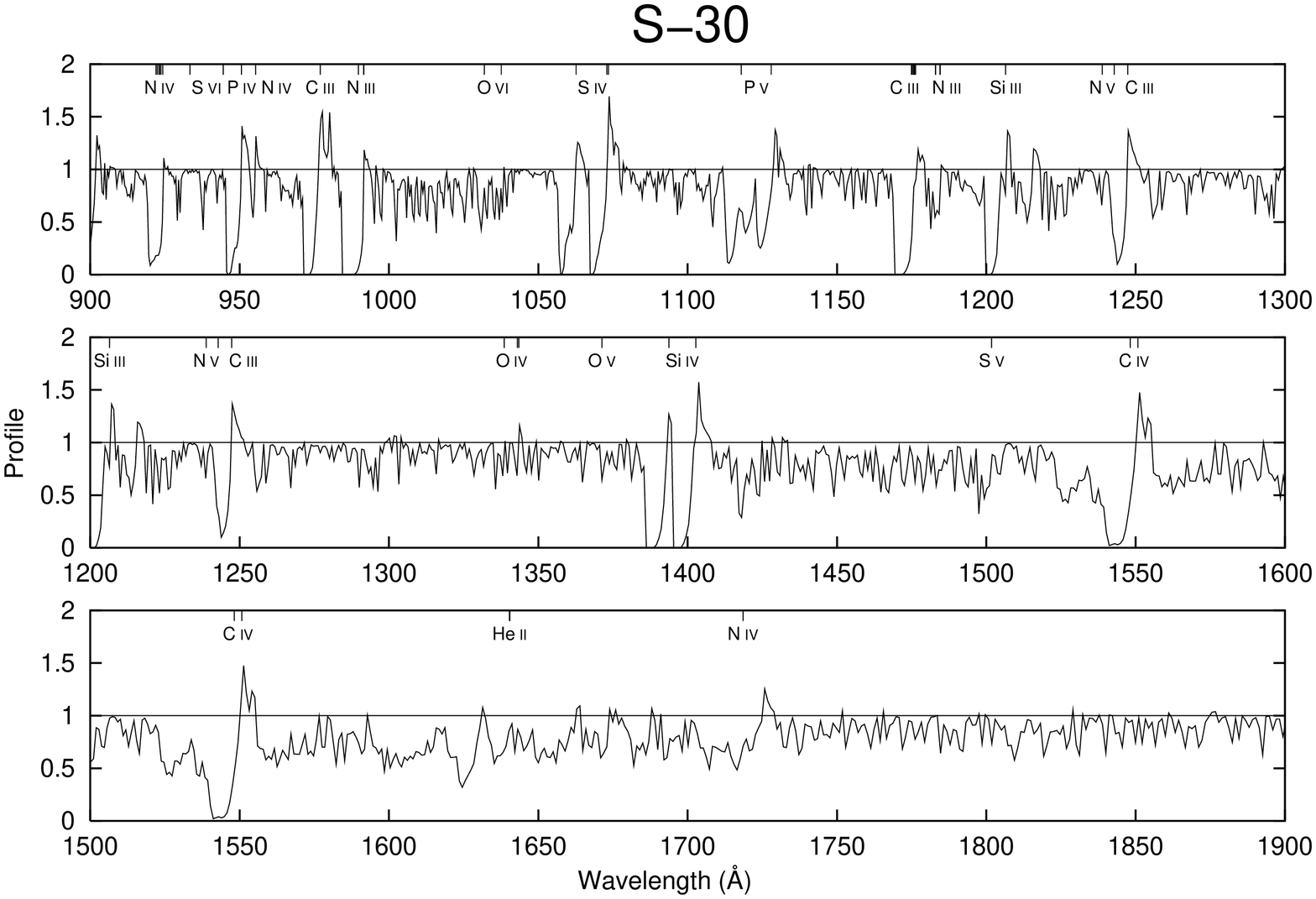} \hfill
\includegraphics[height=\columnwidth,angle=-90]{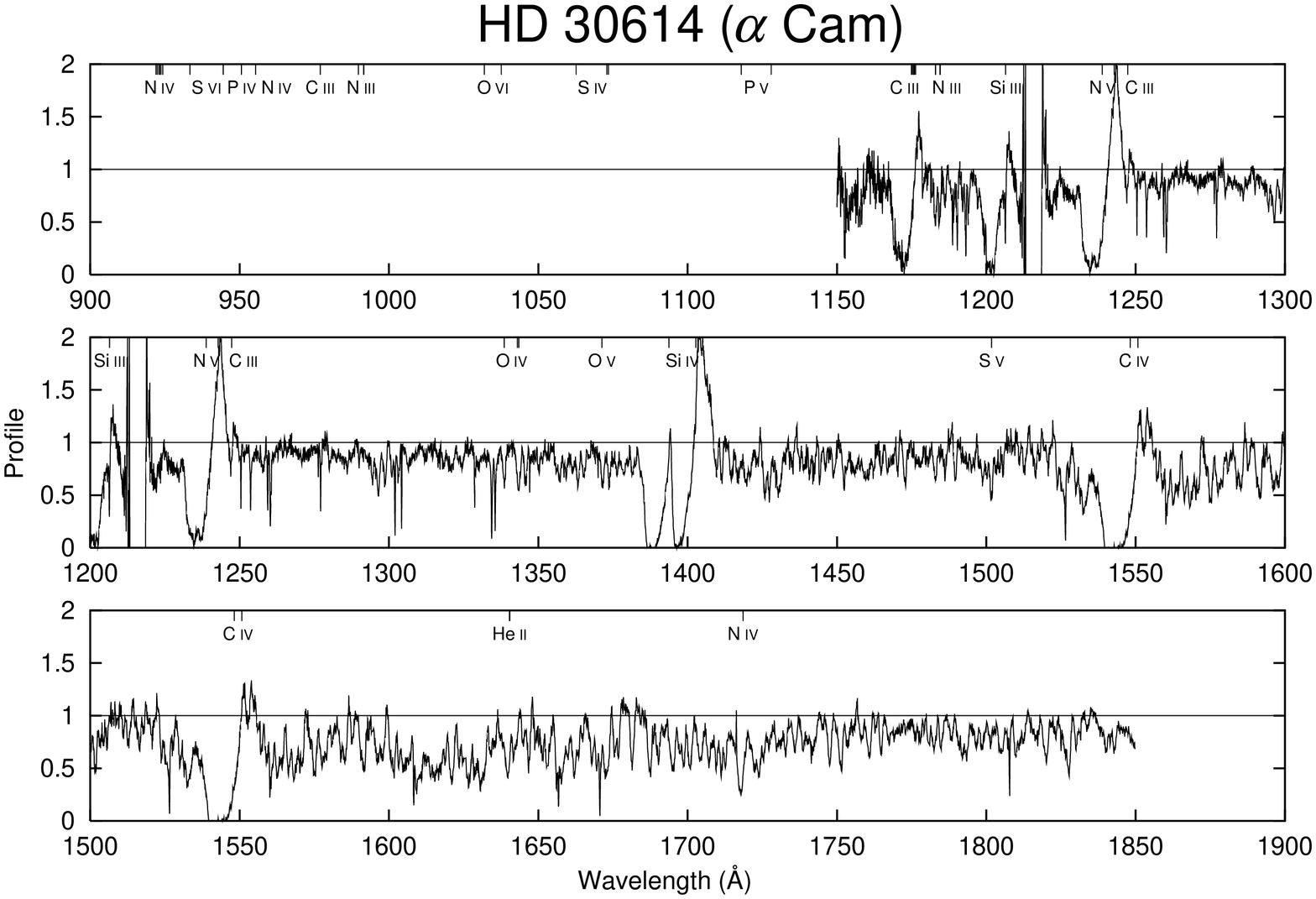}}
\vspace{10mm}
\centerline{
\includegraphics[height=\columnwidth,angle=-90]{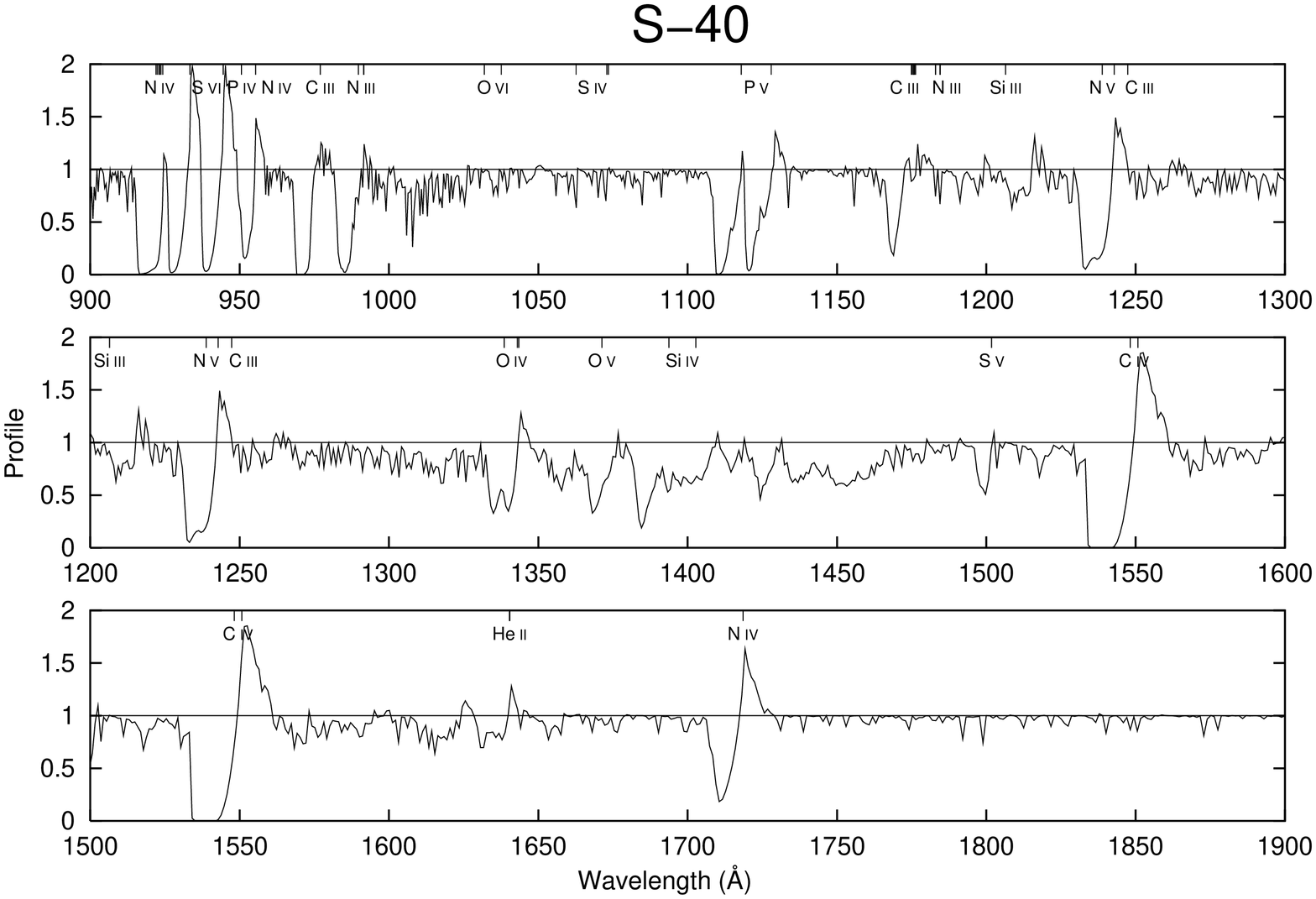} \hfill
\includegraphics[height=\columnwidth,angle=-90]{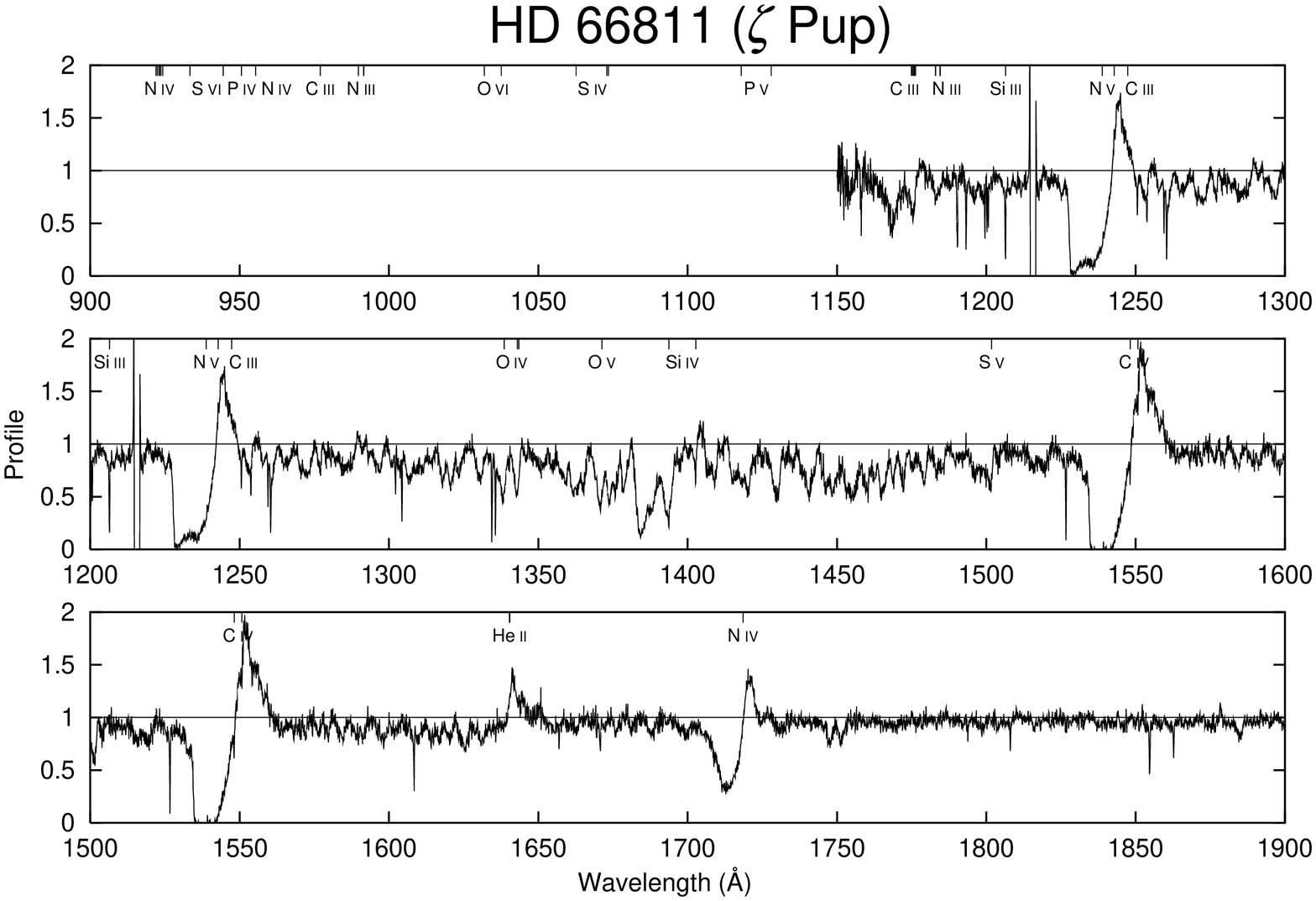}}
\vspace{10mm}
\centerline{
\includegraphics[height=\columnwidth,angle=-90]{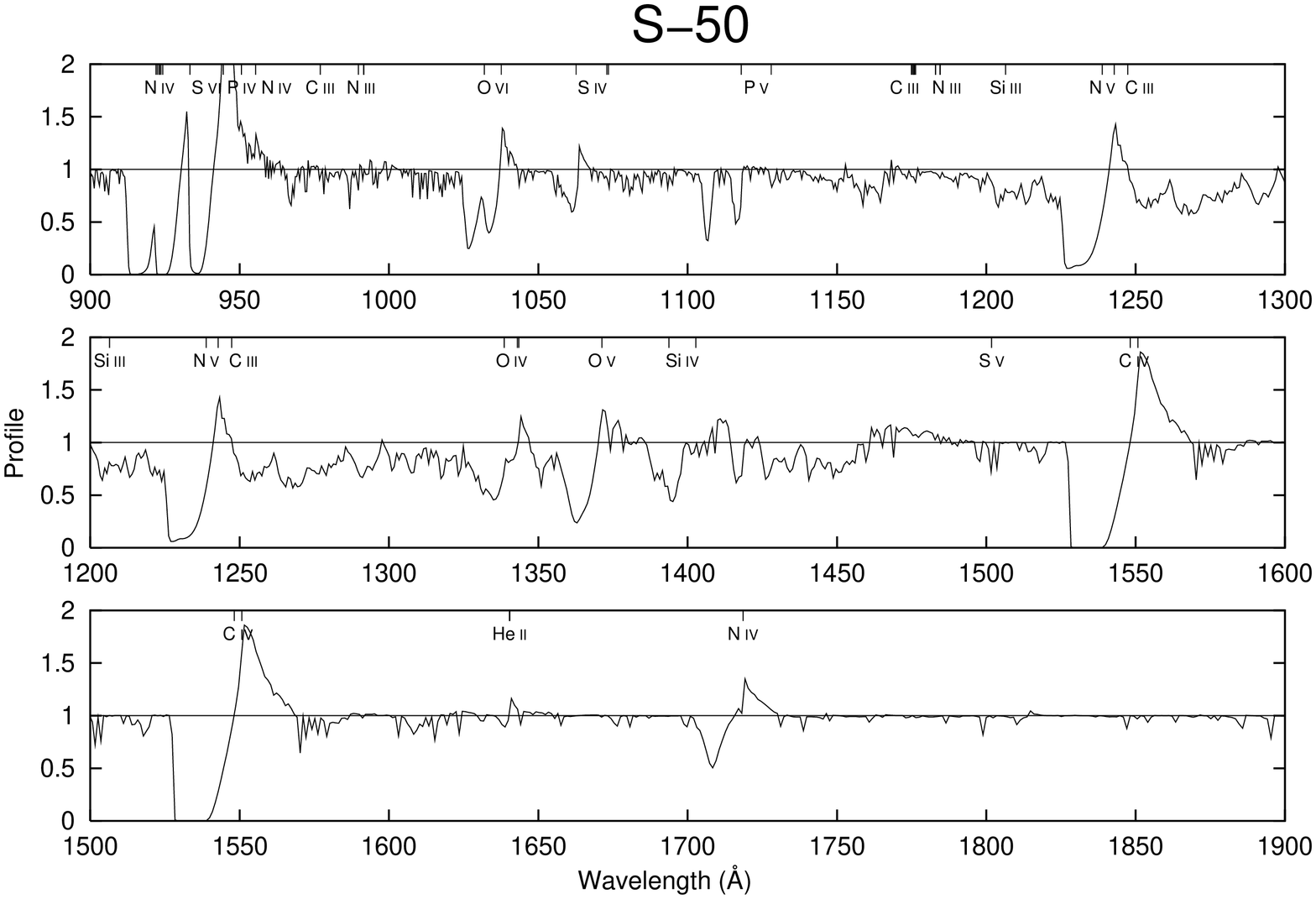} \hfill
\includegraphics[height=\columnwidth,angle=-90]{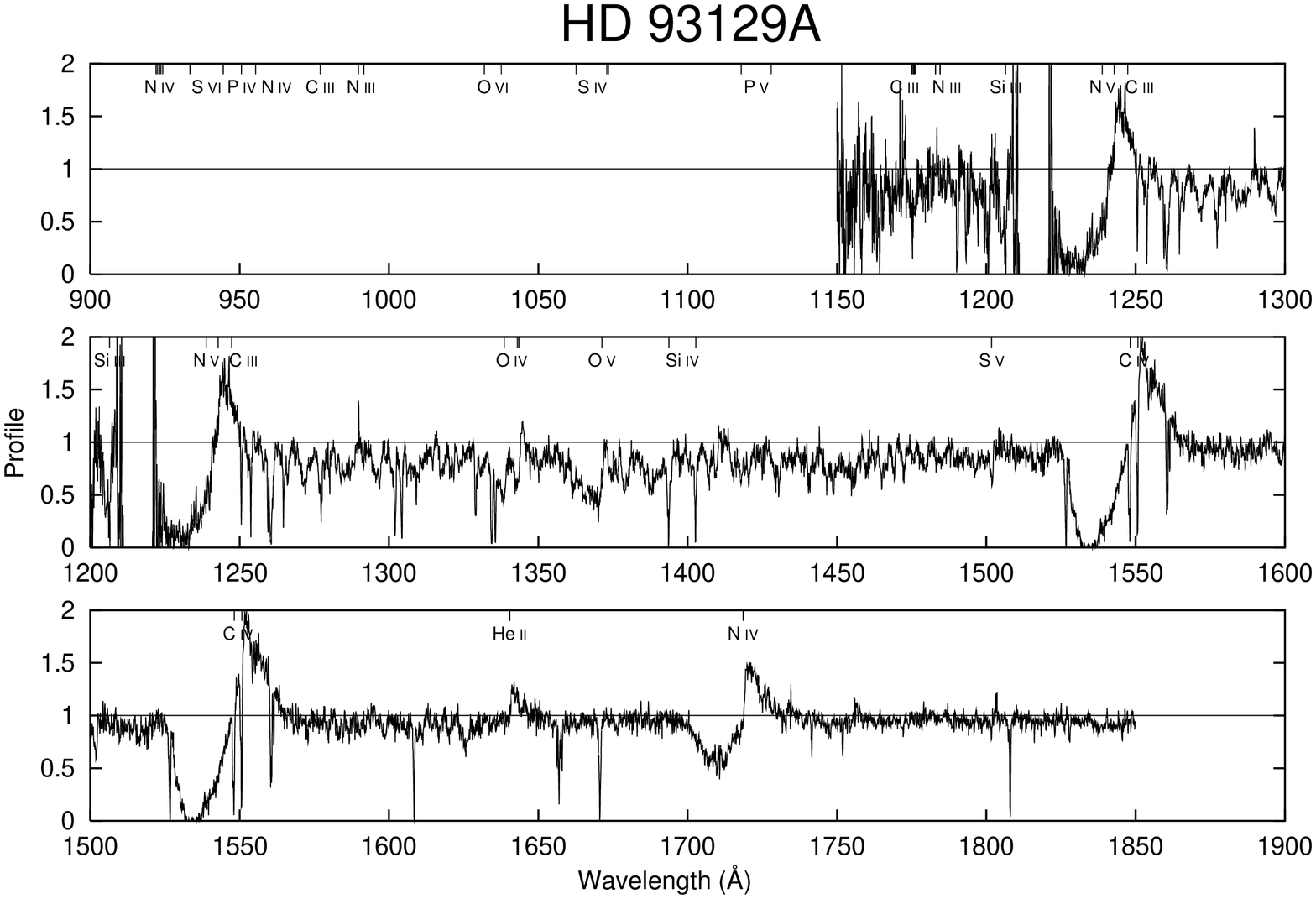}}
\vspace{5mm}
\mycaption{
Calculated UV spectra of the model grid supergiants (left) compared to
observed IUE spectra (right) of stars of similar spectral type.}
\label{fig:spS}
\end{figure*}

Regarding the first point one might speculate that the inclusion of
shock radiation leads to an improvement for the \OV\ line, although
this effect would weaken the \NIV\ line further. As is shown below,
shock radiation cannot solve the problem, as it does not affect the
strength of the \OV\ line at all (cf.~the discussion of the S-50 model
below). Regarding the second point there are three parameters which
could lead to an improvement for both lines. The first one is the
effective temperature which, however, would have to be decreased by at
least 5000\,K. This is on the one hand extremely unrealistic, since
O3 and O4 stars would have almost the same $\Teff$, and it would on
the other hand produce another discrepancy due to an increase of the
strength of the \OIV\ line at 1338\,\AA\ (cf.~the S-40 model in
Fig.~\ref{fig:spS}).  The second parameter is the mass loss rate and
the third one is the abundance. In order to investigate whether a
systematic variation in the mass loss rate can solve the problem we
computed a small model subgrid for this object by changing the mass
loss rate for model D-50, keeping all other parameters the same
(cf.~Table~\ref{tbl:massloss}).

\begin{table}[b]
\begin{center}
\tabcolsep1.8mm
\begin{tabular}{r|ccccc}
\hline Model & D-50-a & D-50 & D-50-b & D-50-c & D-50-d \\ \hline \hline
$\Mdot$ ($10^{-6} M_\odot/{\rm yr}$) & 11.0 & 5.6 & 0.56 & 0.12 & 0.01 \\
\hline
\end{tabular}
\end{center}
\mycaption{Mass loss rates for the models shown in
Figure~\ref{fig:massloss}. All other parameters are identical to those
of the D-50 model shown in Table~\ref{tbl:gridparams}.}
\label{tbl:massloss}
\end{table}

\begin{figure*}
\vspace{10mm}
\centerline{
\includegraphics[height=\columnwidth,angle=-90]{spec-HD_093250.ps} \hfill
\includegraphics[height=\columnwidth,angle=-90]{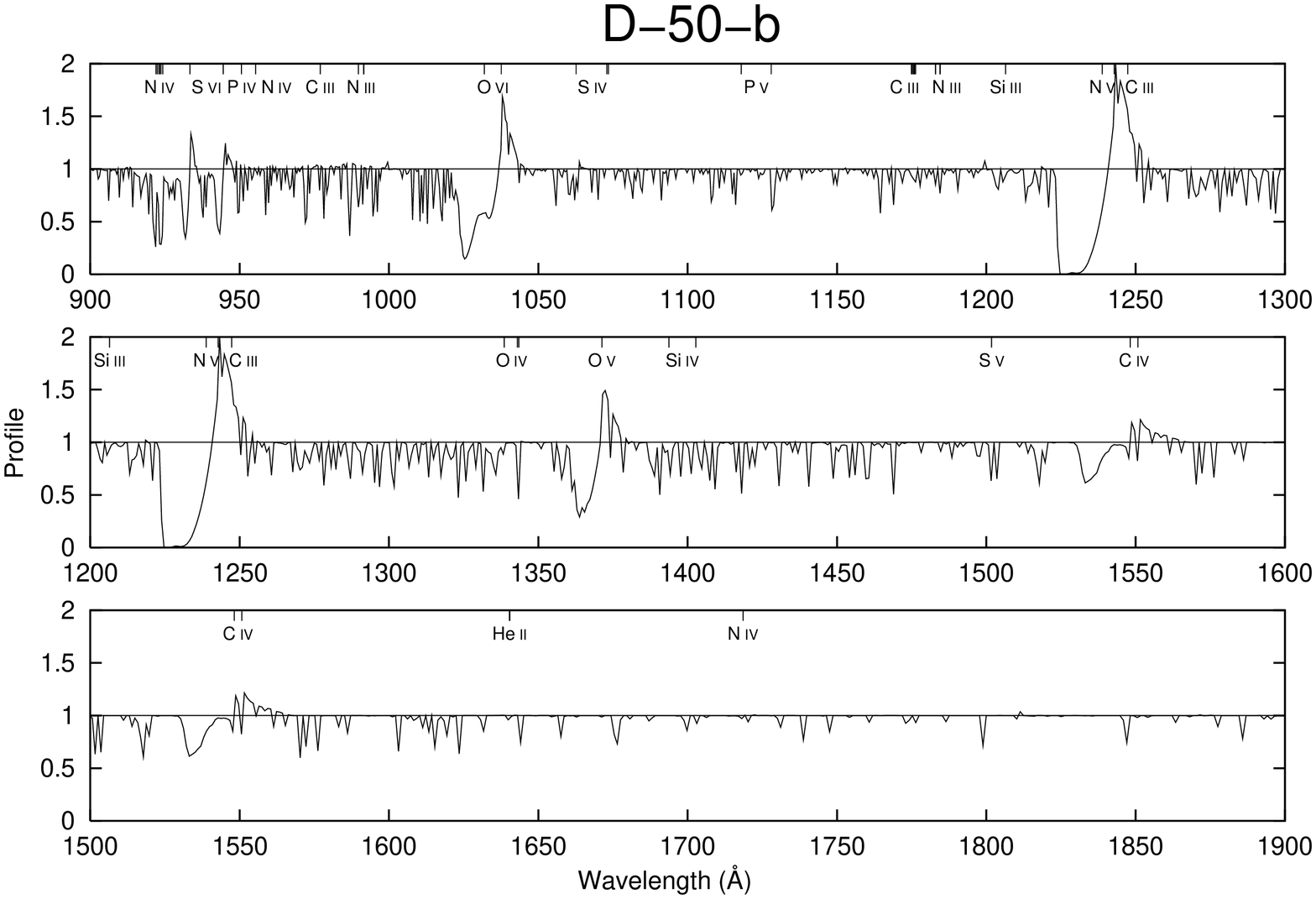}}
\vspace{10mm}
\centerline{
\includegraphics[height=\columnwidth,angle=-90]{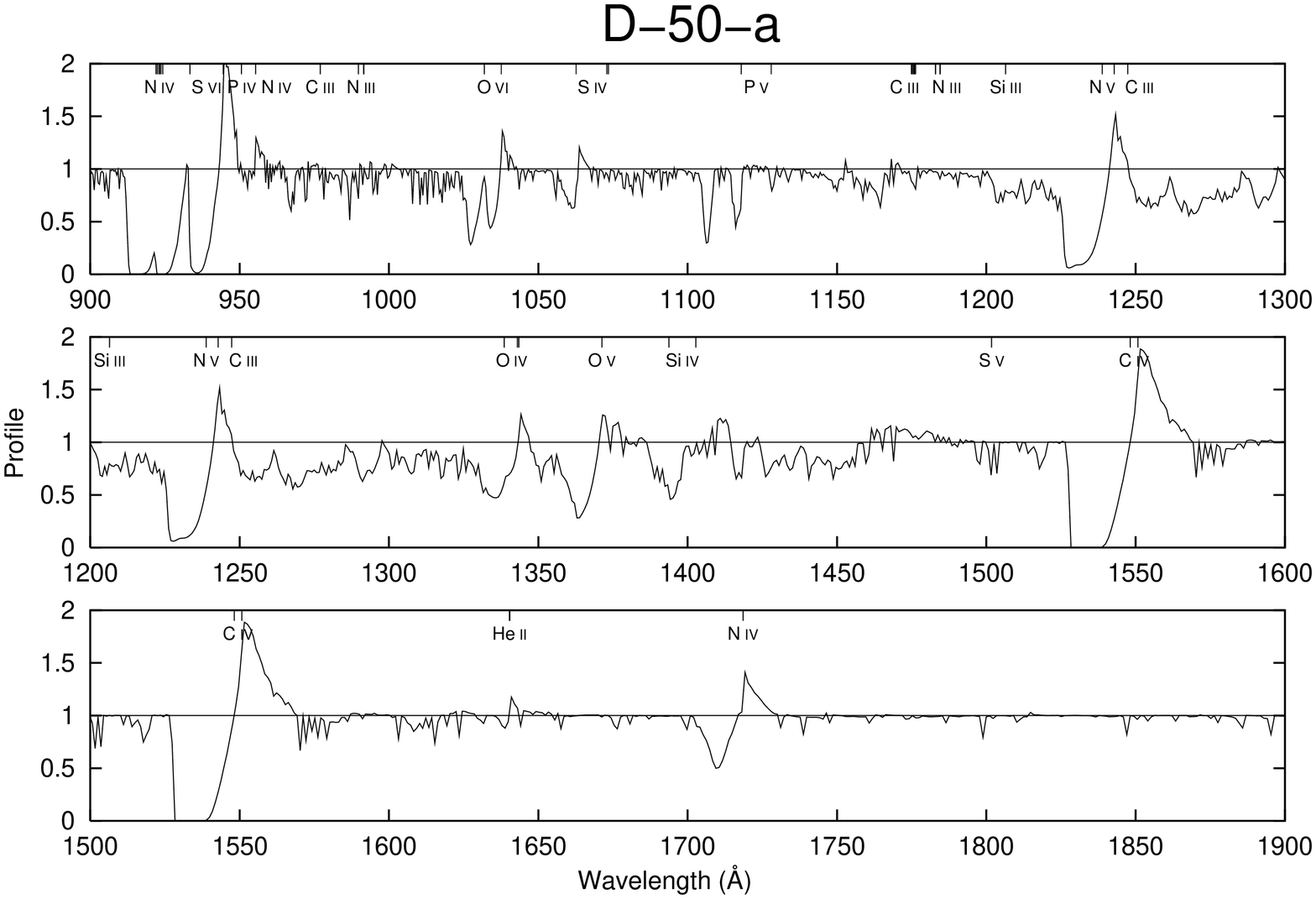} \hfill
\includegraphics[height=\columnwidth,angle=-90]{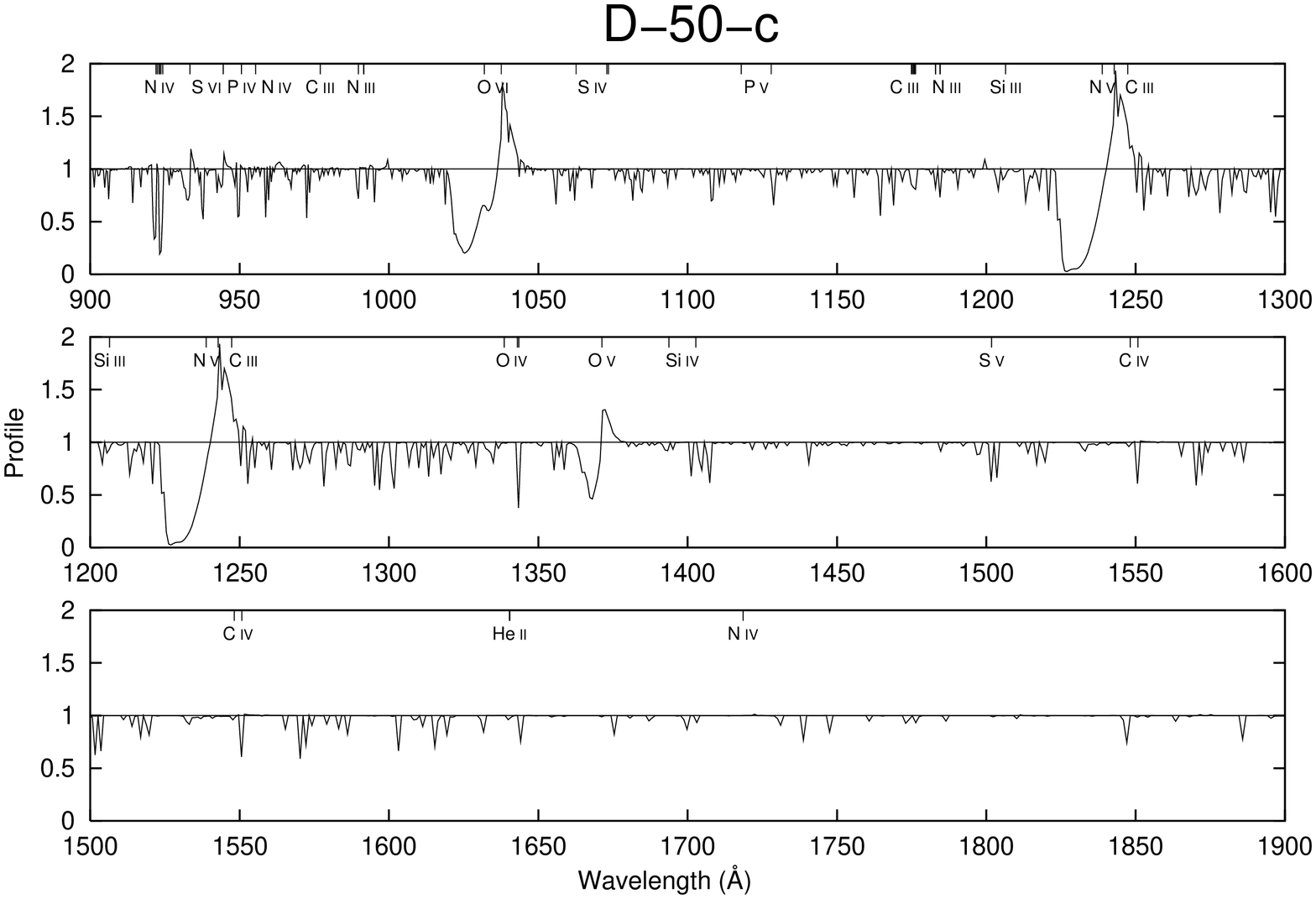}}
\vspace{10mm}
\centerline{
\includegraphics[height=\columnwidth,angle=-90]{spec-D-50.ps} \hfill
\includegraphics[height=\columnwidth,angle=-90]{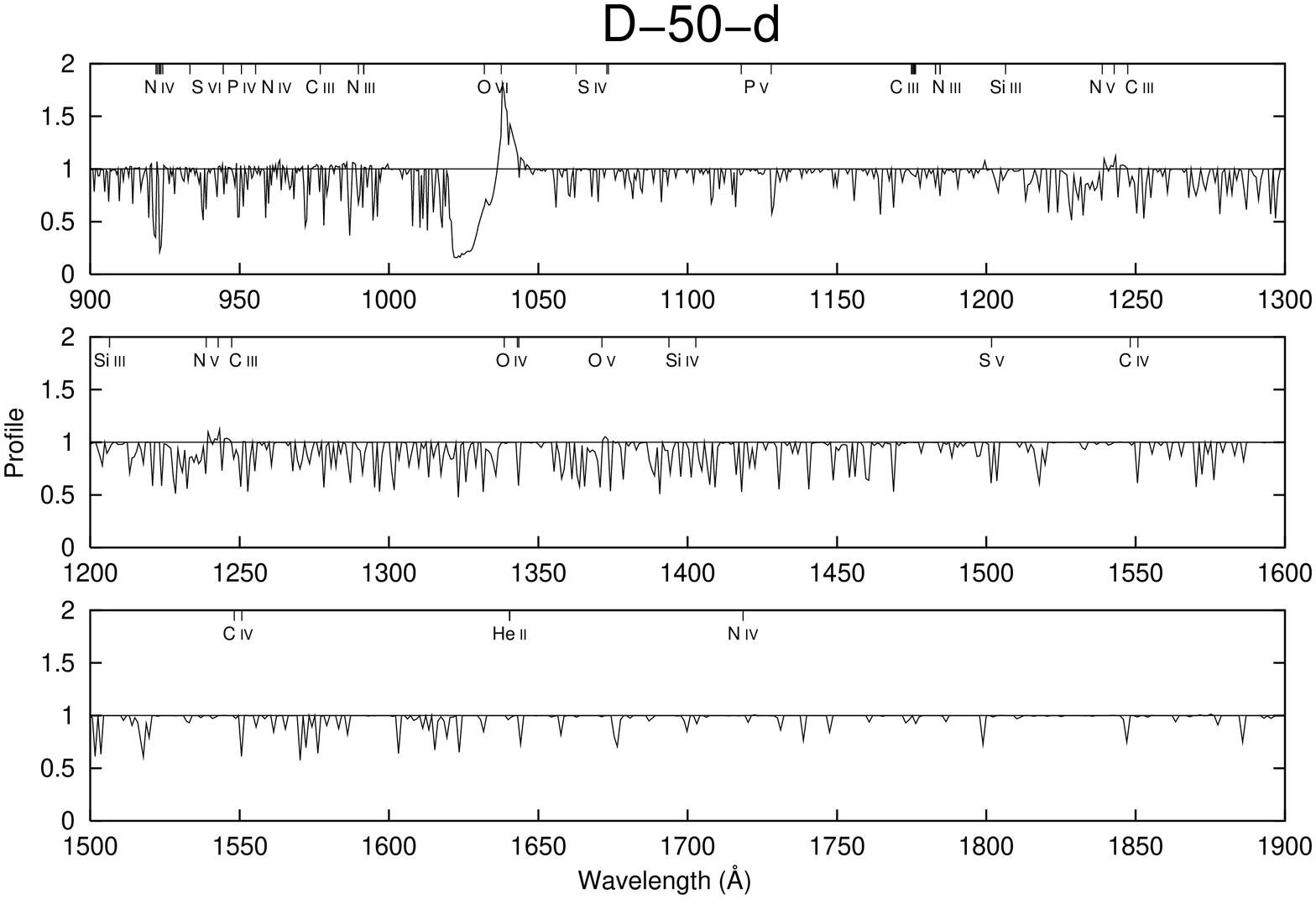}}
\vspace{5mm}
\mycaption{
Spectra obtained for a mass loss rate sequence for the D-50 dwarf
model. For comparison, the IUE spectrum of HD~93250 is also shown.
The mass loss rates of the models are given in
Table~\ref{tbl:massloss}. Lowering the mass loss rate cannot solve the
problem of the too-strong \OV~line, as this is one of the last lines
to disappear with diminishing mass loss rate.}
\label{fig:massloss}
\end{figure*}

\begin{figure*}
\centerline{\includegraphics[height=11cm,angle=-90]{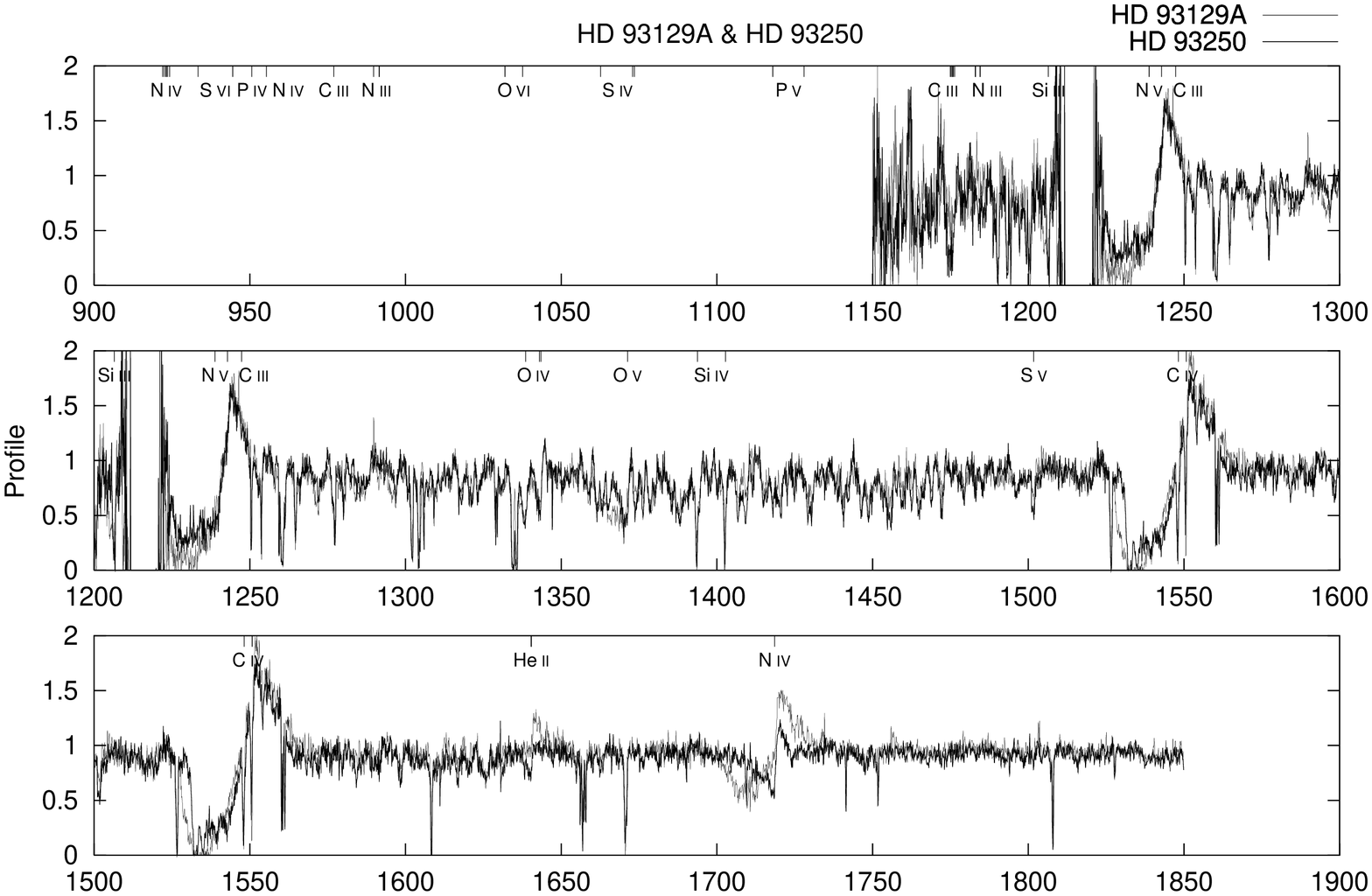}}
\centerline{\includegraphics[height=11cm,angle=-90]{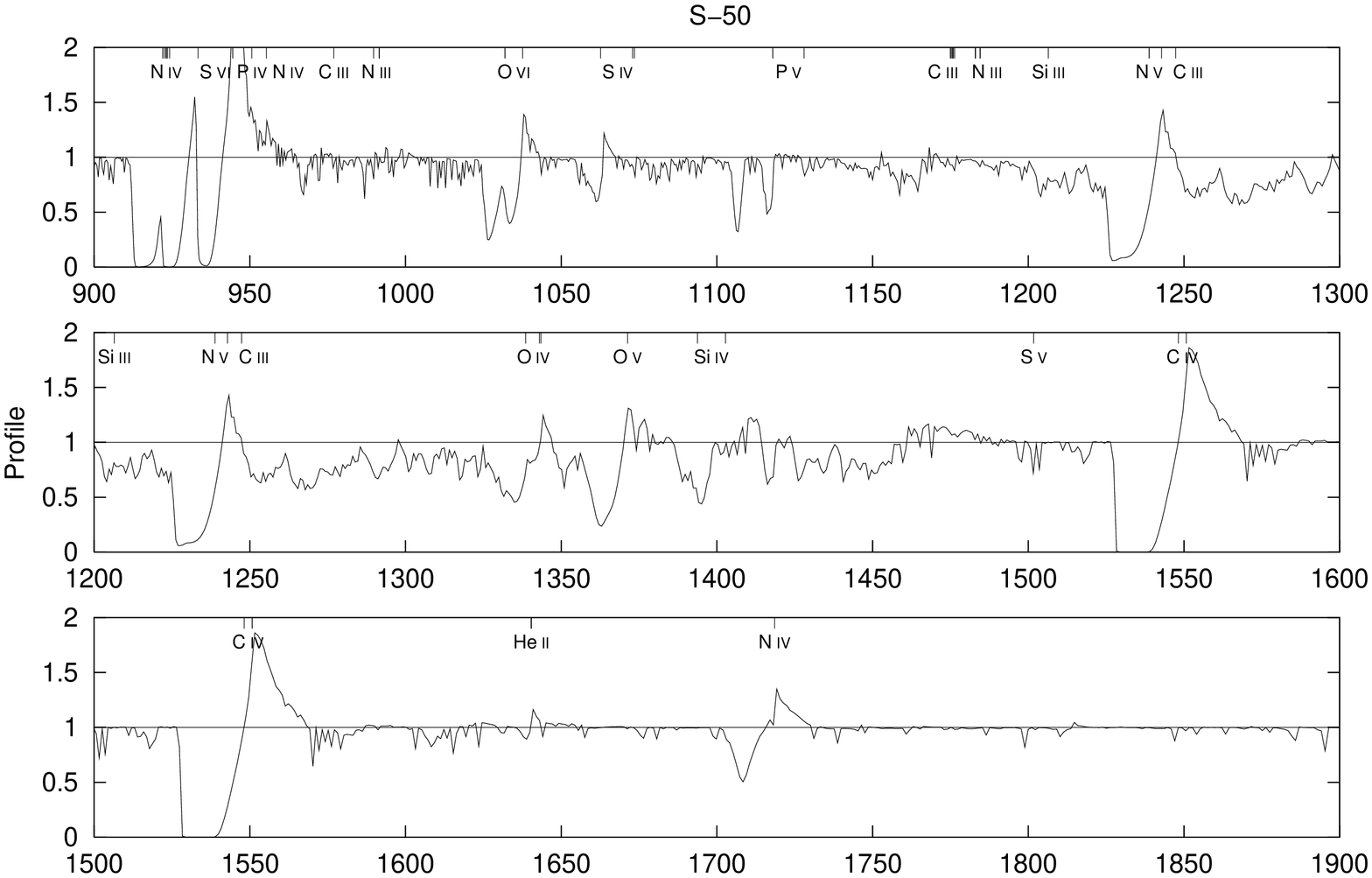}}
\centerline{\includegraphics[height=11cm,angle=-90]{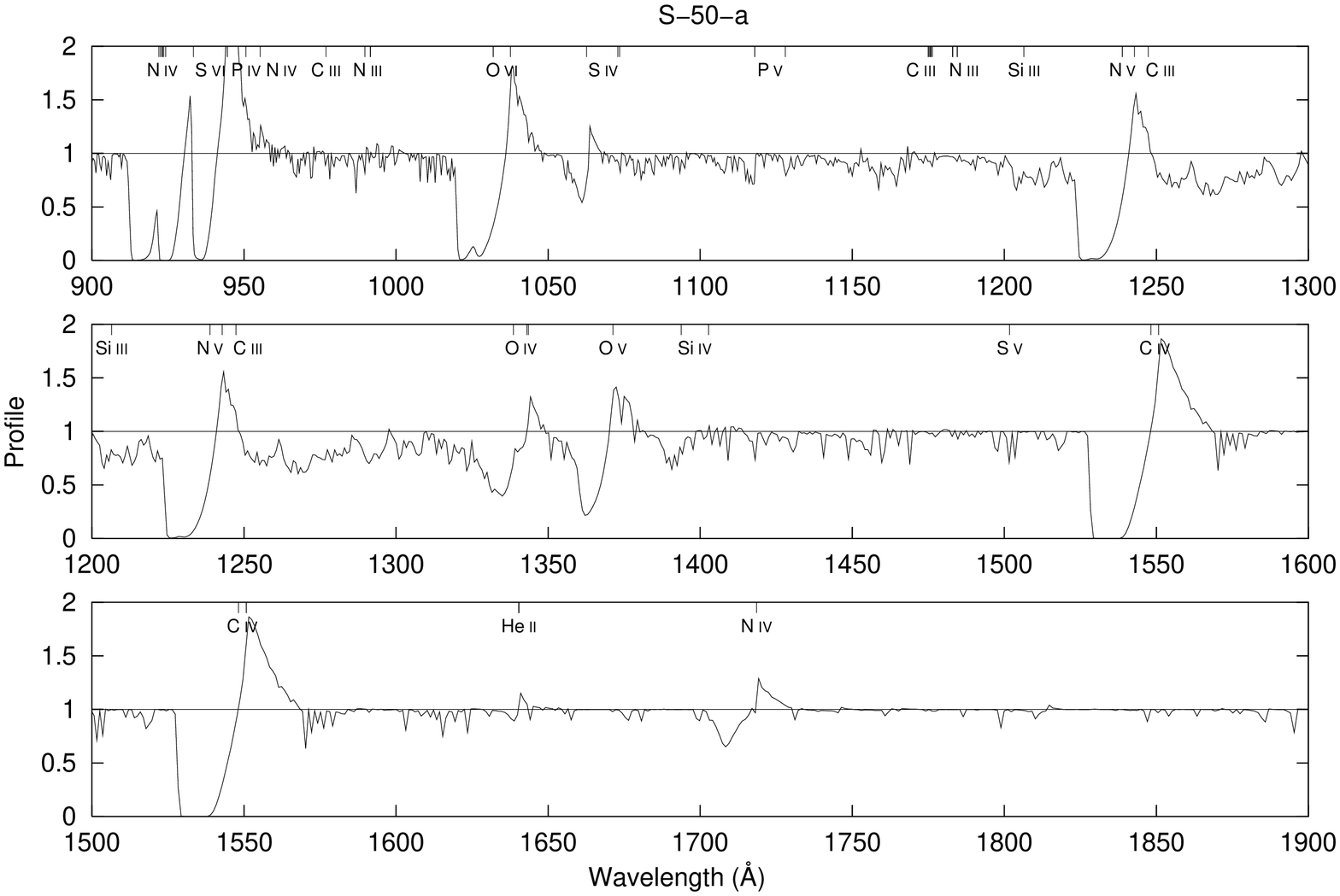}}
\mycaption{ Upper panel: Comparison of the IUE spectra of HD~93129A
(O3\,I\,f) and HD~93250 (O3\,V\,((f))). Apart from the strength of a
few lines of light elements (e.\,g., the \NIV\ line at 1718\,\AA) the
spectra are almost identical. Other panels: Influence of shocks on the
spectrum of the S-50 supergiant model which is compared to the IUE
spectrum of HD~93129A. Middle panel: without shocks, lower panel: with
shocks. The influence of the shocks on the strength of the \OIV\ and
\OV~lines is negligible.}
\label{fig:comp-sd}
\end{figure*}

The synthetic spectra obtained for the mass loss rate sequence are
shown in Figure~\ref{fig:massloss}. As can be seen, lowering the mass
loss rate does not solve the problem of the too-strong \OV~line, as
this is one of the last lines to disappear with diminishing mass loss
rate, whereas the \NIV~line which was already too weak disappears
immediately.  A mass loss rate as low as $10^{-8}\,M_\odot/{\rm yr}$
would be required to reduce the strength of the \OV~line to the
observed case. The only model where the situation has clearly improved
regarding both the \NIV\ and the \OV~line is D-50-a, the one with a
higher mass loss rate. Hence we conclude that the mass loss rate has
to be larger by a factor of $\approx 2$ --- note that this is also
indicated by the stellar parameters which are almost identical to
those of HD~93129A, which means that the wind parameters have to be
almost identical too; note further the strong similarity of the
observed spectra of HD~93129A and HD~93250, which points in the same
direction (cf.~Fig.~\ref{fig:comp-sd}, upper panel); in addition,
the abundances of the CNO elements have to be reduced strongly
compared to the solar values (for oxygen a reduced value of
approximately a factor of 50 is required to weaken the strength of
both the \OIV\ and the \OV~line, as has been inferred from additional
test calculations).  This, however, should not be the case for the
heavier elements (Fe, Ni) since a solar-like abundance of these
elements is required to account for the radiative acceleration
necessary to produce the higher mass loss rate
(cf.~\cite{}Pauldrach,~1987).

\begin{figure*}
\centerline{\includegraphics[height=14cm,angle=-90]{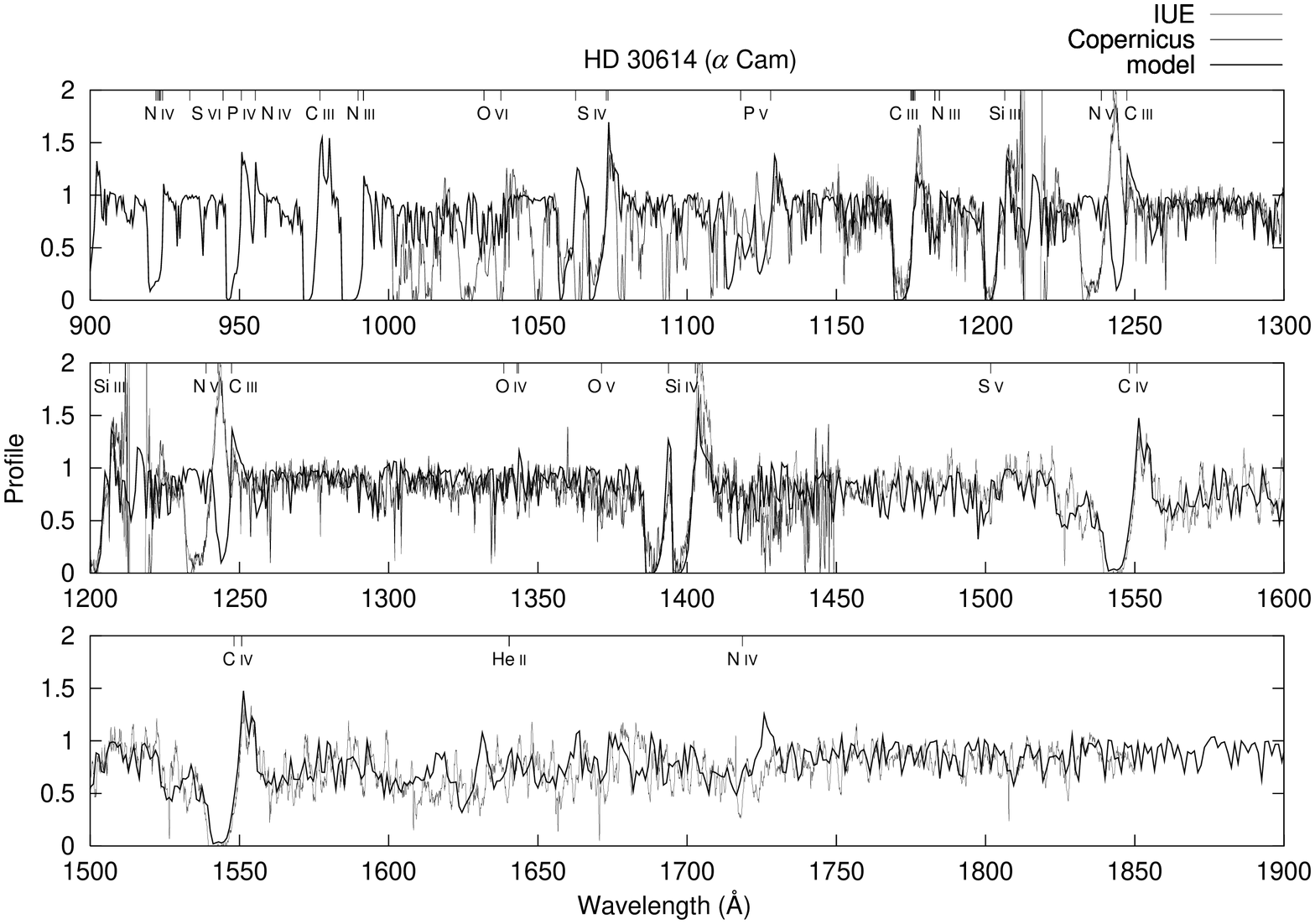}}
\mycaption{
Comparison of the basic S-30 grid model with spectra of $\alpha$~Cam
observed by IUE and Copernicus.}
\label{fig:alphacam-0}
\end{figure*}

Now we inspect the comparison of the supergiant models
(Fig.~\ref{fig:spS}).  As is shown, the observed spectra are
reproduced in principle quite well apart from minor differences which
can be attributed to a change of abundances (note that the discrepancy
of the \NV\ resonance line is due to the omitted shock radiation ---
see below). Again, the most conspicuous difference regards the
\OIV\ and \OV\ subordinate lines which are both too strong, especially
for the S-40 and the S-50 models. From the investigation above it is
already quite clear that the abundance(s) of the (CN)O element(s) has
(have) to be reduced in order to overcome this discrepancy.
Nevertheless, we investigate now whether the inclusion of shock
radiation leads to an improvement. We have therefore computed an
additional model for the S-50 supergiant, model~S-50-a, where the
influence of shocks on the spectrum has been accounted for; the shock
parameters are given in Table~\ref{tbl:s50shock}.

\begin{table}
\mycaption{Shock parameters for the S-50-a model ---
for an explanation of the parameters see Section~4.3.}
\label{tbl:s50shock}
\begin{center}
\begin{tabular}{cccc}
\hline
$\logLx$ & $\vturb$ & $\gamma$ & $m$ \\
\hline
\hline
$-$7.0 & 0.1 & 1 & 1 \\
\hline
\end{tabular}
\end{center}
\end{table}

Fig.~\ref{fig:comp-sd} shows that the influence of the shocks on the
strength of the \OIV\ and \OV~lines, and hence the ionization balance,
is negligible --- just \OVI\ is enhanced selectively, which can be
seen by the strength of the \OVI\ resonance line.  Another line which
is considerably affected by shock emission is the \SiIV\ resonance
line. This is because the soft X-ray radiation field of the shocks
enhances the ionization of \SiV, and thus the recombination to
\SiIV\ is decreased (cf.~\cite{}Pauldrach et al.,~1994).  As can be
inferred from Fig.~\ref{fig:comp-sd} this improves the fit of the
\SiIV\ line significantly. Concerning oxygen we come to the same
conclusion as for the dwarf models, namely that the abundance has to
be reduced strongly compared to the solar value.

\subsection{Detailed analysis of $\alpha$~Cam}

\def\sameasabove{--- \raisebox{-.5ex}{\tt "} ---}
\begin{table*}
\mycaption{Model parameters for $\alpha$~Cam. Abundances not explicitly
mentioned are solar; if given, the number is the factor relative to
the solar value. $\Teff$ is in K, $\Mdot$ in $10^{-6}\,\Msun/{\rm
yr}$.}
\label{tbl:alphacammodels}
\begin{center}
\begin{tabular}{r@{}cccccccccc}
\hline
 & Model & $\Teff$ & $\Mdot$ & $\vturb$ & $\logLx$ & $\gamma$ & $m$ & abundances \\
\hline\hline
 & a  & 30000 & 5 & 0.125 & $-$7.5 & 1   &  1 & solar \\
 & b1 & 30000 & 5 & 0.14  & $-$7.5 & 1   &  1 & solar \\
 & b2 & 30000 & 5 & 0.14  & $-$6.0 & 1   &  1 & solar \\
 & b3 & 30000 & 5 & 0.14  & $-$6.5 & 1   &  1 & solar \\
 & c  & 30000 & 5 & 0.20  & $-$7.5 & 0.5 &  1 &
     C~=~0.1, N~=~10., O~=~1.0, P~=~0.1, S~=~2.0 \\
 & d1 & 30000 & 5 & 0.14  & $-$7.0 & 0.5 &  1 & \sameasabove \\
 & d2 & 30000 & 5 & 0.14  & $-$7.0 & 0.5 & 30 & \sameasabove \\
 & d3 & 30000 & 5 & 0.14  & $-$7.0 & 0.5 & 40 & \sameasabove \\
 & d4 & 30000 & 5 & 0.14  & $-$7.0 & 0.5 & 60 & \sameasabove \\
 & e  & 30000 & 5 & 0.25  & $-$6.5 & 0.5 & 30 &
     C~=~0.1, N~=~2.0, O~=~0.3, P~=~0.05, S~=~1.0 \\
 & f  & 29000 & 5 & 0.25  & $-$6.5 & 0.5 & 30 & \sameasabove \\
best fit $\rightarrow$
 & g  & 29000 & 5 & 0.25  & $-$6.5 & 0.5 & 30 &
     C~=~0.05, N~=~1.0, O~=~0.3, P~=~0.05, S~=~1.0 \\
 & h  & 28000 & 5 & 0.25  & $-$6.5 & 0.5 & 30 & \sameasabove \\
 & i  & 28000 & 2.5 & 0.25  & $-$6.5 & 0.5 & 30 & \sameasabove \\
 & j  & 28000 & 10 & 0.25  & $-$6.5 & 0.5 & 30 & \sameasabove \\
\hline
\end{tabular}
\end{center}
\end{table*}

In this section we provide, using our S-30 grid model as a starting
point, a detailed determination of the abundances and stellar wind
properties, and verification of the stellar parameters of
$\alpha$~Cam.  Special emphasis is given to the shock radiation needed
for a fit of the \NV\ and \OVI\ resonance lines.  Although preliminary
results from our new shock description look very promising (see
section~\ref{sec:shocks}), this new method is not yet fully
implemented in WM-basic. For this investigation we therefore use the
method based on isothermal shocks. See Pauldrach et al.~(1994) for a
detailed explanation of the shock parameters and the rationale behind
the parametrization.

We wish to point out here that we have not attempted an exact
determination of $\logg$ and stellar radius, but have rather kept
the values of our S-30 grid model. The reason is that in contrast to
the hydrodynamics the UV spectrum depends only marginally on these
parameters, the main influence being due to $\Teff$ and $\Mdot$
(i.\,e., density). The radius and the surface gravity ($\logg$) can in
principle be determined from a selfconsistent calculation of the
hydrodynamics and the NLTE model, in which both values would be
adapted in such a way that the hydrodynamics, with consistent force
multiplier parameters from the NLTE occupation numbers (in turn again
dependent on the hydrodynamic solution), would yield the mass loss
rate and the terminal velocity deduced from $H_\alpha$ and the
observed UV spectrum (cf.~Pauldrach et al.,~1994, 1994a). This latter
procedure has however not yet been implemented in WM-basic.

In Fig.~\ref{fig:alphacam-0} we compare the spectrum of our basic S-30
grid model with the spectra of $\alpha$~Cam observed with IUE and
Copernicus.  With the exception of a few strong lines, notably the
\NV\ resonance doublet at 1238, 1242~\AA\ and the subordinate \CIII\
line at 1247~\AA, the agreement is very good for a first step,
especially with respect to the iron and nickel ``forest'' between 1400
and 1600~\AA.  As the two lines mentioned above are mainly affected by
shock radiation, we will first attempt to fit the shock parameters.
We note in passing that the strong lines from 1000 to 1100~\AA\ (with
the recurring pattern), that complicate an exact fit of the \OVI\
resonance line, are absorption by interstellar molecular hydrogen.

\begin{figure*}
\centerline{\includegraphics[height=11.7cm,angle=-90]{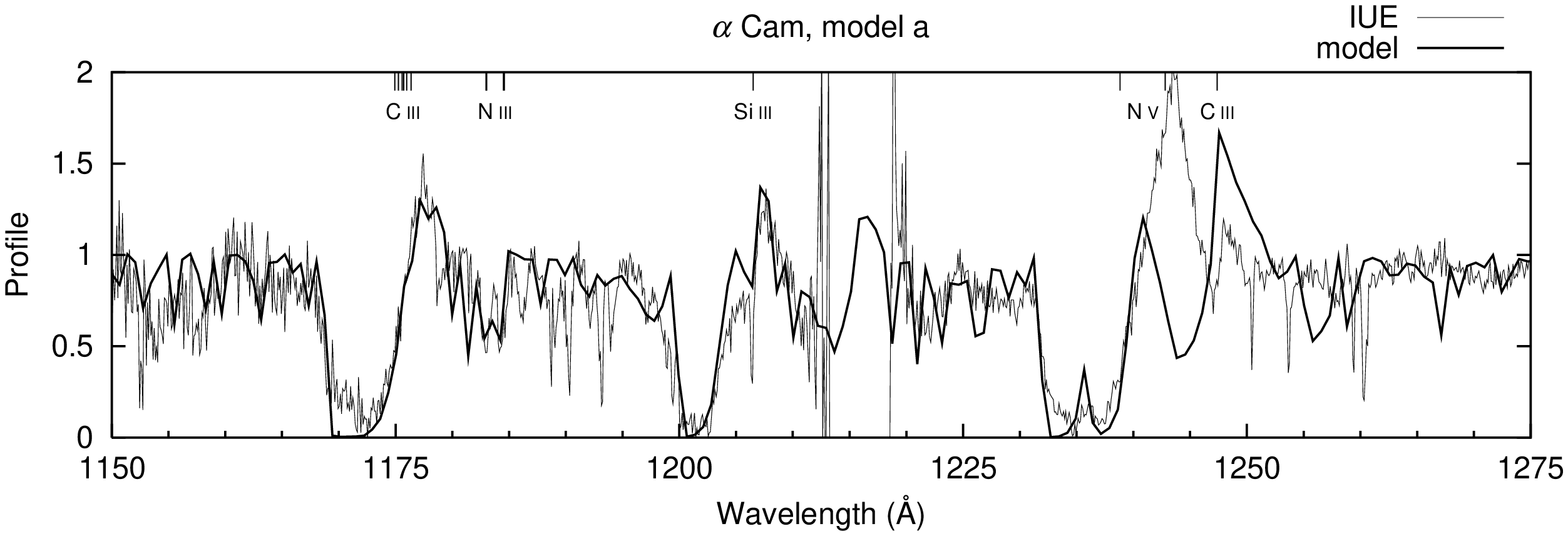}}
\mycaption{Model~a, using our ``first guess'' values for the shock
parameters. \CIII~$\lambda$1247 is much too strong.}
\label{fig:alphacam-a}
\centerline{\includegraphics[height=11.7cm,angle=-90]{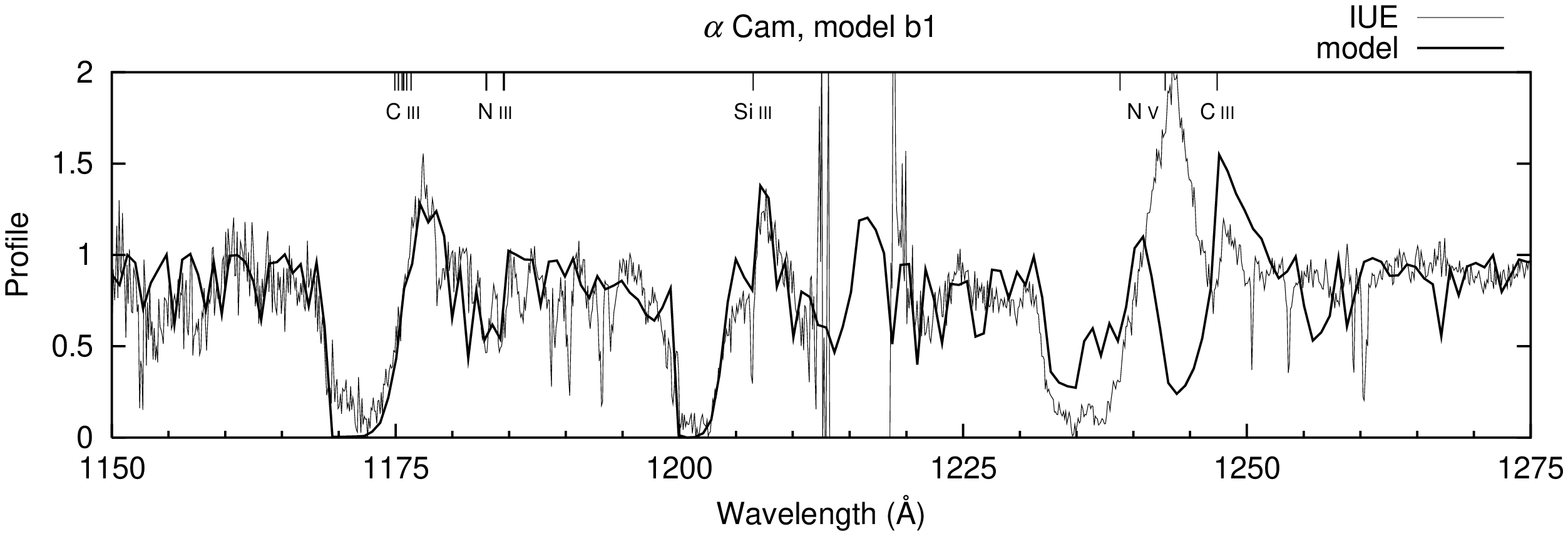}}
\centerline{\includegraphics[height=11.7cm,angle=-90]{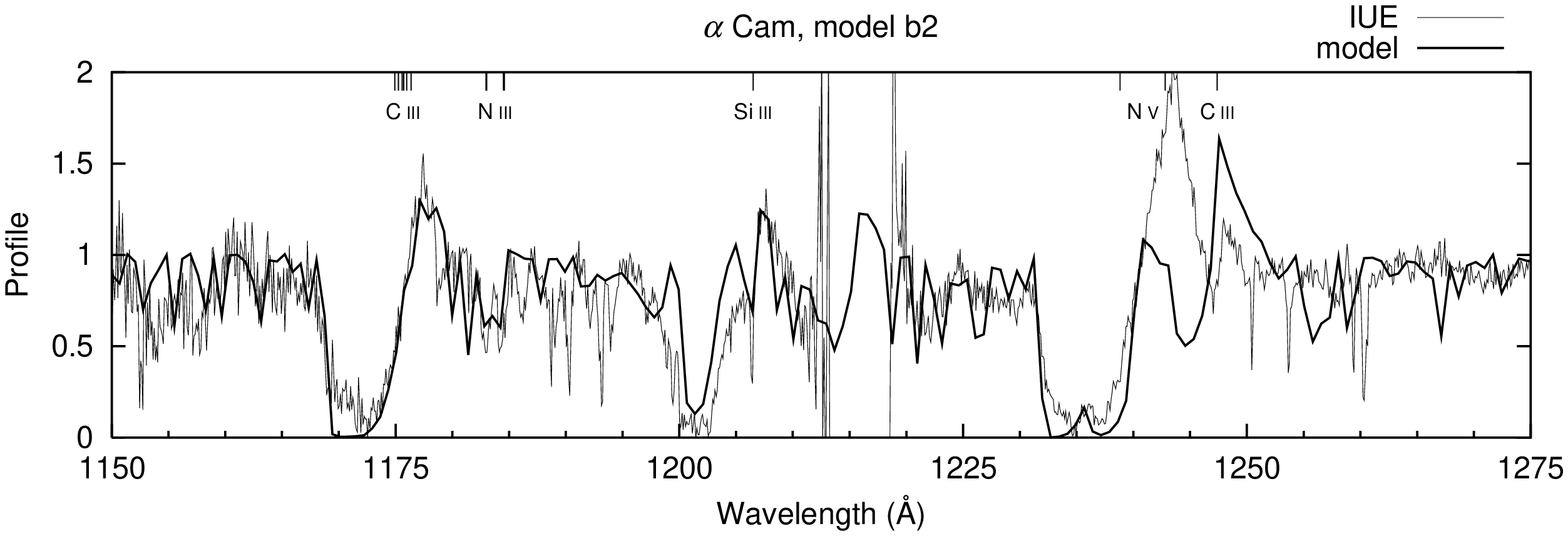}}
\centerline{\includegraphics[height=11.7cm,angle=-90]{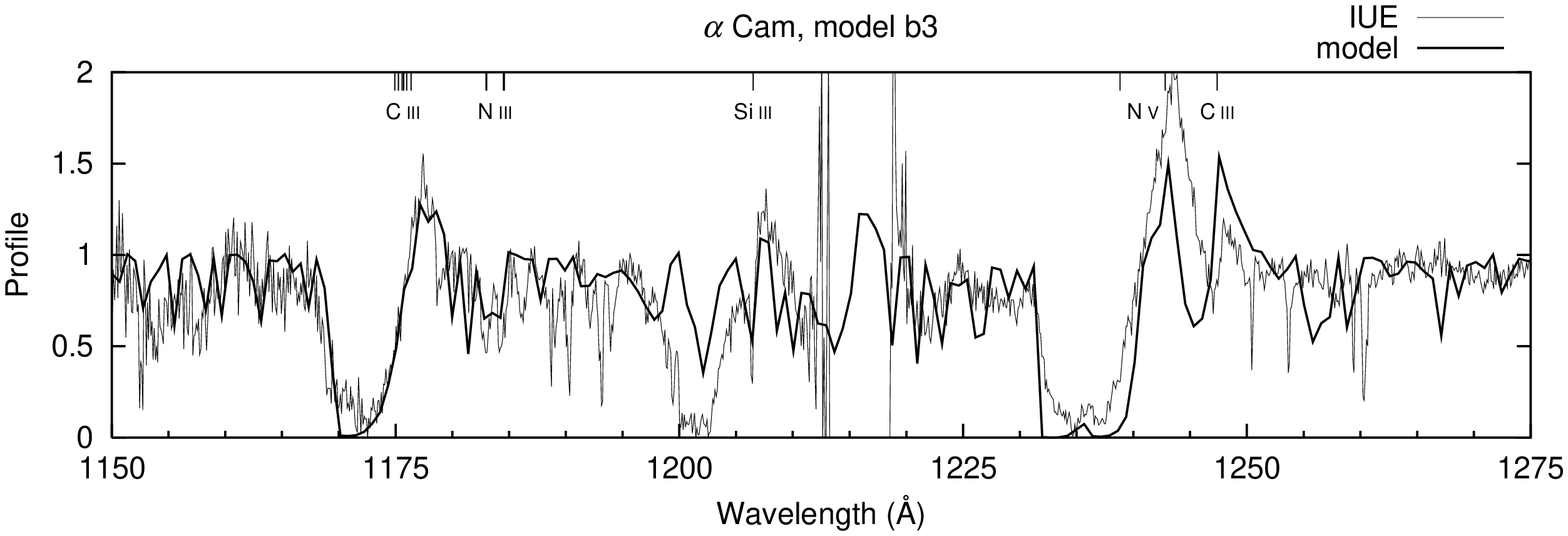}}
\mycaption{Models~b1--b3, a sequence showing the influence of increasing
$\logLx$. Note that \SiIII\ disappears before \CIII\ has decreased to
its observed strength.}
\label{fig:alphacam-b}
\centerline{\includegraphics[height=11.7cm,angle=-90]{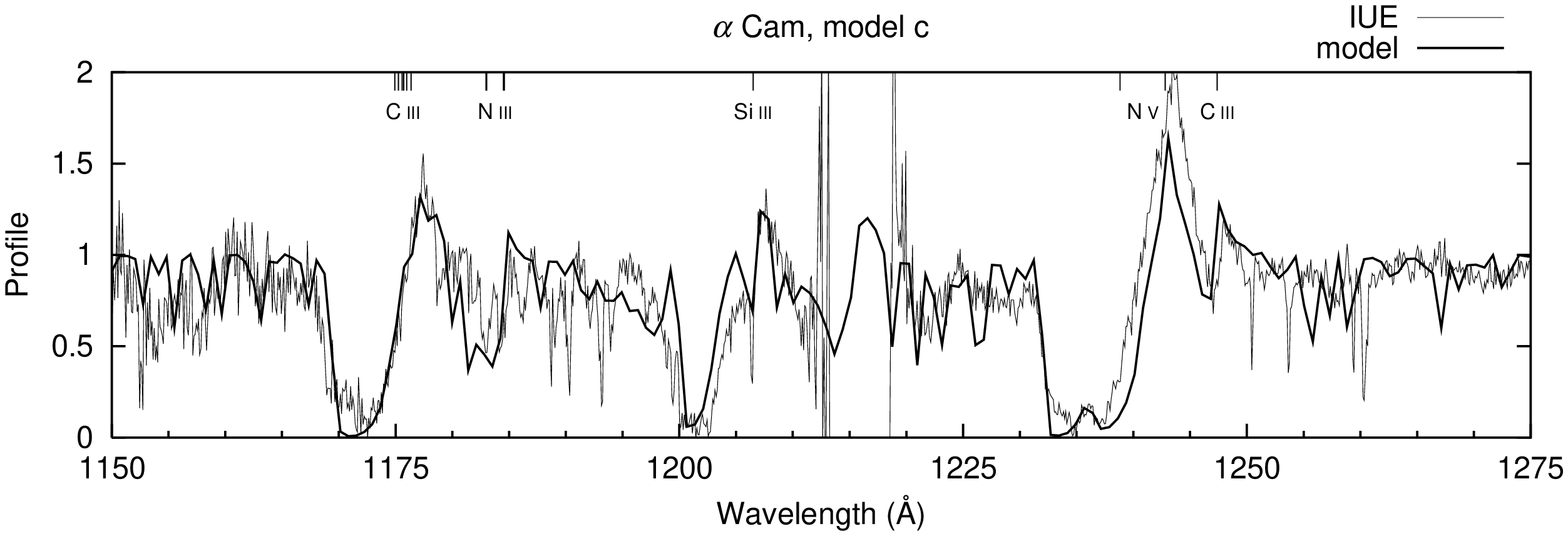}}
\mycaption{Model~c. Here we have adapted the shock parameters and
abundances (see Table~\ref{tbl:alphacammodels}) to provide for a
reasonable fit of \SiIII\ and \CIII.}
\label{fig:alphacam-c}
\end{figure*}

A model with reasonable ``first guess'' values for the shock
parameters (model~a, Table~\ref{tbl:alphacammodels}) already gives a
very good fit to the \NV\ resonance line, as shown in
Fig.~\ref{fig:alphacam-a}, the emission, however, completely decimated
by the much too strong \CIII~$\lambda$1247 line.  Increasing the X-ray
luminosity ($\logLx$) to reduce the \CIII\ occupation unfortunately
also tends to ionize \SiIII\ to \SiIV, thus reducing the strength of
the \SiIII\ line. Modifying $\vturb$ (the ratio of the maximum jump
velocity to the terminal velocity, where $v_{\rm t}$ characterizes the
immediate post-shock temperature; see Pauldrach et al.,~1994) does
not change this fact, as we have confirmed by calculating a grid of
models with values of $\logLx$ ranging from $-$8 to $-$6.5 and
$\vturb$ ranging from 0.1 to 0.2. For example, a model sequence
(models~b1--b3) is shown in Fig.~\ref{fig:alphacam-b}, in which we
vary $\logLx$ from $-$7.5 to $-$6.5, keeping all other parameters
constant (cf.~Table~\ref{tbl:alphacammodels}).  As can bee seen, the
\SiIII\ line already begins to weaken, while \CIII\ still remains too
strong. Adjusting the $\gamma$-parameter (which controls the strength
of the shocks relative to the local velocity --- cf.~Pauldrach et
al.,~1994) does not help either, as both the \CIII\ and the \SiIII\
line are formed in the inner part of the wind, thus both being subject
to the same radiation field.

From this we conclude that solar abundances cannot reproduce the
observed spectrum. A model, however, with a reduced carbon abundance
of one tenth solar (model~c) can indeed give a good fit of the line,
as shown in Fig.~\ref{fig:alphacam-c}. We will use this abundance for
the following calculations, unless stated otherwise. As carbon thus
shows indications of the CNO-process, we have for reasons of
consistency also increased the nitrogen abundance by a factor of 10,
but such a large factor is not compatible with the fit of the nitrogen
lines; from the final models we determined the nitrogen abundance to
be approximately solar.

Despite the good fit, this model is still not satisfactory, as it
shows no signs of \OVI. Since the \CIII/\SiIII-balance strongly
constrains the shock strength in the inner regions, this cannot be
achieved simply by increasing the X-ray luminosity. However, the onset
of shocks can be adapted with the $m$-parameter, which gives the ratio
of outflow to sound velocity where shocks start to form, and as the
jump velocity is correlated to the outflow velocity, the corresponding
radius (cf.~Pauldrach et al.,~1994). Increasing this parameter allows
shocks in the outer regions (\OVI\ appears close to $\vinf$), while
leaving the inner regions (where \SiIII\ is present) largely
undisturbed. This behaviour of the shocks is already an important
result from our analysis.

\begin{figure*}
\centerline{\includegraphics[height=11.8cm,angle=-90]{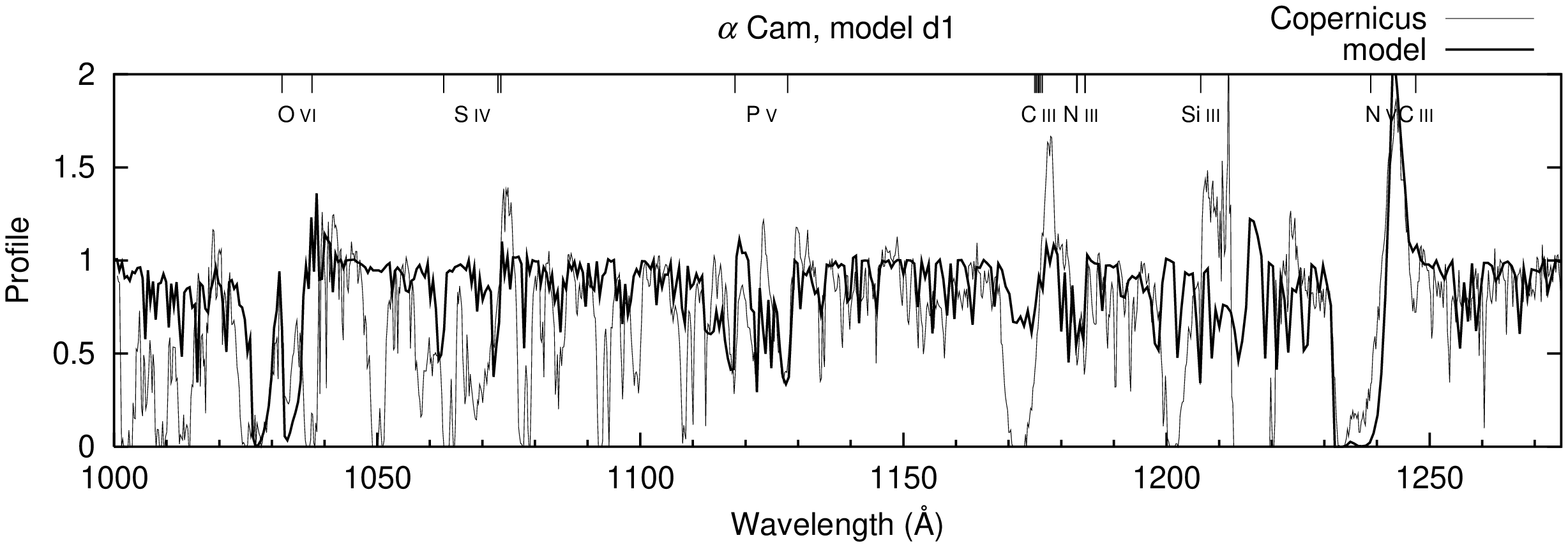}}
\centerline{\includegraphics[height=11.8cm,angle=-90]{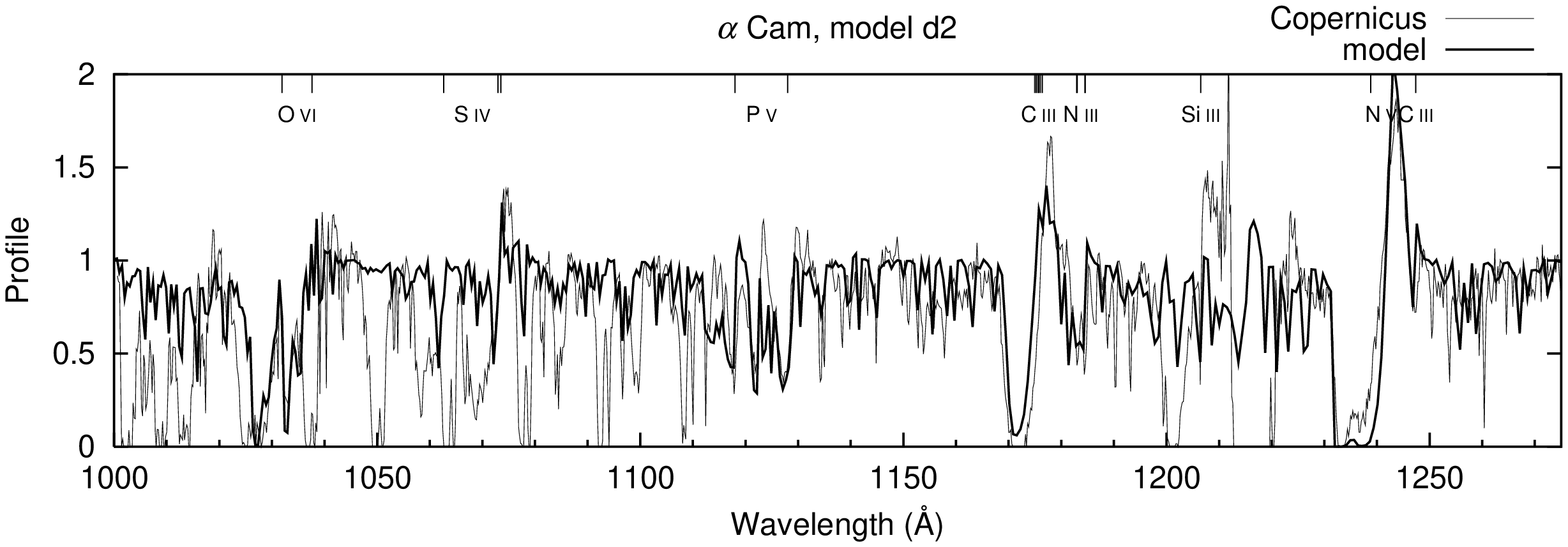}}
\centerline{\includegraphics[height=11.8cm,angle=-90]{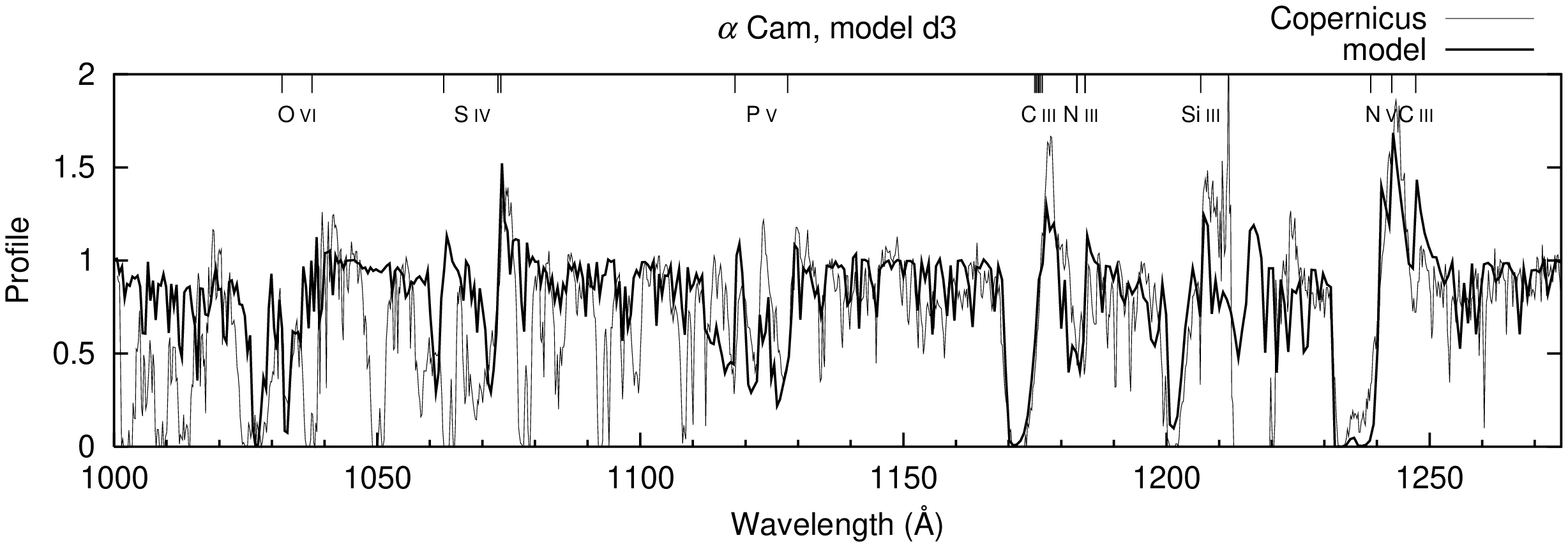}}
\centerline{\includegraphics[height=11.8cm,angle=-90]{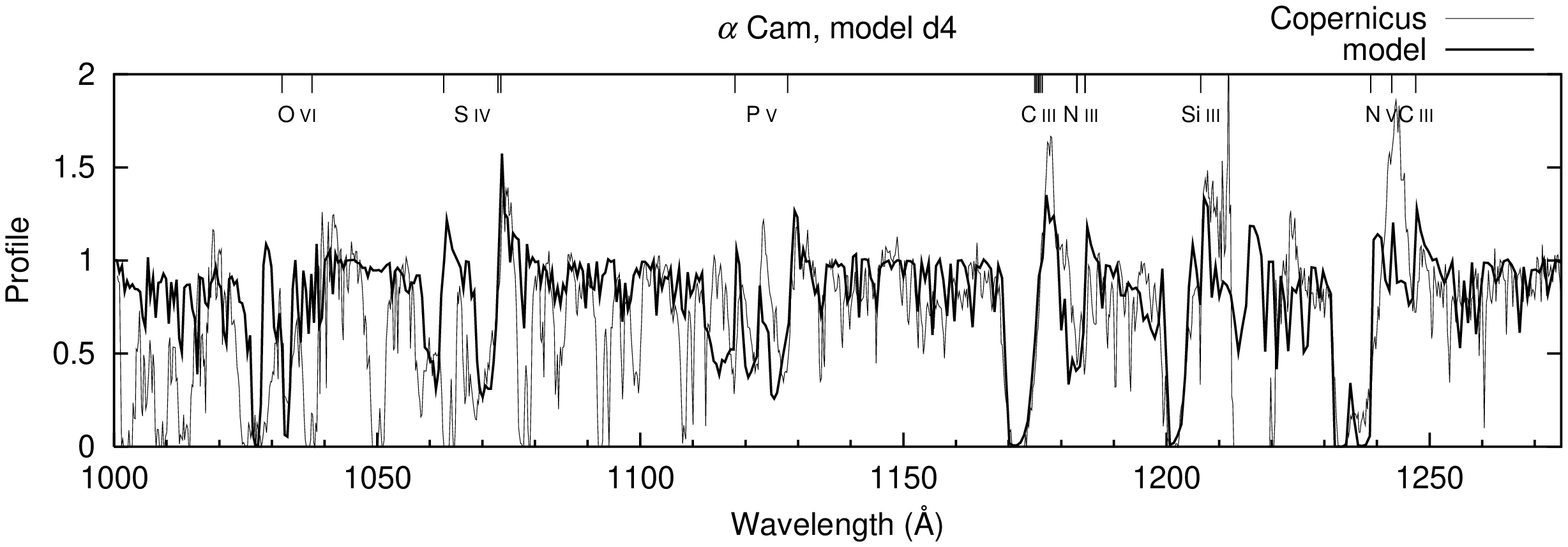}}
\mycaption{Models~d1--d4, a sequence in which the radius at which shocks
start to form has been successively moved outwards, keeping all other
parameters equal. In model d4, where \SiIII\ remains fairly unaffected
by shocks, the P~Cygni emission of \OVI\ and \NV\ has completely
disappeared.}
\label{fig:alphacam-d}
\centerline{\includegraphics[height=11.8cm,angle=-90]{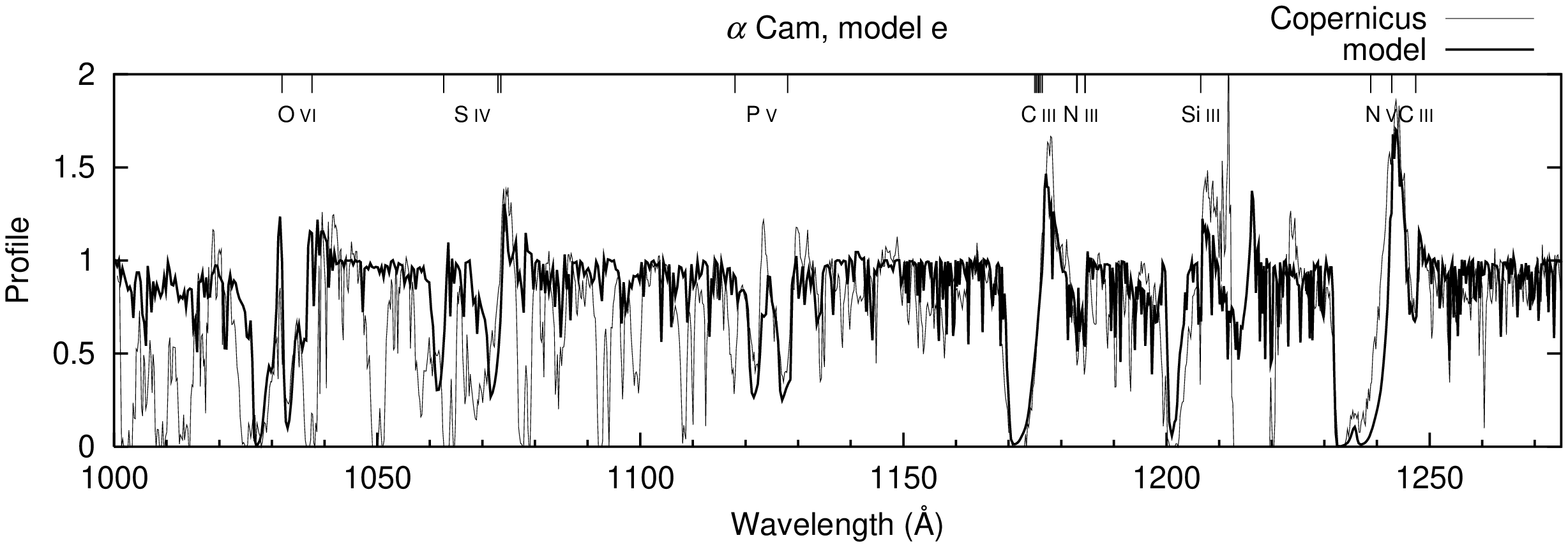}}
\mycaption{Model~e. Increasing the shock jump velocity provides
for harder shock radiation that has a stronger influence on \OVI\ and
\NV\ than on \SiIII. The abundances of C, N, O, and P have also been
adapted.}
\label{fig:alphacam-e}
\end{figure*}

\begin{figure*}
\centerline{\includegraphics[height=15cm,angle=-90]{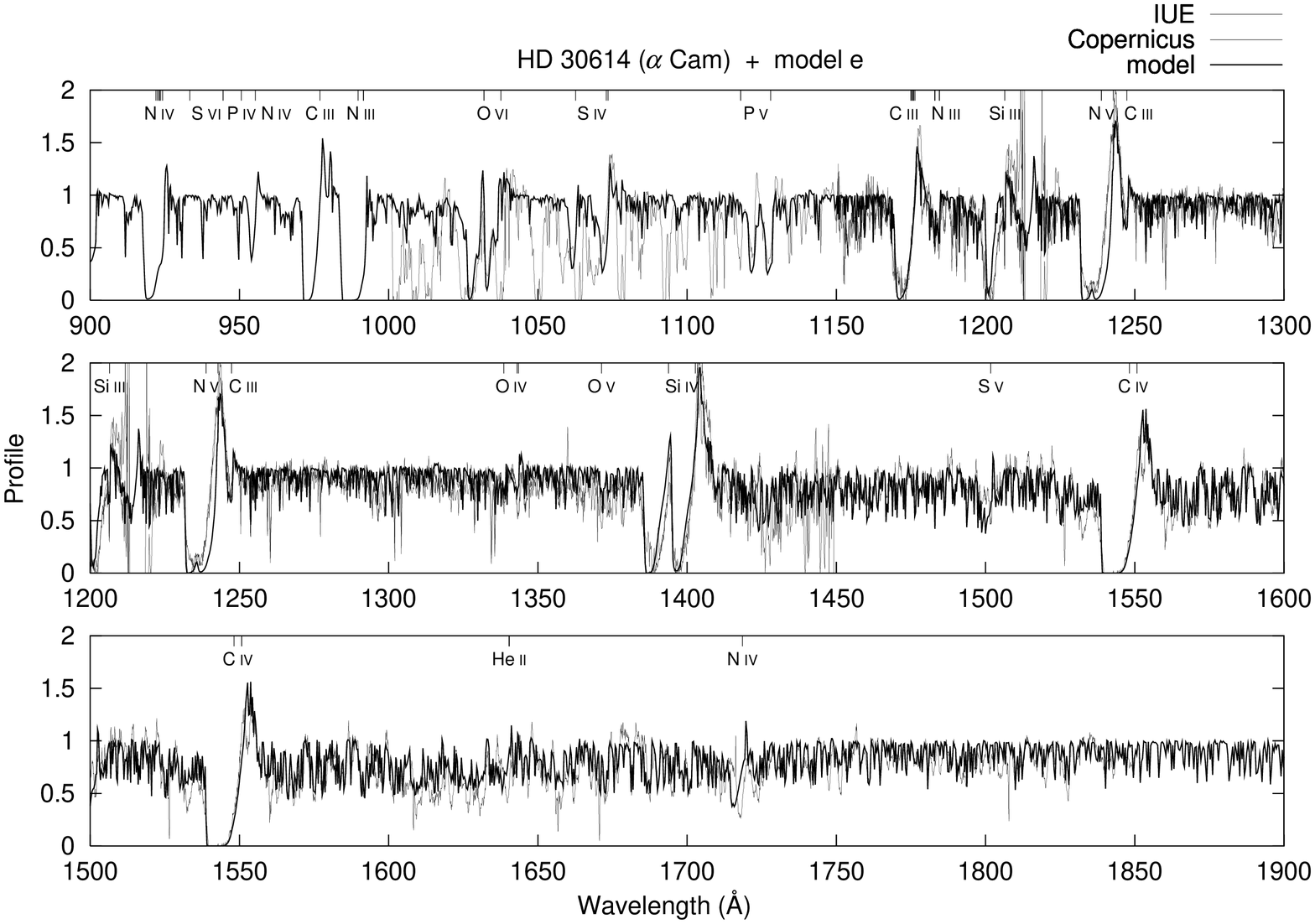}}
\mycaption{The complete EUV spectrum of model~e ($\Teff=30000\,{\rm K}$).}
\label{fig:alphacam-e1}
\centerline{\includegraphics[height=15cm,angle=-90]{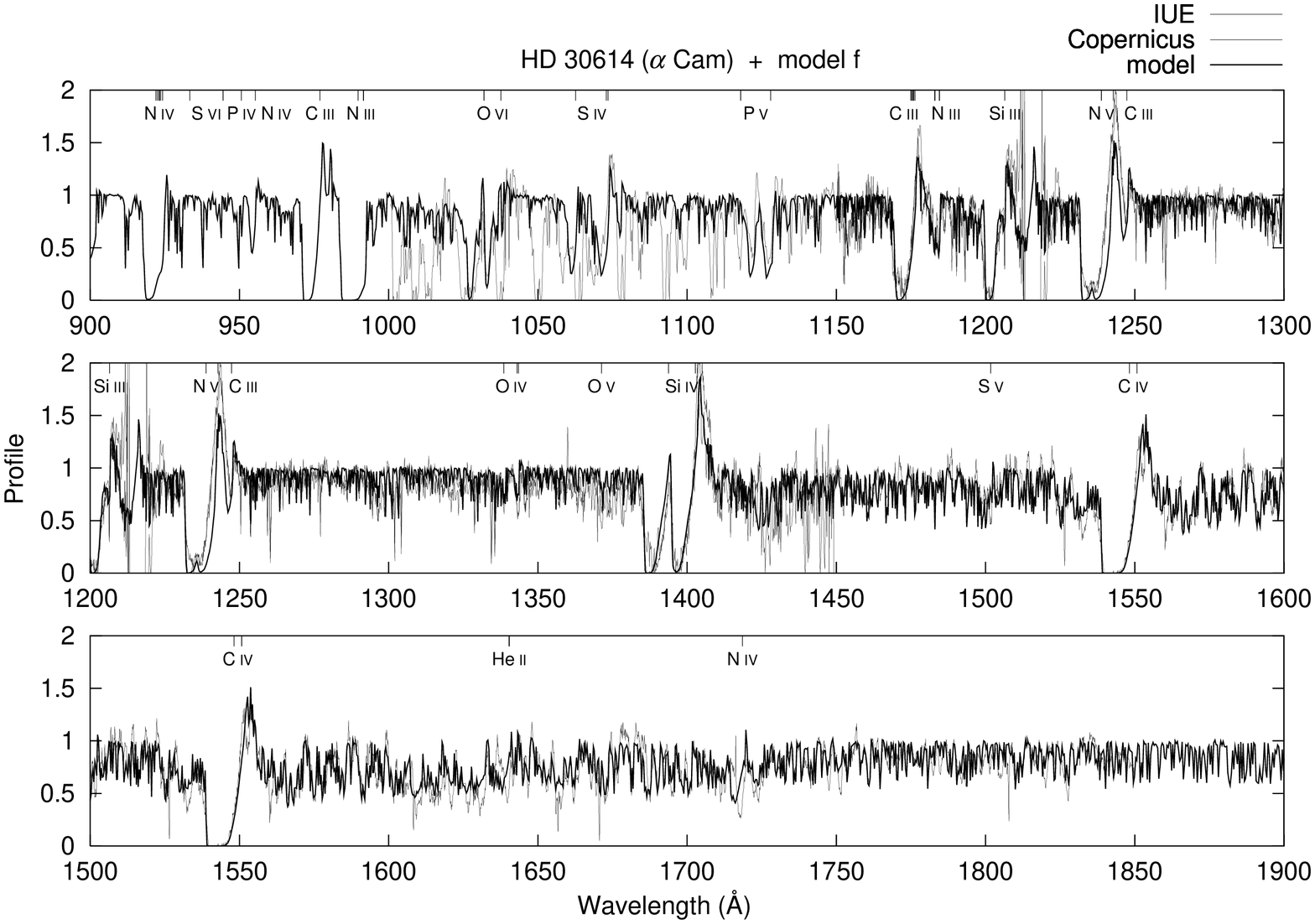}}
\mycaption{Model~f. Like model~e, but with a $\Teff$ of $29000\,{\rm K}$.
Note the better fit of \SiIII\ and \OIV.}
\label{fig:alphacam-f}
\end{figure*}

\begin{figure*}
\centerline{\includegraphics[height=15cm,angle=-90]{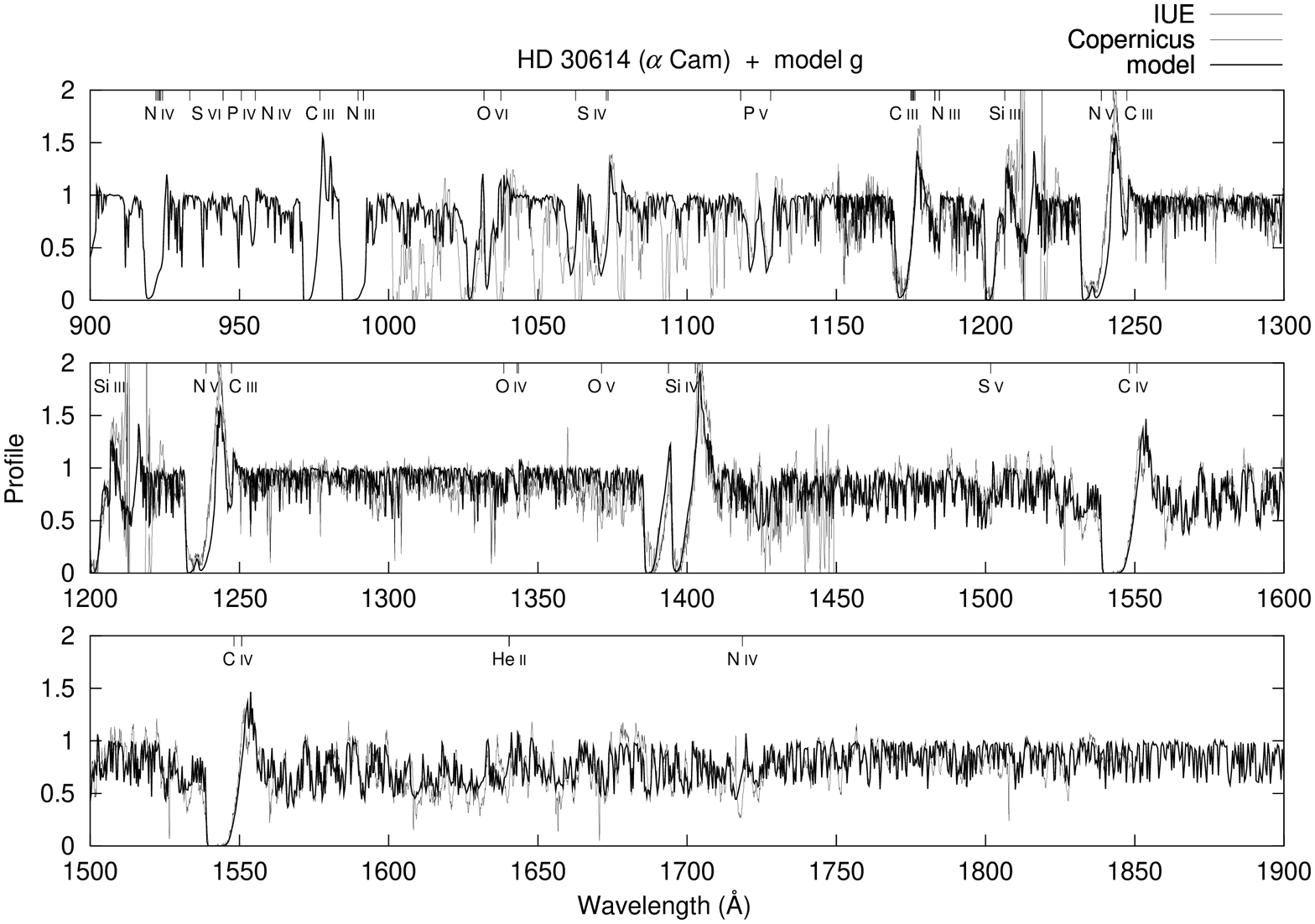}}
\mycaption{Model~g. Like model~f, but with adapted abundances (see
Table~\ref{tbl:alphacammodels}). This is our best fit.}
\label{fig:alphacam-g}
\centerline{\includegraphics[height=15cm,angle=-90]{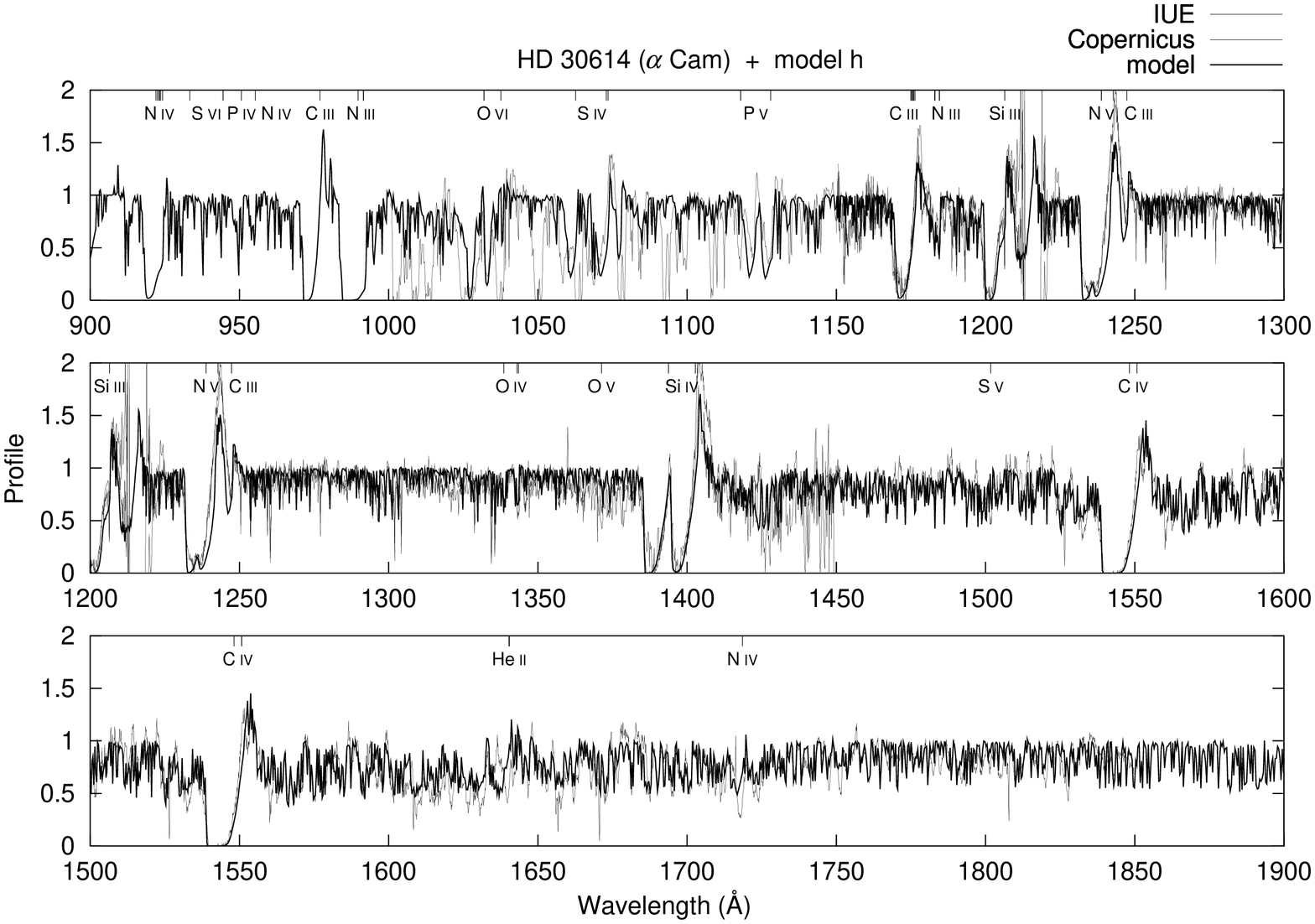}}
\mycaption{Model~h. Same as model~g, but with $\Teff=28000\,{\rm K}$}
\label{fig:alphacam-h}
\end{figure*}

\begin{figure*}
\centerline{\includegraphics[height=15cm,angle=-90]{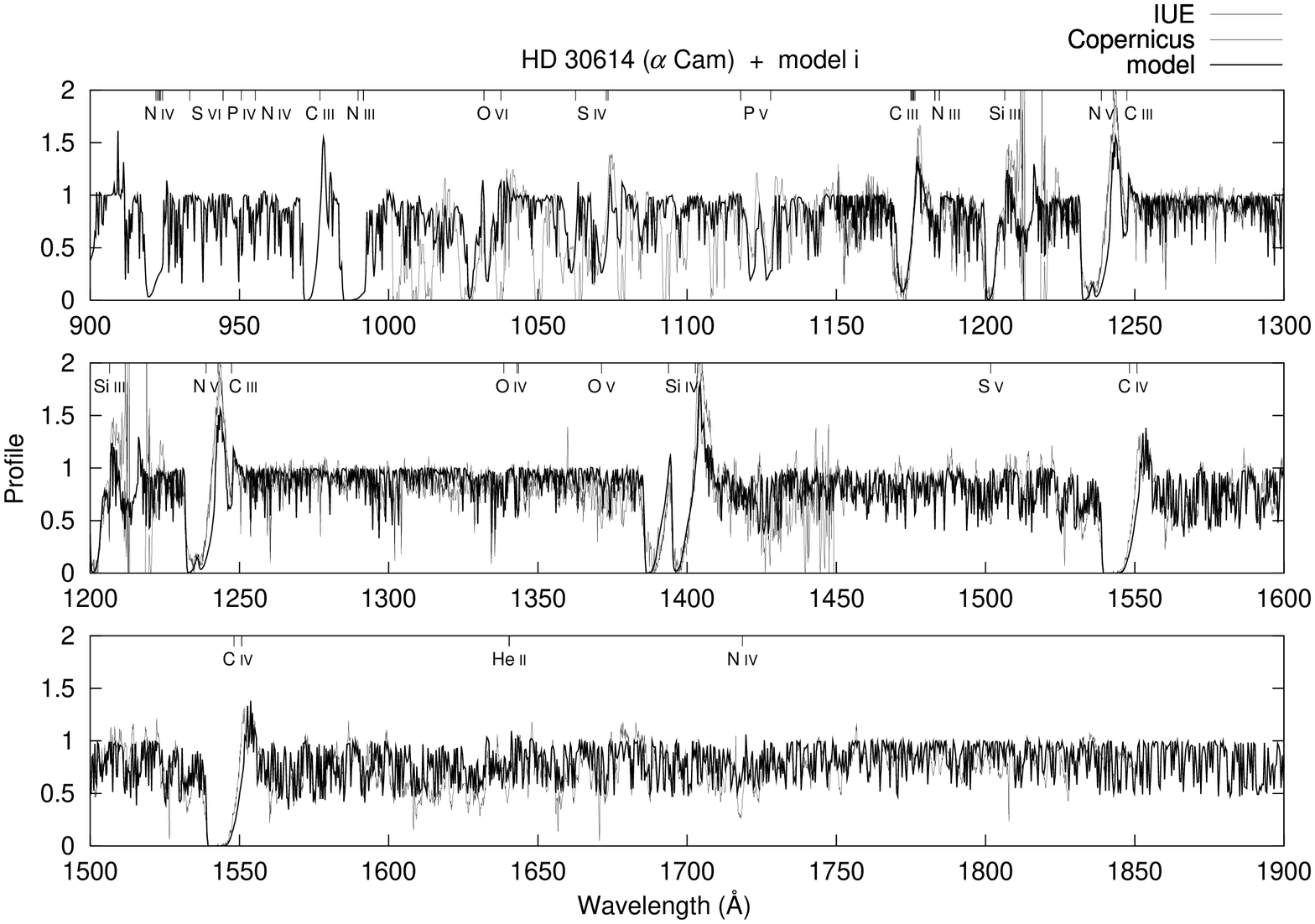}}
\mycaption{Model~i. As model~h, but with a lower mass loss rate
($\Mdot=2.5\times 10^{-6}\,\Msun/{\rm yr}$ compared to
$\Mdot=5\times 10^{-6}\,\Msun/{\rm yr}$ for model~h).}
\label{fig:alphacam-i}
\centerline{\includegraphics[height=15cm,angle=-90]{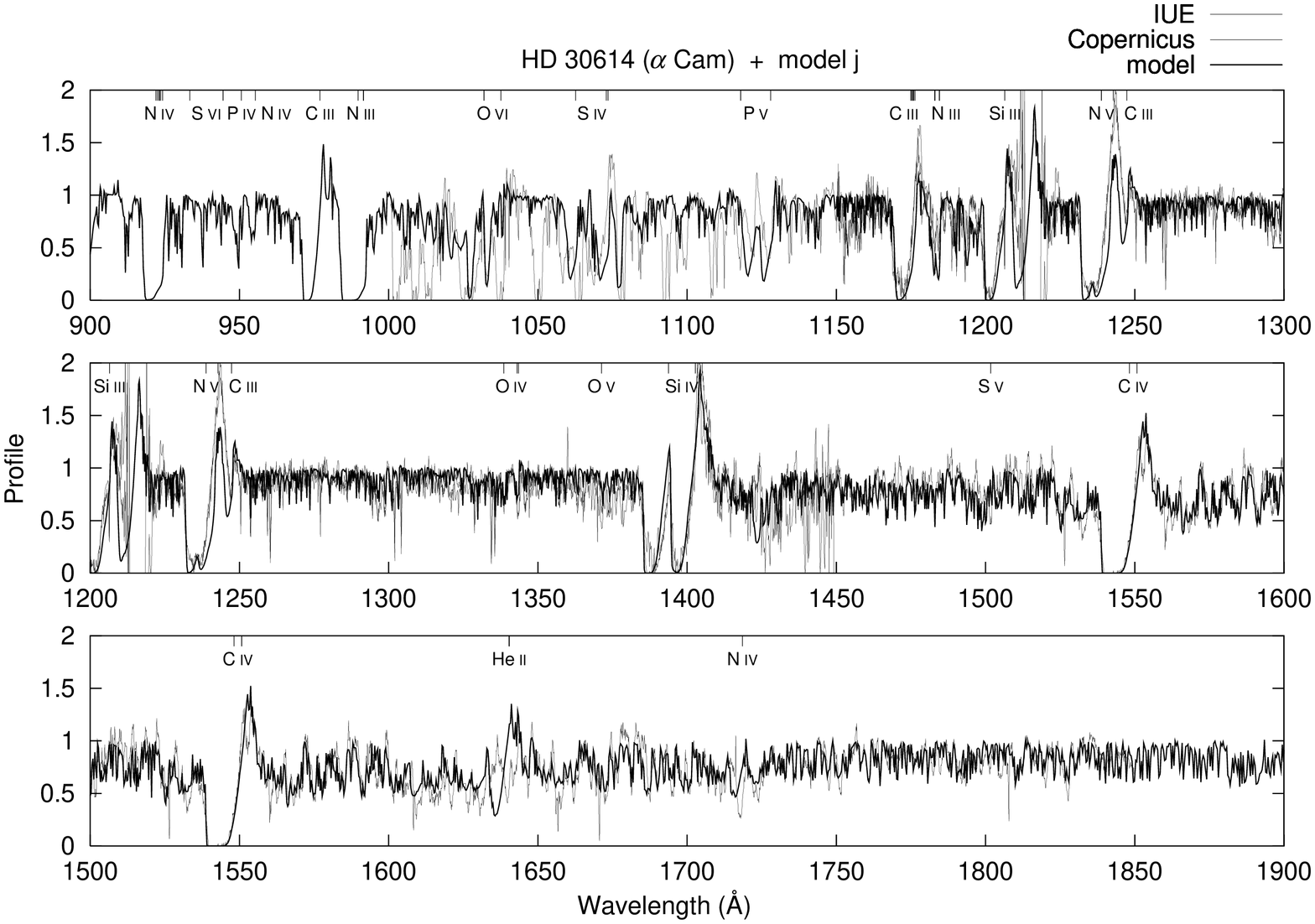}}
\mycaption{Model~j. As before, but with a higher mass loss rate
($\Mdot=10\times 10^{-6}\,\Msun/{\rm yr}$).
Note the change in the strength of the \HeII\ line.}
\label{fig:alphacam-j}
\end{figure*}

\begin{figure*}
\centerline{\includegraphics[height=15cm,angle=-90]{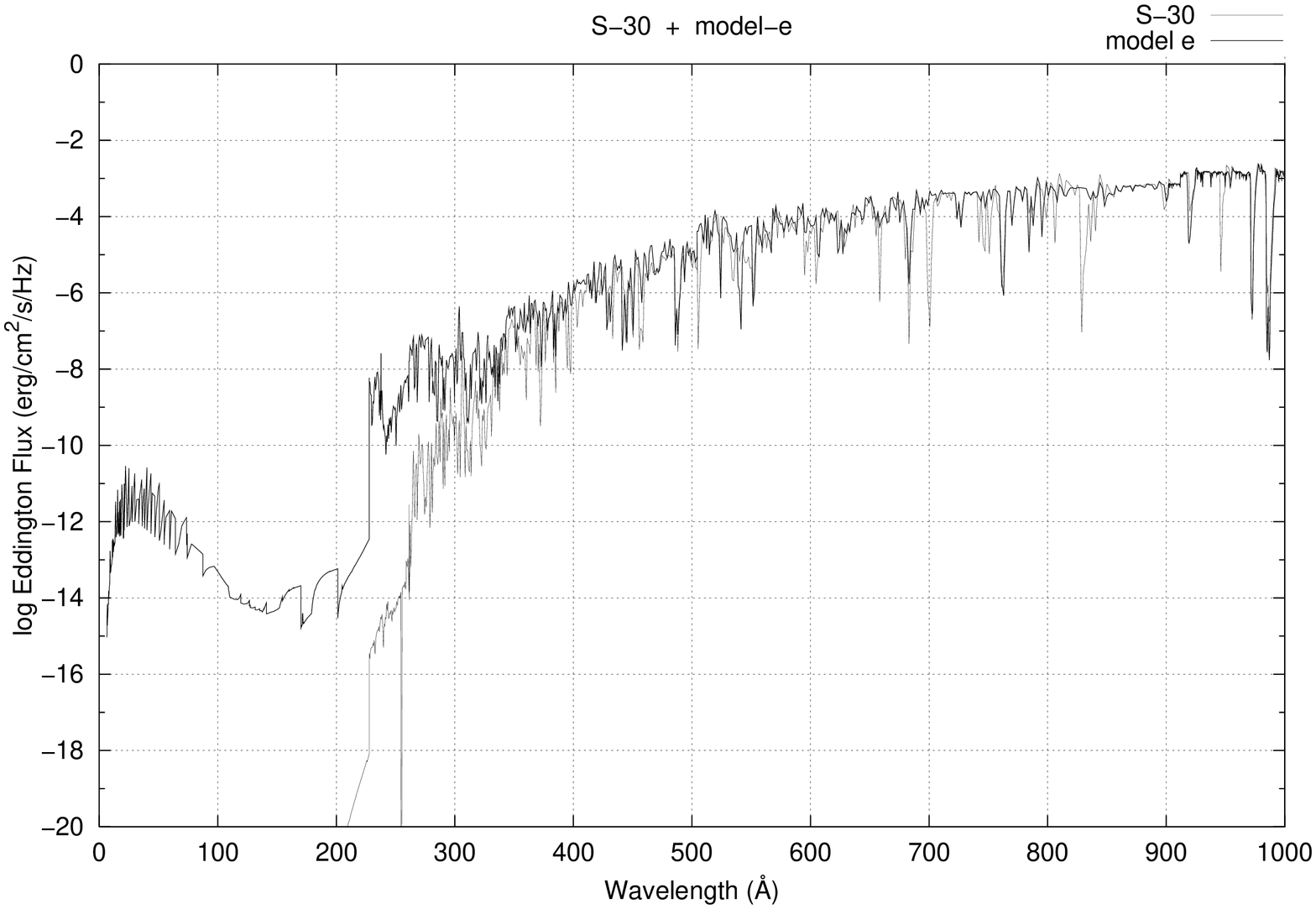}}
\mycaption{
Comparison of the ionizing flux of our S-30 grid star, where shocks
have been neglected, to that of model~e, showing the non-negligible
influence of shocks on the ionizing fluxes.}
\label{fig:compareshocks}
\end{figure*}

Fig.~\ref{fig:alphacam-d} shows a model sequence in which $m$ is
increased from 1 to 60; $\logLx$ has been increased compared to
model~c to provide for sufficient \OVI.  It can bee seen that in
model~d4 ($m=60$), where \SiIII\ retains its correct strength and the
absorption of \OVI\ and \NV\ is of the correct magnitude, the emission
of both latter lines has disappeared completely. This is because the
emission of P~Cygni lines due to resonance scattering arises to a
large part from the inner regions of the wind, where the stellar
radiation field is still strong; in this model however, \NV\ and \OVI\
hardly exist in these regions, since the shock radiation responsible
for the presence of these ions is only produced far out in the wind
and the ionization continuum is still optically thick at the
corresponding photoionization thresholds. We conclude that the onset
of shocks must lie further in, and, correspondingly, the shock
radiation must be harder (maximum at smaller wavelengths) than
attainable with a $\vturb$ value of 0.14, so as to have a
comparatively larger influence on \NV\ and \OVI\ via Auger-ionization
than on \SiIII\ via photoionization.

This reasoning is confirmed by model~e (Fig.~\ref{fig:alphacam-e}),
in which we have reduced $m$ to 30 and increased $\vturb$ to 0.25
(corresponding to a maximum shock temperature of $2.0\times 10^6\,{\rm
K}$). The shock parameters of this model yield the best overall
agreement of the most important UV spectral lines. Note that for this
model we have in addition reduced the phosphorus abundance to 0.05
times solar, to obtain a fit of the \PV\ line.  As it concerns only a
single element and since evidence for an underabundance of phosphorus
has already been encountered from an analysis of another hot star,
namely the O4f-star $\zeta$ Puppis (cf.~Pauldrach et al.,~1994; and
Section~\ref{sec:shocks}, this paper), it is most likely that the
discrepancy between the observed and calculated \PV\ resonance line is
caused by the proposed abundance effect. We further note that the
imbalance of \SIV\ and \SV\ is most likely due to a combination of
abundances and imperfect atomic data, as our \SIV\ atomic model has
not yet reached the quality of those of other ionization stages and is
still in a stage of rather incomplete description
(cf.~Table~\ref{tab1}). Note that we have also reduced the oxygen
abundance, thereby improving the fit of both \OIV\ and \OVI.

Having thus constrained the strength and distribution of the shocks,
we still need to check to what extent the effective temperature and
the mass loss rate can be constrained further through our analysis of
the UV spectrum.  For this purpose, we have computed a model with
$\Teff$ of 29\,000\,K but otherwise same parameters (model~f), whose
spectrum is shown in Figure~\ref{fig:alphacam-f}. Comparing this to
the spectrum of model~e (30\,000\,K) (Fig.~\ref{fig:alphacam-e1}) we
note a marginally better fit of the spectral region from 1450 to
1650~\AA, and a slight improvement in \SiIV\ and \OIV.  More
significant is the extreme sensitivity of the \SiIII\ and the \CIII\
line to this small change in temperature, which reveals that these
lines can be utilized as temperature indicators in this spectral
range.  We concede that the fit of \NV\ is still somewhat imperfect,
but point out that the shock model used is not in a final stage (see
above); we expect this to improve with our new method.  In a next step
we have reduced the carbon abundance further (to 0.05 solar) and
nitrogen to solar abundance, as this improved not only the \NV/\CIII\
fit, but also the \NIII\ line at 1183,1185~\AA\
(Fig.~\ref{fig:alphacam-g}).  Note also that the saturated \CIV\
resonance line is not affected by this change of the carbon abundance.

Lowering the effective temperature further to 28\,000\,K worsens the
fit, as the iron and nickel lines around 1500~\AA\ become too strong
(Fig.~\ref{fig:alphacam-h}).  As the strength of these lines depends
also on the mass loss rate, it is conceivable that an adaptation of
this parameter can again improve the fit. However, $\Mdot$ is strongly
constrained by the strength of the \HeII\ line, as demonstrated in
figures~\ref{fig:alphacam-i} ($\Mdot=2.5$) and~\ref{fig:alphacam-j}
($\Mdot=10$). We have confirmed in test calculations that the iron
abundance must be solar ($Z=1.0$) to reproduce the observed spectrum
and to account for the radiative acceleration needed to produce the
observed mass loss rate (cf.~\cite{}Pauldrach,~1987), a finding
compatible with our earlier statement concerning the relative
abundances of iron to the light CNO elements.

Finally, we wish to demonstrate the influence of the shocks on the
ionizing fluxes.  In Figure~\ref{fig:compareshocks} we compare the
ionizing flux of our S-30 supergiant grid model, where shocks have
been neglected, to that of our model~e for $\alpha$~Cam.  (We have
used model~e for this comparison, and not our final model~g, since
model~e and the grid star S-30 both have the same effective
temperature of 30\,000\,K.) Due to the lower optical depth redwards of
the \HeII\ edge, the shock radiation enhances the flux as far as
redwards as 400~\AA;  this will have important implications concerning
the solution of the \NeIII-problem (cf.~Sellmaier et al.~1996),
pending further investigation.  Note: the Zanstra integral for \HeII\
is increased from $\log Q_{\rm He^+}=6.39$ for the S-30 model to 15.07
for model~e!

\section{Conclusions}

After a long period of work in the areas of non-LTE radiative
transfer, hydrodynamics, and atomic physics we have now developed a
fast numerical model code for expanding atmospheres which incorporates
for the first time the physics required --- rate equations for all
ions using detailed atomic physics, the equations of radiation
hydrodynamics, the energy equation, the radiative transfer equation
including the effects of overlap of numerous spectral lines of
different ions, and shock emission of instabilities in the stellar
wind flow. We have shown that modelling hot star atmospheres involves
replicating the tightly interwoven mesh of these physical processes,
the most complicating effect in this system being the overlap of
thousands of spectral lines of different ions. Especially concerning
this latter point we have made significant progress; the decisive
factor has been to relax some rather severe approximations concerning
the correct treatment of Doppler-shifted line radiation transport and
the corresponding coupling with the radiative rates in the rate
equations. We have demonstrated that these modifications to the models
lead to changes in the energy distributions, ionizing continua, and
line spectra with much better agreement with the observed spectra.
This has important repercussions for the quantitative analysis of hot
star spectra. With this new method in hand we have already established
a new basic model grid of stars of solar metallicity that can be used
as input for the analysis of spectra of emission line nebulae.  The
qualitative diagnostic investigation performed on basis of the model
grid revealed further that the oxygen abundance is considerably
reduced compared to the solar value, at least for objects younger than
O5.

From the first detailed analysis of the O9.5 supergiant $\alpha$~Cam
we conclude that our spectrum synthesis technique does, in principle,
allow the determination of effective temperature and abundances --- in
fact, a determination of the effective temperature to within $\pm
1000\,{\rm K}$ and of the abundances to within a factor of 2 seems not
unreasonable.  Carbon and phosphorus show clear signs of an
underabundance on the order of one tenth its solar value, as does
oxygen with about 0.3 solar, whereas the abundance of iron must be
roughly solar to reproduce the spectrum of the numerous \FeIV\ and
\FeV\ lines.  To produce the ionization balance observed in the
lighter elements C, N, O, and Si, the influence of shock radiation
must start at larger radii where for shorter wavelengths the largest
shock temperatures dominate. Thus, the way the X-ray spectral region
selectively affects the ionization balance of different elements,
observable through the lines in the EUV spectrum, provides constraints
on the lower shock temperatures; we have determined maximum shock
temperatures on the order of $2.0\times 10^6\,{\rm K}$.  Especially
the \SiIII\ and \CIII\ lines have been found to be invaluable
diagnostic instruments for this purpose.

Our detailed analysis of the UV spectrum and the shocks needed to
reproduce the observed lines has led to a significant difference in
the ionizing flux of this model. Thus we conclude that this type of
analysis is indispensable and must be regarded as the ultimate test
for the accuracy of ionizing fluxes from models.

Our research plan for the future has three major objectives. First, we
will have to implement further improvements to the model atmosphere
code, especially concerning the planned analyses of optical lines. For
this purpose Stark-broadening has to be included for the affected
spectral lines (e.\,g., H and \HeI\ lines) and concerning the rate
equations instead of using the Sobolev-plus-continuum method some of
these lines should be treated in the comoving frame, if they are used
for diagnostic purposes (cf.~\cite{}Sellmaier et al.,~1993, who found
that compared to the comoving frame treatment the
Sobolev-plus-continuum approximation leads to a non-negligible change
of the strength of some H and He lines). In connection with this some
minor approximations will also have to be checked in detail.

Second, we plan to apply the model atmosphere code to a comprehensive
sample of stars of different metallicities for modelling ionizing
fluxes and line spectra, the qualitative review of the model grid
presented in this paper being just a first step for a detailed
quantitative analysis of ionizing fluxes and quantitative verification
of the accuracy and reliability of the models. Furthermore, spectral
analysis in the UV and the optical range of individual stellar objects
in Local Group galaxies and in galaxies as distant as the Virgo
Cluster will be performed.

Third, with the new diagnostic tool of detailed expanding model
atmospheres presented, and the observations of the HST space
observatory and the ESO VLT ground-based telescope which are already
available, the concept of using luminous hot stars for quantitative UV
spectral analyses for determining the properties of young populations
in galaxies is not only reasonable, but first steps in this direction
have already been taken (see \cite{}Mehlert et al.,~2000, who found
that the spectra of galaxies they had observed at high redshifts
($z\sim 3$) displayed the typical features usually found in the UV
spectra of hot stars; first diagnostic investigations have shown that
these spectra, considering proper reddening, can be well fitted with
synthetic spectra from our hot star models).  Thus, the calculation of
integrated spectra for the determination of stellar abundances and the
physical properties of the most UV luminous stars in star forming
galaxies even at high redshifts using the technique of population
synthesis is feasible.

  \acknowledgements{ We wish to thank our colleagues Dr.~J.~Puls,
Dr.~R.-P.~Kudritzki and Dr.~K.~Butler for helpful discussions and
Dr.~S.~Becker for help in improving the atomic data. It is a special
pleasure to thank Dr.~J.~Puls also for providing us with a basic
solver routine for the formal integral. This research was supported by
the Deutsche Forschungsgemeinschaft in the ``Gerhard Hess Programm''
under grant Pa~477/2-3.}

  \appendix
  \section{A new concept for a fast solution of the Rybicki-method}
\label{app:rybick}

Here we present for the solution of the Rybicki-scheme a concept which
is compared to the standard procedure 10 times faster on a vector
processor and 3 to 5 times faster on a scalar processor (see Sec.~3.3).

We start from the final system for the solution of the mean intensity
--- the vector $\vec J$ describes its depth variation (the number of
depth points is $N$):
\begin{equation}
\vec J=W^{-1}\vec Q
\end{equation}
with ($N'$ is the number of $p$-rays)
\begin{equation} \label{wq}
W = \tilde W -{\bf 1} = -{\bf 1} -\sum_{j=1}^{N'} T_j^{-1} U_j ,\quad
\vec Q = -\sum_{j=1}^{N'} T_j^{-1} \vec K_j
\end{equation}
where
\begin{equation}
T_j = T'_j w_j^{-1},\qquad
w_j=\left(\begin{array}{ccc}
w_{1,j} & & \\
        & \ddots & \\
        &        & w_{N_j,j}
\end{array}\right),
\end{equation}
and $N_j$ is the number of radius points for the $j\/$th $p$-ray
(cf.~Fig.~\ref{fig:pzgeo}).
$w_j$ is the diagonal matrix of the integration weights,
and $T'_j$ is a tri-diagonal matrix defined by the
coefficients of the difference equation of transfer
\begin{equation}
\label{eq:appatrf}
a_{i,j}u_{i-1,j}+b_{i,j}u_{i,j}+c_{i,j}u_{i+1,j}-\beta_i J_i= (1-\beta_i)S_i
\end{equation}
where $a_{i,j}$, $b_{i,j}$, and $c_{i,j}$ are the coefficients as in
eq.~\ref{eq:trfcoeff} (section~\ref{sec:opasamp}) for the $j\/$th $p$-ray,
\begin{equation}
T'_j=\left(\begin{array}{ccc}
b_{1,j} & c_{1,j}   & \\
\ddots  & \ddots    & \ddots \\
        & a_{N_j,j} & b_{N_j,j}
\end{array}\right).
\end{equation}
The variables $u_{i,j}$ are the symmetric averages of the intensities,
the coefficients $\beta_i$ are the ratios of Thomson-opacities to
total opacities, and the diagonal matrix $U_j$ and the vector $\vec
K_j$ are defined as
\begin{equation}
U_j=\left(\begin{array}{ccc}
-\beta_1 & & \\
         & \ddots & \\
         &        & -\beta_{N_j}
\end{array}\right)
\quad
\vec K_j=\left(\begin{array}{c}
(1-\beta_{\>1\>\>})S_{\>1\>\>} \\ \vdots \\ (1-\beta_{N_j})S_{N_j}
\end{array}\right).
\end{equation}
The usual, but time consuming solution method is to calculate the
inverse matrices $T_j^{-1}$ --- by a forward-elimination and
back-substitution procedure --- and from these, obtaining $\vec Q$ and
$W$, and, finally, $\vec J$.

Due to the diagonal character of the matrix $U_j$ this is, however,
not necessary, because from the solution of the first column of
$\tilde W$ obtained by using just the first column of $U_j$ in
eq.~\ref{wq} the solutions of the remaining columns of $\tilde W$ can
be generated. Hence, the elimination procedure only has to be applied
to the following two sets of equations
($N'_i$ is the maximum number of $p$-rays for the $i\/$th radius point):
\begin{equation}  \label{wq1}
Q_i = -\sum_{j=1}^{N'_i} \tilde K_{j,i} 1
\end{equation}
\begin{equation}  \label{wq2}
\tilde W_{i,1}= -\sum_{j=1}^{N'_i} \tilde U_{j,i}=
                -\sum_{j=1}^{N'_i} \tilde B_{j,i} V_{j,1}
\end{equation}
with
\begin{equation}
\tilde{\vec K_j} = T_j^{-1} \vec K_j ,\qquad
\tilde{\vec B_j} = T_j^{-1} \vec B_j
\end{equation}
where
\begin{equation}   \label{wb1}
V_{j,1}=\left(\begin{array}{c}
-\beta_1  \\ \vdots  \\ -\beta_1
\end{array}\right)
\qquad
\vec B_j=\left(\begin{array}{c}
1 \\ 0 \\ \vdots \\ 0
\end{array}\right).
\end{equation}
Note that since the solution procedure is not recursive with respect
to the index $j$ the elimination can be performed simultaneously for
all $p$-rays. Note further that the structure of the sums in
eqs.~\ref{wq1} and~\ref{wq2} are equivalent to the operations $x =
A\cdot y$ where $A$ is a triangular matrix. Thus, {\sc blas} level-2
routines can be applied.

For the construction of the remaining columns of $W$ we now make use
of the already calculated matrix $\tilde B_{j,i}$ and two auxiliary
matrices obtained as a byproduct during the forward-elimination
procedure by solving eq.~\ref{wq2}.

The first one is:
\begin{equation}  \label{wq3}
F'_{j,i} = F'_{j,i-1}f_{j,i},
\qquad
i=1,\dots,N-1
\end{equation}
with
\begin{equation}  \label{wq4}
F'_{j,0} = 1,
\qquad
f_{j,i} = -a_{j,i+1}/\tilde b_{j,i}.
\end{equation}
Here $\tilde b_{j,i}$ is the updated value of $b_{j,i}$ obtained after
the forward-elimination step by solving eq.~\ref{wq2} ($\tilde
b_{j,i+1}=b_{j,i+1}+f_{j,i}c_{j,i}$).

The second one is:
\begin{equation}  \label{wq5}
F''_{j,i} = F''_{j,i+1}g_{j,i},
\qquad
i=N-1,\dots,1
\end{equation}
with
\begin{equation}  \label{wq6}
F''_{j,N} = 1,
\qquad
F''_{N'_i,i} = 1,
\qquad
g_{j,i} = -c_{j,i}/\tilde b_{j,i}.
\end{equation}

The final step --- construction of the columns $l=2,\dots,N$ of $W$
(where $l$ denotes the position of the unity value for the vector
$\vec B_j$ in eq.~\ref{wb1}) --- consists of two parts, where the
first part replaces the forward-elimination and the second part the
back-substitution procedure:

In the first part the components $i=l,\dots,N$ of the $l\/$th column of
$\tilde W$ are determined by
\begin{equation}  \label{wq7}
\tilde W_{i,l}= -\sum_{j=1}^{N'_i} \tilde B_{j,i} V'_{j,l}
\end{equation}
where
\begin{equation}  \label{wq8}
V'_{j,l} = V_{j,l}/F'_{j,l-1}.
\end{equation}
Again, a triangular matrix has simply to be multiplied by a vector using
a {\sc blas} level-2 routine.

In the second part the components $i=l-1,\dots,1$ of the $l\/$th column
of $\tilde W$ are determined by
\begin{equation}  \label{wq9}
\tilde W_{i,l}= -\sum_{j=1}^{N'_l} F''_{j,i} V''_{j,l}
\end{equation}
where
\begin{equation}  \label{wq10}
V''_{j,l} = V'_{j,l} \tilde B_{j,l}/F''_{j,l}.
\end{equation}
Now, a rectangular matrix has simply to be multiplied by a vector
again using a {\sc blas} level-2 routine.  Hence, instead of a number
of operations $\sim N^3$ a number of operations only $\sim N^2$ must
be performed, and as extremely fast routines are used to solve the
non-recursive system, the solution of the Rybicki-scheme is now almost
as fast as the solution of the moments equation.

  \section{A simple demonstration of the failure of the
standard $\tau$ discretization of the equation of transfer}
\label{app:discret}

Let us assume a one-dimensional, sharply bounded cloud of material of
moderate opacity,
\begin{equation}
\chi = \left\{
\begin{array}{ll}
1,        & \quad |z| < 5 \\
10^{-10}, & \quad |z| \ge 5
\end{array}
\right.
\end{equation}
with a symmetric boundary condition at $z=0$ (the middle of the
cloud) and no influx from the outside (i.\,e., $I^-=0$ at $z=10$).

\begin{figure}
\centerline{\includegraphics[height=\columnwidth,angle=-90]{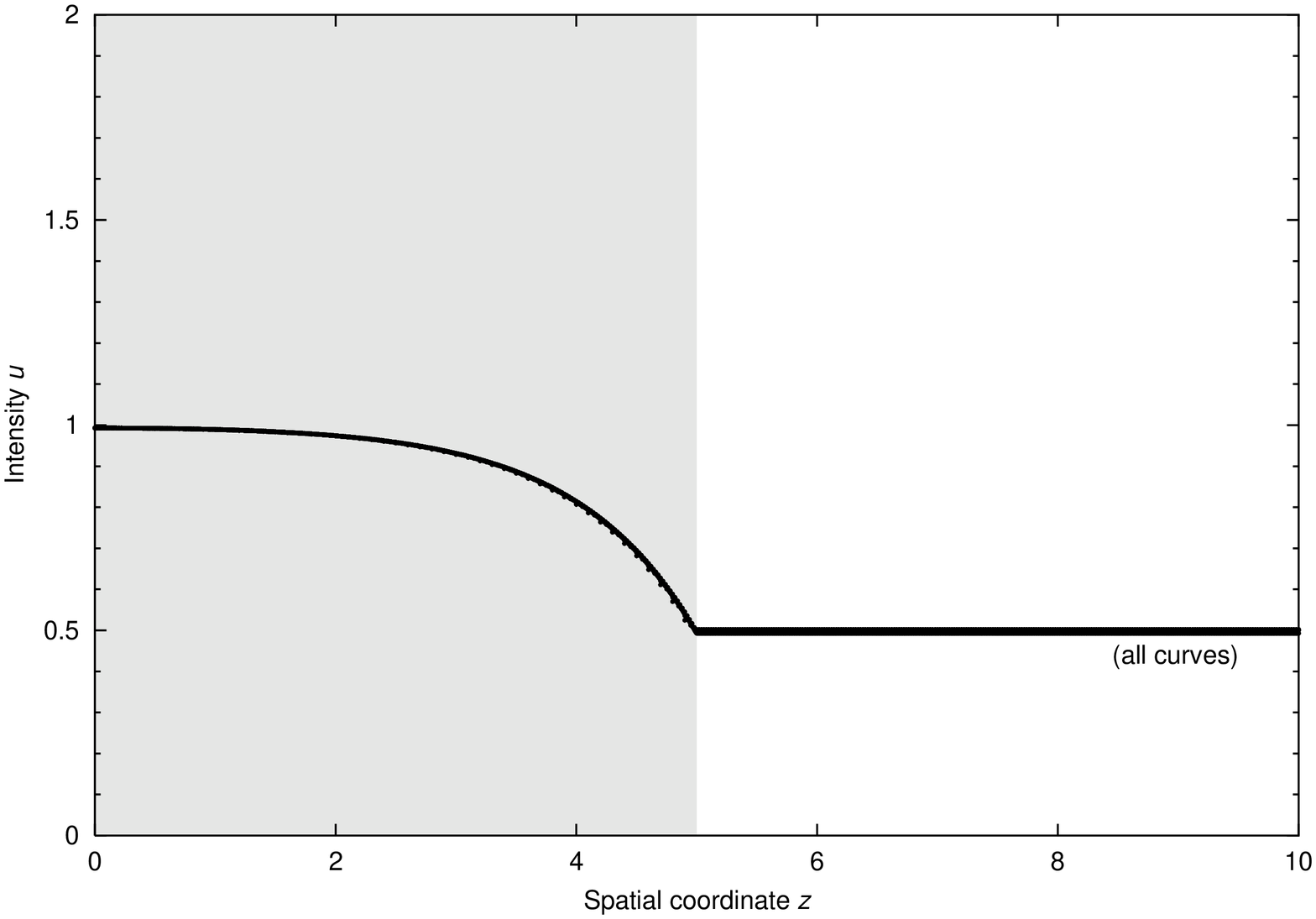}}
\centerline{\includegraphics[height=\columnwidth,angle=-90]{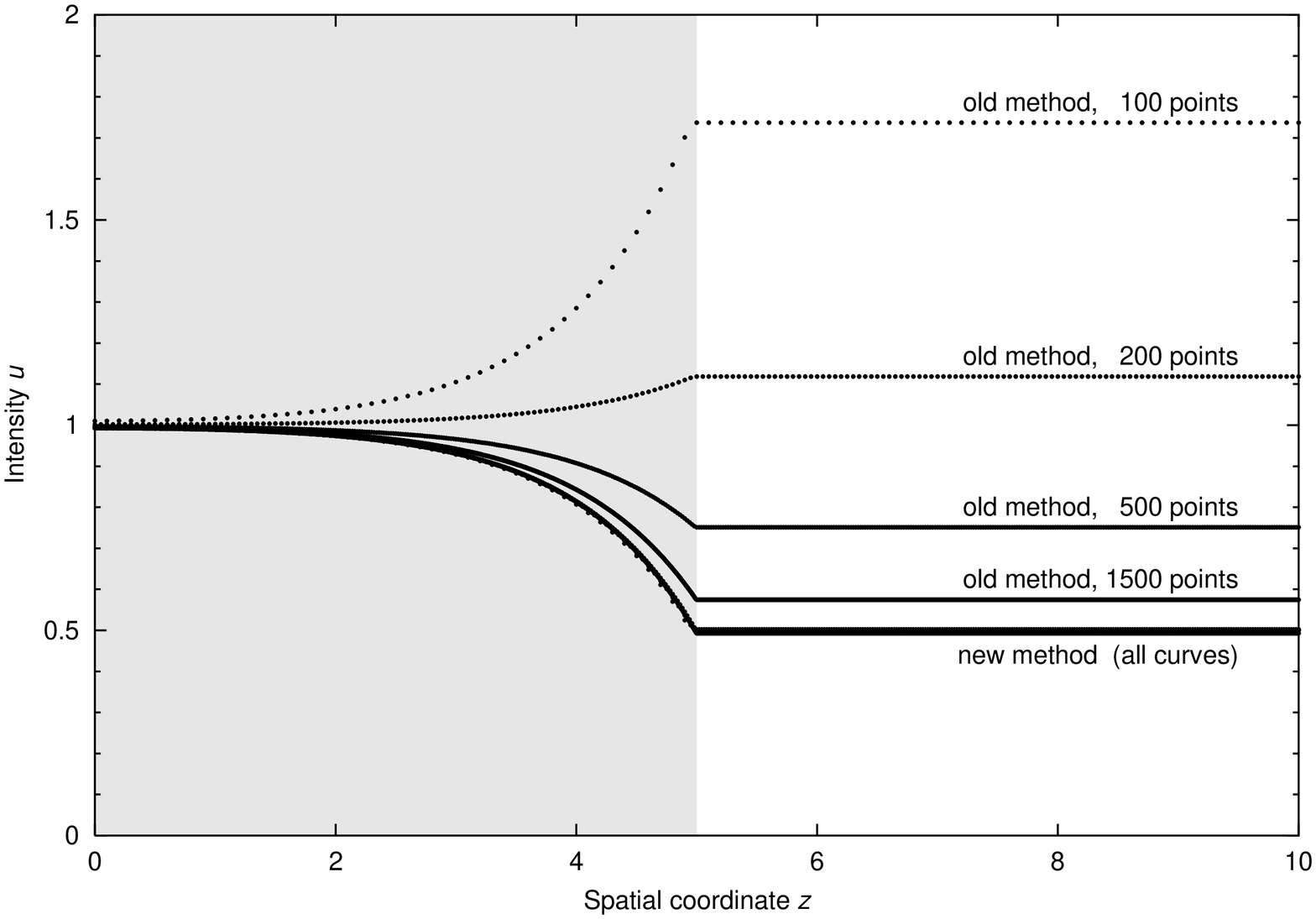}}
\mycaption{
{\bf Upper panel:}
Mean intensity in a one-dimensional nebula; constant source function,
high opacity on the left (shaded in gray), low opacity on the right.
Both discretization methods give the same results.
{\bf Lower panel:}
If the first grid point with low opacity is given a high source
function, the old method produces artificial emission. The new method
does not show this behavior.}
\label{fig:discretdemo}
\end{figure}

With a constant source function $S=1$ everywhere, both the old and the
new discretization of the transfer equation yield essentially the same
radiation field (upper panel of Figure~\ref{fig:discretdemo}), a
solution which is immediately obvious: The center of the cloud ($z=0$)
is optically thick, the intensity there thus equal to the source
function. Towards the edge of the cloud, more and more radiation
escapes, and the intensity decreases.  On the outside of the cloud,
essentially no emission is produced (nothing is absorbed, either); only
the radiation emitted by the cloud contributes to the local intensity
(since we have no influx from the right), therefore $u={1\over2}S$.

The results change considerably if we adopt a high source function of
$S=100$ for the {\em single grid point} at $z=5$, the first grid point
{\em outside} of the cloud.
As illustrated in the lower panel of Figure~\ref{fig:discretdemo}, the
old method produces extra emission on a scale of
\begin{eqnarray}
\Delta I \, = \, {1\over2} S \cdot \Delta\tau
& \approx & {1\over2}100 \cdot {1\over2}(1+10^{-10})\Delta z\\
& \approx & {1\over4} 100 \Delta z.
\end{eqnarray}
For the grid with 100 points, $\Delta z=0.1$, giving us an intensity
\begin{equation}
u \, \approx \, 0.5+{1\over2}\cdot{1\over4}100\cdot 0.1 \, = \, 1.75
\end{equation}
outside the cloud.
($0.5$ is from the normal emission of the cloud as before, and the
factor $1/2$ in the second term is again due to the fact that
$u={1\over2}(I^++I^-)$, with $I^-=0$ on the outside of the cloud, so
that $u={1\over2}I^+$.) The extra emission is proportionally reduced
in the finer grids as $\Delta z$ gets smaller.

The new method, on the other hand, is essentially indifferent to the
grid spacing in this particular configuration, since the emission is
only computed on the basis of the {\em local} opacity, which is low
for the point in question.  If we were to choose a point inside the
cloud to have a high source function, then both methods again produce
similar results, also dependent on grid spacing, since in this case
the point in question {\em does} have a high opacity, and therefore
produces substantial emission proportional to the interval length.

Concluding, it can be said that the discretization methods differ in
the assumptions that are made about the emissivity (in the old method,
based on average values of opacity and source function assuming linear
run between the grid points; in the new method, extrapolating the
local value) --- information not present given just the values at the
grid points, and which {\em must} therefore be supplied separately
through the choice of discretization coefficients. It just so happens
that the new method produces results which are more compatible with
our detailed formal integral.

\flushleft

\end{document}